\documentclass[a4paper,11pt]{article}
\usepackage{jcappub} % for details on the use of the package, please see the JINST-author-manual
\usepackage{lineno}
\usepackage[T1]{fontenc}
\usepackage{graphicx}
\usepackage{amsmath}
\usepackage{amssymb}
\usepackage{mathtools}
\usepackage[dvipsnames]{xcolor}
\usepackage{dsfont}
\usepackage{microtype}
\usepackage{booktabs}
\usepackage{mleftright}
\usepackage{subcaption}
\usepackage{placeins}
\usepackage{siunitx}
\usepackage[ISO]{diffcoeff}
\usepackage{dashbox}
\usepackage{empheq}
\usepackage{enumitem}
\usepackage{changepage}
\usepackage{tabularx}

\usepackage{hyperref}
\usepackage[capitalise, noabbrev]{cleveref}

\makeatletter
\input{aas_macros.sty}
\let\jnl@style=\relax
\makeatother

\arxivnumber{2504.21133}

\title{\boldmath Analytic Model for Covariance Matrices of the 2-, 3-, and 4-Point Correlation Functions in the Gaussian Random Field Approximation}

\author[1]{Jessica Chellino,}
\author[1]{Alessandro Greco,}
\author[2]{Simon May,}
\author[1,3]{and Zachary Slepian}
\affiliation[1]{Department of Astronomy, University of Florida,\\211 Bryant Space Science Center, Gainesville, FL 32611, USA}
\affiliation[2]{Perimeter Institute for Theoretical Physics,\\31 Caroline Street North, Waterloo, ON N2L 2Y5, Canada}
\affiliation[3]{Physics Division, Lawrence Berkeley National Laboratory,\\1 Cyclotron Road, Berkeley, CA 94709, USA}

\emailAdd{jchellino@ufl.edu}

\makeatletter
\let\ftype@table=\ftype@figure
\makeatother

\DeclareSIUnit{\year}{yr}
\DeclareSIUnit{\Gyr}{\giga\yr}
\DeclareSIUnit{\pc}{pc}
\DeclareSIUnit{\kpc}{\kilo\pc}
\DeclareSIUnit{\Mpc}{\mega\pc}
\DeclareSIUnit{\Gpc}{\giga\pc}
\DeclareSIUnit{\hHubble}{\text{\ensuremath{h}}}
\DeclareSIUnit{\Msun}{\text{\ensuremath{M_\odot}}}

\crefname{section}{\S}{\S}
\crefformat{section}{\S#2#1#3}
\crefmultiformat{section}{\S#2#1#3}{ and~\S#2#1#3}{, \S#2#1#3}{ and~\S#2#1#3}
\crefname{equation}{equation}{equations}
\newcommand*{\eqlabelleft}{(}
\newcommand*{\eqlabelright}{)}
\NewDocumentCommand{\pcref}{m}{%
    \begingroup%
    \renewcommand*{\eqlabelleft}{}%
    \renewcommand*{\eqlabelright}{}%
    \cref{#1}%
    \endgroup%
}
\creflabelformat{equation}{\eqlabelleft#2#1#3\eqlabelright}

\renewcommand{\vec}[1]{\boldsymbol{#1}}

\newcommand{\dD}{\delta_{\mathrm{D}}^{[1]}}
\newcommand{\kD}{\delta^{\mathrm{K}}}

\newcommand{\vs}{\vec s}

\newcommand{\tj}[3]{\begin{pmatrix} {#1} & {#2} & {#3}\\ 0 & 0 & 0\end{pmatrix}}
\newcommand{\sj}[6]{\begin{Bmatrix} {#1} & {#2} & {#3}\\ {#4} & {#5} & {#6}\end{Bmatrix}}
\newcommand{\nj}[9]{\begin{Bmatrix} {#1} & {#2} & {#3}\\ {#4} & {#5} & {#6}\\ {#7} & {#8} & {#9} \end{Bmatrix}}

\newcommand{\rom}[1]{\MakeUppercase{\romannumeral #1}}
\newcommand{\twoFone}{{}_2{F_1}}
\newcommand{\threeFtwo}{{}_3{F_2}}

\newcommand{\MeijerGTwoTwoThreeThree}[7]{G^{2,2}_{3,3}\mleft({#1} \middle| \begin{matrix}{#2}, & {#3}, & {#4} \\ {#5}, & {#6}, & {#7} \end{matrix}\mright)}
\newcommand{\MeijerGTwoOneTwoTwo}[5]{G^{2,1}_{2,2}\mleft({#1} \middle| \begin{matrix}{#2}, & {#3}\\ {#4}, & {#5}\end{matrix}\mright)}

\definecolor{midgreen}{RGB}{0,150,0}

\definecolor{newpurple}{RGB}{128,0,188}

\definecolor{neworange}{RGB}{255,105,0}

\abstract{Analyses of the galaxy N-Point Correlation Functions (NPCFs) have a large number of degrees of freedom, meaning one cannot directly estimate an invertible covariance matrix purely from mock catalogs, as has been the standard approach for the 2PCF and power spectrum. Instead, templates are used based on assuming a Gaussian Random Field density with the true, Boltzmann-solver-computed power spectrum. The resulting covariance matrices are sparse but have notable internal structure. To understand this structure better, we seek a fully analytic, closed-form covariance matrix template, using a power law power spectrum $P(k) \propto 1/k$ and including shot noise. We obtain a simple closed-form solution for the covariance of the 2PCF, as well as closed-form solutions for the fundamental building blocks (termed ``$f$-integrals'') of the covariance matrices for the 3PCF, 4PCF, and beyond. We achieve single-digit percent level accuracy for the $f$-integrals, confirming that our power spectrum model is a suitable alternative to the true power spectrum. In the $f$-integrals, we find that the greatest contributions arise when closed triangles may be formed. When $f$-integrals are multiplied together, as needed for the covariance, the number of non-vanishing configurations reduces. We use these results to present a clearer picture of the covariance matrices' structure and sparsity, which correspond to triangular and non-triangular regions. This will be useful in guiding future NPCF analyses with spectroscopic galaxy surveys such as DESI, Euclid, Roman, and SPHEREx.}

\begin{document}

\maketitle
\flushbottom

\section{Introduction}
\label{sec:intro}

Recent years have seen significant interest in the use of higher-order spatial clustering statistics, such as N-Point Correlation Functions, to extract cosmological information from the large-scale distribution of galaxies. An NPCF characterizes the excess over a random Poisson noise in the spatial clustering of sets of $N$ galaxies at a time. The 2PCF and 3PCF (along with their Fourier space counterparts, the power spectrum and bispectrum) have been used to constrain the cosmic expansion history and growth rate \cite{gilmarin_boss_bao_growth_rate, gil_marin_rsd_pk_bispec, gil_marin_pk_bispec_sdss, beutler_rsd_pk, philcox_boss_dr12_pk_bispec, reid_boss_growth, massara_simbig_marked_pk, ivanov_bispec_boss, estrada_cosmology_behind_mask, veropalumbo_23pcf_vipers, mohammad_vimos_growth_rate, pezzotta_vimos_growth_rsd, beutler_6df_growth_rate, gagrani_bispec, novellmasot_fullshape_desi, qin_zspace_pk_sdssv}, reveal information about primordial non-Gaussianity \cite{brown_png_23pcf, damico_png_boss, rezaie_png_desi, jeong_png_bispec, ross_boss_png, giri_fnl, shirasaki_png_bispec, gualdi_bispec_anisotropic, goldstein_fnl_bispec, novell_masot_geofpt, rossiter_png_bispec}, detect baryon acoustic oscillations (BAO) for use as a standard ruler \cite{gilmarin_boss_bao_growth_rate, xu_2pcf_cov, SE_3PCF_BAO, bao_bispectrum, matt_bao, percival_bao_sdss, anderson_bao_sdss, ross_bao_sdss, abbott_des_bao, moon_bao_desi, behera_bao_bispec, cuesta_sdss_bao_in_cf, hou_sdss_bao_rsd, slepian_constraining_baryon_dm, slepian_largescale_3pcf_sdss, slepian_signature_baryon_dm, moresco_c3_bao, eisenstein_detection_bao, beutler_6df_bao, child_bispec_bao_interferometer}, probe the relationship between galaxies and halos \cite{gal_halo_connection, squeezed3PCF, hikage_constraining_hod_growth_pk, fosalba_3pcf_halomodel, gao_desi1_galhalo, yuan_assembly_bias, hadzhiyska_galhalo, scoccimarro_galaxy_fit_halo, neyrinck_halo_pk, neyrinck_info_content2, guo_clustering_sdss, wang_3pcf_hod, gao_elg_correlations, eke_pk_dmhalo}, and constrain the neutrino mass \cite{farshad_neutrino, bayer_neutrino, garay_neutrino, cuesta_neutrino, mv_bispectrum_1, mv_bispectrum_2, demnuni, mv_tree_level_bispec}. The 3PCF and 4PCF (or their Fourier counterparts) can also be used to quantify magnetohydrodynamic turbulence within the interstellar medium \cite{portillo_ism, obrien_ism, burkhart_ism, victoria_mhd_ism_4pcf}. The connected 4PCF produced by gravitationally-induced non-Gaussianity \cite{bartolo_nongaussianity} has also recently been detected \cite{connected}. 

Various methods have been developed to compute NPCFs (or polyspectra) efficiently. These include using Fourier Transforms (FTs) \cite{ZS_23pcf_FTs, zs_los, sarabande, NPCF_Ddim, conker, fast_accurate_comp, guidi_3pcf_fftlog, philcox_polybin3d, lee_ang_trispec}, spherical Fourier-Bessel modes \cite{benabou_bispec_sfb_basis, wen_pk_sbf, khek_sfb_lss, grasshorn_sfb_pk_ezmocks, semenzato_sfb_pk_gr}, graph databases \cite{graph_database, gramsci}, probability maps \cite{demina_2pcf_probmap}, weighted pair counts \cite{computing_small_scale_pk, faster_ft}, neural networks \cite{pk_neuralnetwork, bai_fremu, matryoshka, matryoshka2, pkann1, pkann2}, multipole moments, spherical harmonics, and isotropic basis functions \cite{ZS_3pcf_N2, umeh_optimal_3pcf, sugiyama_fft_bispec, sugiyama_aniso_23pcf, wang_triumvirate, slepian_practical_3pcf, encore}. On the modeling side, efforts have been made to evaluate NPCFs with perturbation theory \cite{novell_masot_geofpt, zs_model_3pcf, bertolini_trispec_eftlss, ortola_leonard_4pcf, ivanov_precision_bispec, matsubara_integrated_pt, hashimoto_bispec_pt_1loop, wang_pt_remixed, wang_pt_remixed2}, including effective field theory \cite{steele_trispec_eft, alkhanishvilli_pt_bispec, baldauf_bispec_eft, damico_oneloop_bispec_eft, steele_1loop_bispec_eft, bertolini_eft_lss, baldauf_2loop_bispec, calderon_primordial_features_eftlss, zhang_fullshape_pk_recon}. Efficiently using a varying line of sight has also been explored \cite{karol_los_3pcf, beyond_yamamoto}.

A major challenge in higher-order spatial clustering statistics has been the dimensionality of the observables, which ranges from hundreds to thousands. The typical method of determining the covariance matrix (or its inverse \cite{padmanabhan_sparse_precision, friedrich_precision_expansion}) from mock catalogs breaks down for analyses with many degrees of freedom, since several mocks are needed per degree of freedom to render the matrix smoothly invertible \cite{looijmans_shrinkage, jackknife_mocks, grieb_gauss_cov_aniso, hartlap_unbiased_inverse_cov, joachimi_shrinkage, pope_shrinkage, sellentin_heavens_parameter_inference, taylor_preicision_cosmology_covariance, sellentin_lost_info_cov}. The covariance matrix is a critical piece of any analysis as uncertainties within the covariance propagate to estimates of cosmological parameters \cite{sellentin_heavens_parameter_inference, dodelson_cov_error, percival_clustering_cov_matrix_errors, taylor_joachimi_estimating_cov}. This becomes particularly important for analyses investigating detection of a novel signal, such as large-scale parity violation with the 4PCF \cite{cahn_parity, hou_parity, philcox_parity, kendrick_no_parity}.

Many efforts have been made to efficiently compute covariance matrices. Often, the covariance is estimated with jackknife resampling \cite{jackknife_mocks, jack_boot_estimate, large_cov_matrices, rascalc, 2pcfcov_fewer_mocks, op_estimating_cov}, shrinkage estimation \cite{looijmans_shrinkage, joachimi_shrinkage, pope_shrinkage}, the response approach \cite{barreira_response_pk_cov, barreira_cov_response, barreira_bispec_cov_response}, or perturbation theory \cite{wadekar_pt, lacasa_cov_pk, kobayashi_cov_pk, bertolini_cov_pk_eft, mohammed_pt_cov_pk, hikage_cov_pk, sugiyama_pt, zhao_cov_pk_bao_recon}. The integrals required for covariance calculations may be accelerated with Fast Fourier Transforms (FFTs) \cite{kobayashi_cov_pk, fang_2dfftlog}. 

There have also been a number of approaches to reduce noise in the covariance. The covariance can be made less noisy with data compression techniques such as binning the observables \cite{fewer_mocks_less_noise}, down-weighting the off-diagonal covariance elements \cite{cov_taper}, using neural networks \cite{cov_ML, adamo_nn_pk_cov}, defining data compression vectors \cite{massive_data_compression}, and expanding the covariance in eigenmodes \cite{chan_bao_ang_clustering}. Noise may also be reduced by smoothing the residual between a model and mock covariance \cite{oconnell_residual_smoothing} or by incorporating survey geometry with a smooth model covariance \cite{oconnell_large_cov_smooth_model}. By making use of a model with few parameters \cite{fumagalli_fitting_cov_models, pearson_pk_cov_fewer_mocks}, analytically computing the large-scale portion of the covariance \cite{howlett_2pcf_cov_nextgen}, using a Bayesian framework to combine model and mock covariances \cite{hall_bayesian_cov}, or determining the 2PCF from data \cite{rashkovetskyi_cov_2pcf_desi}, the covariance may be estimated with fewer mock catalogs than would typically be needed. Similarly, the computational cost of the full covariance may be reduced by analytically calculating the disconnected covariance \cite{li_disconnected_cov, li_covdisc}. The anisotropic analytic covariance has also been studied \cite{grieb_gauss_cov_aniso}.

An analytic template can be computed for the leading-order covariance by assuming the density field is a Gaussian Random Field (GRF). This was pioneered for the 2PCF using a maximum likelihood approach \cite{xu_2pcf_cov} and for the 3PCF with a multipole moment decomposition \cite{ZS_3pcf_N2}. This approach was extended to NPCFs \cite{cov} using the isotropic basis functions \cite{iso, jess_gen_func}. These analytic covariance templates all rely on numerical integrals of the galaxy power spectrum, which characterizes the clustering of matter as a function of length scale (wavenumber) and is the Fourier-space analog of the 2PCF. The power spectrum is often modeled with perturbation theory, and evaluated with power-law decompositions or reduced to convolutions with FFTs \cite{beyond_yamamoto, mcewen_conv_int, fast_pt_2, fang_angpowspec, simonovic_pt, assassi_angstat, schmittfull_pt, chen_fft, aniso_pk_estimator, beyond_trad_los, fast_est_zspace, mohammed_analytic_pk_cov, matsubara_integrated_pt, wadekar_pt, cohn, slepian_decouplingpt, taule_2loop_pk}. However, it is desirable to have a fully analytic solution for the power spectrum integrals to enable better understanding of the covariance matrix's structure.

In this work, we first obtain a simple yet accurate power law model for the linear matter power spectrum (\cref{sec:pk_model}). We compare this model to the true power spectrum, then use each to numerically compute the covariance matrix of the 2PCF. We compare these covariance matrices and show that the covariance computed using the model power spectrum matches fairly well to that computed using the true power spectrum (\cref{sec:half_inv}). We then use the model power spectrum to seek an analytic closed-form covariance matrix for the 2PCF, and examine its structure (\cref{sec:2pcf_overview}). In \cref{sec:higher_order_matrices}, we outline the covariance matrices of higher-order correlation functions. We then use our model power spectrum to analytically evaluate the fundamental building blocks ($f$-integrals) of these higher-order covariance matrices (\cref{sec:f_int}). In \cref{sec:sparse_matrices}, we find that the greatest contributions to the $f$-integrals arise when closed triangles may be formed. The number of non-vanishing configurations decreases when $f$-integrals are multiplied together; as the covariance depends on products of $f$-integrals, this explains the sparse structure of the true covariance. Finally, we explore how our analytic results can allow for easy inversion of the covariance with the addition of a correction term (\cref{sec:inverse_cov}) as well as provide a computationally efficient method for determining the covariance (\cref{sec:computational_complexity}). We summarize our findings in \cref{sec:conclusion}. Throughout this work, the formulae for the covariance capture the GRF contribution, which is always the leading-order piece.

\section{Power Law Model for the Power Spectrum}
\label{sec:pk}
In \cref{sec:pk_model}, we first outline a model for the power spectrum and compare this to the true power spectrum. Then, in \cref{sec:half_inv}, we use each power spectrum to compute the covariance matrix of the 2PCF, and compare these matrices. 

\subsection{Inverse Wavenumber Power Spectrum Model}
\label{sec:pk_model}
A crucial piece for the GRF approximation for the covariance matrices of NPCFs is the power spectrum. We model the power spectrum as
\setlength{\fboxrule}{0.5pt}
\begin{empheq}[box=\fbox]{align}
\label{eqn:pk}
    P(k) &= \frac{A}{k}+\frac{1}{\bar{n}},
\end{empheq} 
where $A$ is an amplitude and $\bar{n}$ is the average number density of galaxies. The first term of \cref{eqn:pk} is the physical part of the power spectrum and stems from galaxy clustering; the second term is shot noise (Poisson noise) from the discrete nature of the galaxies. As the 2PCF exhibits a $1/s^2$ dependence, and the power spectrum is the Fourier transform of the 2PCF, we use a $1/k$ model for the physical power spectrum. Using this power law model enables analytic evaluation of the spherical Bessel function (sBF) integrals that enter the NPCF covariance matrices, as will be shown. We undertake this in the following sections. Here, we first explore how well \cref{eqn:pk} captures the true power spectrum.

The true linear power spectrum is obtained from the Code for Anisotropies in the Microwave Background (\textsc{camb}) \cite{camb}, which is a Boltzmann code built on \textsc{cmbfast} \cite{cmbfast} and is a precursor to \textsc{class} \cite{CLASS}. We use 10,000 sample points in $k$, uniform in logarithmic space, from $k_{\mathrm{min}} = \SI{e-3}{\hHubble\per\Mpc}$ to $k_{\mathrm{max}} = \SI{10}{\hHubble\per\Mpc}$. We use a geometrically flat $\Lambda$CDM cosmology with parameters from \textit{Planck} 2018 \cite{planck2018} (using the \textsc{Plik} likelihood) at $z=0.57$ (the effective redshift of the Constant Mass (CMASS) sample of the Sloan Digital Sky Survey Baryon Oscillation Spectroscopic Survey (SDSS BOSS) \cite{boss_sdss3}). The Hubble constant is $H_0 = \SI{67.36}{\km\per\s\per\Mpc}$, the baryon density (in units of the critical density) is $\Omega_{\mathrm{b}} h^2 = 0.02237$, the CDM density is $\Omega_{\mathrm{c}} h^2 = 0.1200$, and the scalar spectral index is $n_{\mathrm{s}} = 0.9649$. $\sigma_8$, which quantifies matter density fluctuations on spheres of $\SI{8}{\per\hHubble\Mpc}$, is 0.5880.

For our numeric analysis, we include the linear bias, $b_1$, on the physical part of the power spectrum model in \cref{eqn:pk} to account for the variation between the distributions of baryonic and dark matter:
\begin{align}
\label{eqn:pk_smooth}
    P_{\mathrm{bias}}(k) &= b_1^2\frac{A}{k} + \frac{1}{\bar{n}}.
\end{align}
We set $b_1=2$, as was found in many analyses \cite{SE_3PCF_BAO, ntelis_cosmic_homogeneity, ereza_uchuuglamboss_ebosslrg_lightcones, slepian_constraining_baryon_dm, slepian_largescale_3pcf_sdss, gil_marin_rsd_pk_bispec, gil_marin_rsd_los_pk} of the Luminous Red Galaxies (LRGs) in CMASS \cite{boss_sdss3}. We also include linear bias and shot noise with the true linear power spectrum from \textsc{camb} (this is \pcref{eqn:pk_smooth}, with $A/k$ replaced by the \textsc{camb} power spectrum). We have found that setting the amplitude to be $A = \SI{277}{\per\hHubble\squared\Mpc\squared}$ gives good agreement between the model and true power spectra.

We also use a Gaussian damping $\exp[-(\sigma k)^2]$ for the model and true power spectra to prevent the Gibbs phenomenon ringing that would otherwise arise from a sharp cutoff in $k$ when we take the inverse Fourier transform. The damping scale $\sigma = \SI{1}{\per\hHubble\Mpc}$ alters the power spectrum at higher wavenumbers than those for which we expect the linear power spectrum to be valid.

\begin{figure}
    \centering  
    \includegraphics[width=\textwidth]{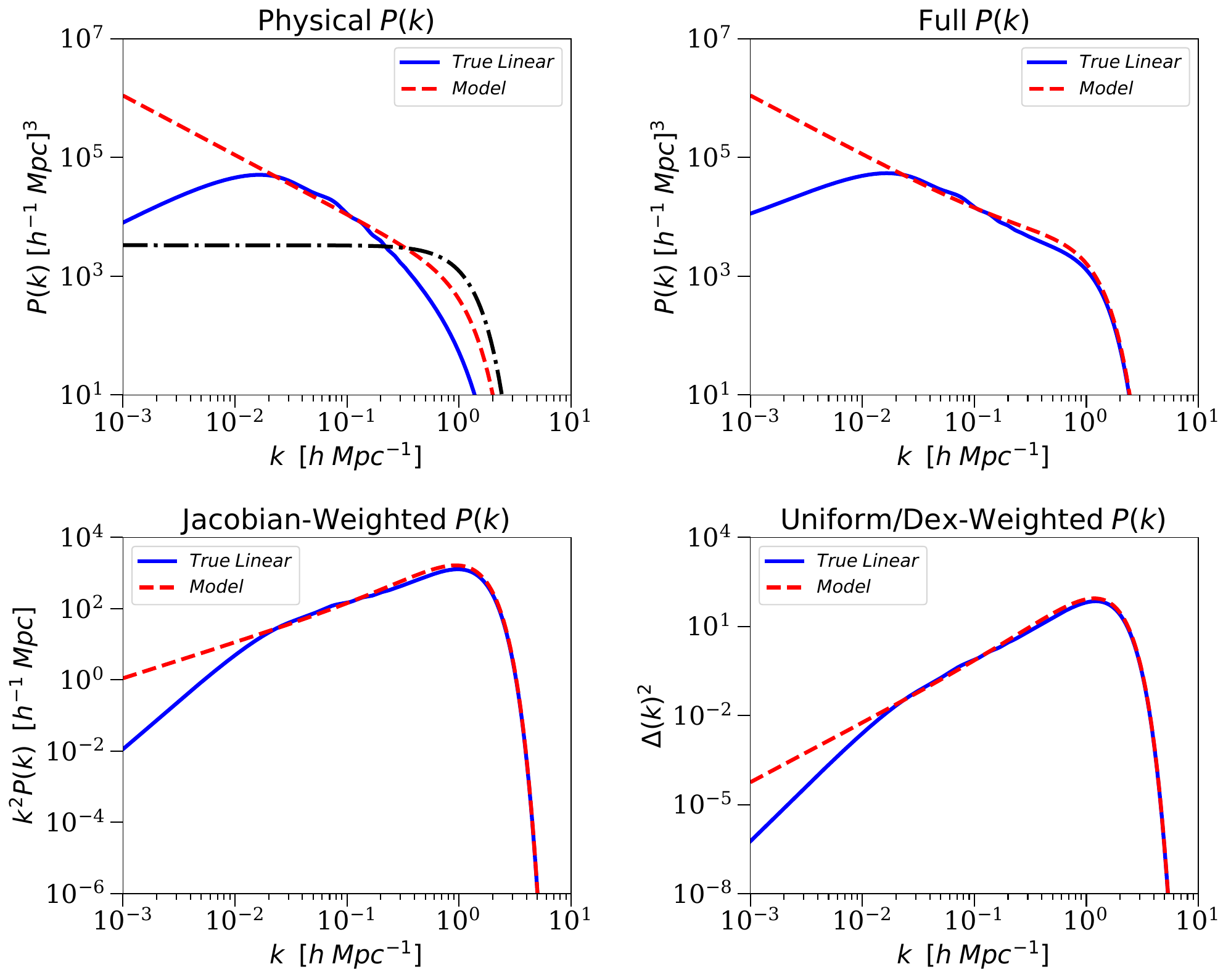}

    \caption{Here, we compare the model power spectrum (\pcref{eqn:pk_smooth}, dashed red curve) to the \textsc{camb} power spectrum (solid blue curve). Gaussian damping is included on each of the power spectra and on the shot noise. The model power spectrum has an amplitude set to $A = \SI{277}{\per\hHubble\squared\Mpc\squared}$. The \textsc{camb} power spectrum has cosmological parameter values set by \textit{Planck} 2018 \cite{planck2018}. The \textsc{camb} power spectrum includes BAO; our model does not. In all panels, there is a deviation between the model and true power spectra at low $k$. However, this disagreement will not greatly affect the covariance matrix, as described in \cref{app_trunc_pk}. At high $k$, the physical piece of the power spectrum is less dominant than the shot noise piece, which is exactly equal between the model and true power spectra. \textit{Top left panel:} The physical part of the linear matter power spectra from the model and \textsc{camb} are shown. The black dash-dotted curve is the shot noise (with $\bar{n} = \SI{3e-4}{\hHubble\cubed\per\Mpc\cubed}$). \textit{Top right panel:} The model and \textsc{camb} power spectra, including the physical and shot noise pieces, are displayed. \textit{Bottom left panel:} The true and model power spectra (both including shot noise) are weighted by $k^2$, as $k^2 \, \dl k$ is the Jacobian over which the power spectrum must be integrated to obtain the covariance matrices of the NPCFs. \textit{Bottom right panel:} In our numerical integration for the covariance, we use a logarithmic grid evenly spaced in $k$, motivating the Jacobian to be rewritten as $k^3\, \dl \ln(k)$ (using $\dl k = k\,\dl \ln(k)$). Here, we display the dimensionless power spectrum $\Delta(k)^2 \equiv k^3P(k)/2\pi^2$, and include shot noise.}
    \label{fig:pk_vs_k}
\end{figure}

In \cref{fig:pk_vs_k}, the number density is set to $\bar{n} = \SI{3e-4}{\hHubble\cubed\per\Mpc\cubed}$, which is the number density of LRGs in CMASS \cite{boss_sdss3}. The \textit{top left panel} shows the power spectrum from our model and the true linear power spectrum from \textsc{camb}, both excluding shot noise. In the \textit{top right panel}, we display the model and true power spectra, both including shot noise. In all panels of \cref{fig:pk_vs_k}, our model deviates from the true power spectrum at low $k$; a broken power law model could remedy this. 

However, as will be shown, the power spectrum must be integrated against the three-dimensional spherically symmetric Jacobian, proportional to $k^2 \, \dl k$, to obtain the covariance matrix. Doing so reduces the deviation between the model and true power spectra; we display $k^2P(k)$ in the \textit{lower left panel} of \cref{fig:pk_vs_k}. In the \textit{lower right panel}, we display the dimensionless power spectrum $\Delta(k)^2 \equiv k^3P(k)/(2\pi^2)$.

\subsection{Half-Inverse Test}
\label{sec:half_inv}
As a second means of assessing the validity of our model power spectrum, we examine the covariance matrix of the 2PCF, which is an average over all pairs of galaxy pairs. The first pair has separation $r$ between the constituent galaxies and the second pair has separation $r'$. The pairs themselves are separated by $s$, which we integrate over (to obtain this average). In practice, for signal-to-noise considerations as well as algorithmic efficiency, the 2PCF is generally computed on spherical shell bins in pair separation. We use separations from \SIrange{0}{200}{\per\hHubble\Mpc}, with bins of width \SI{10}{\per\hHubble\Mpc}. The elements $C_{ij}$ of the binned 2PCF covariance matrix are derived in \cite{xu_2pcf_cov}; we replicate them below:
\begin{align}
\label{eqn:2pcf_cov_binned}
    &C_{ij} = \frac{2}{V}\int_0^{\infty}\frac{k^2\dl k}{2\pi^2}\;\Delta j_1(kr_i)\Delta j_1(kr^{\prime}_j)P(k)^2,
    \\
\label{eqn:delta_j1}
    &\Delta j_1(kr_i) \equiv \left(\frac{3}{r_{i2}^3-r_{i1}^3}\right)\frac{r_{i2}^2j_1(kr_{i2}) - r_{i1}^2j_1(kr_{i1})}{k}.
\end{align}
In \cref{eqn:2pcf_cov_binned}, $V$ is the survey volume, $P(k)$ is the power spectrum (including shot noise), and $\{i,j\}$ are the radial bin indices. The subscripts $\{1,2\}$ in \cref{eqn:delta_j1} represent, respectively, the lower and upper bounds for $r_i$ in a given bin. $j_1(kr_i)$ is the first-order spherical Bessel function (sBF).

We compute the true covariance matrix of the 2PCF from \cref{eqn:2pcf_cov_binned}, where $P(k)$ is replaced with the \textsc{camb} power spectrum (including linear bias). Similarly, to compute a model 2PCF covariance matrix, we replace $P(k)$ in \cref{eqn:2pcf_cov_binned} with our physical model, $b_1^2A/k$. In both cases, we set the volume to \SI{2}{\per\hHubble\cubed\Gpc\cubed} to match the rough effective volume found in analyses of CMASS (\textit{e.\,g.,} \cite{slepian_largescale_3pcf_sdss, SE_3PCF_BAO, cuesta_sdss_bao_in_cf}). We also include Gaussian damping $\exp[-(\sigma k)^2]$ on the physical and shot noise pieces of both the model and true power spectra that the covariance depends on.

In analyses relying on covariance matrices, the data are weighted by the inverse covariance matrix. We are thus interested in ensuring the inverse {of the} covariance matrix (rather than the covariance matrix itself) is correct, as will be shown below. 

We quantify the similarity of the model and true covariance matrices of the 2PCF using a half-inverse test \cite{SE_3PCF_BAO, cov}:
\begin{align}
\label{eqn:half_inv}
    \mathbf{S} &\equiv \mathbf{C}_{\mathrm{model}}^{-1/2}\mathbf{C}_{\mathrm{true}}\mathbf{C}_{\mathrm{model}}^{-1/2} - \mathbf{\mathds{1}},
\end{align}
where $\mathbf{C}_{\mathrm{model}}$ is the covariance from the model power spectrum, $\mathbf{C}_{\mathrm{true}}$ is the covariance computed using the \textsc{camb} power spectrum,\footnote{$\mathbf{C}_{\mathrm{true}}$ is the covariance from the true power spectrum, but is not necessarily the true covariance.} and $\mathbf{\mathds{1}}$ is the identity matrix. The fractional matrix power is computed using a Schur–Padé algorithm \cite{fractional_matrix_power}.

If $\mathbf{C}_{\mathrm{model}}$ = $\mathbf{C}_{\mathrm{true}}$, which is the ideal case, then $\mathbf{S}=\mathbf{0}$. Thus, finding $\mathbf{S}$ close to the null matrix demonstrates good agreement between the inverse model and true covariance matrices. To quantify the disparity between the half-inverse test and the null matrix, we compute the standard deviation (SD) of elements of $\mathbf{S}$.

\begin{figure}
    \centering
    \vspace{-1.4cm}
    \begin{subfigure}[b]{0.45\textwidth}
        \centering
        \includegraphics[width=\textwidth]{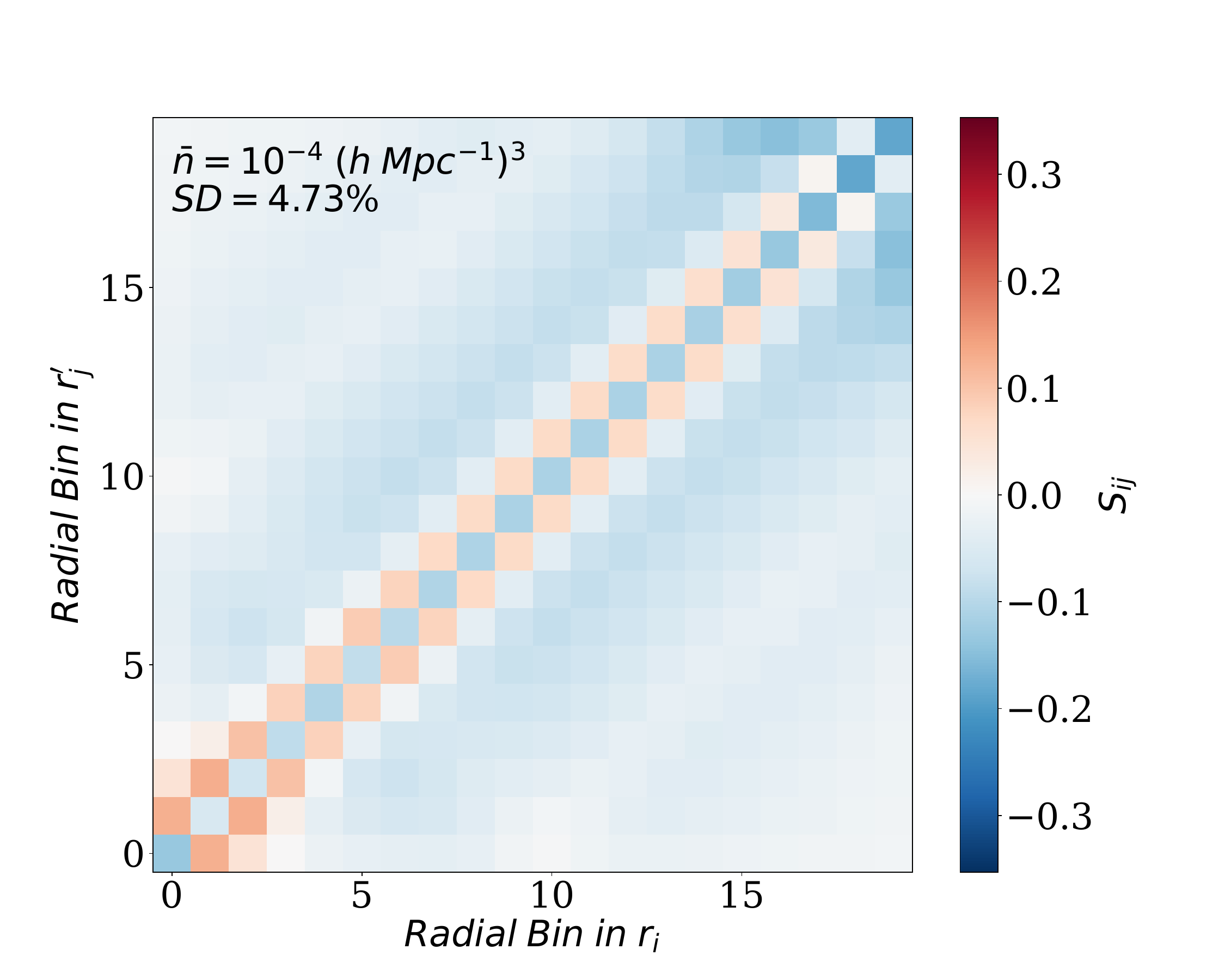}
    \end{subfigure}
    \hspace{-0.25cm}
    \begin{subfigure}[b]{0.45\textwidth}
        \centering
        \includegraphics[width=\textwidth]{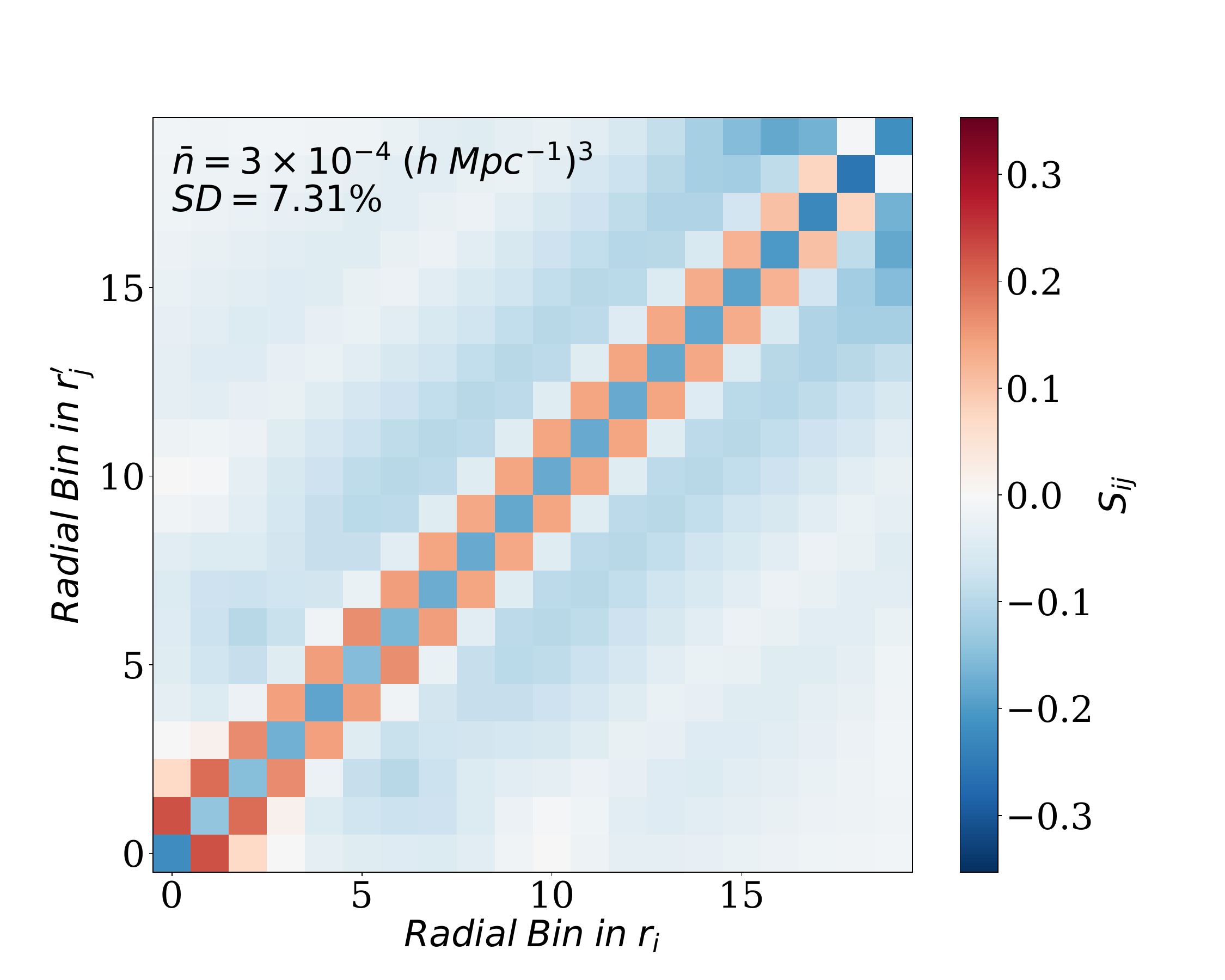}
    \end{subfigure}

    \vspace*{-1.2ex}

    \begin{subfigure}[b]{0.45\textwidth}
        \centering
        \includegraphics[width=\textwidth]{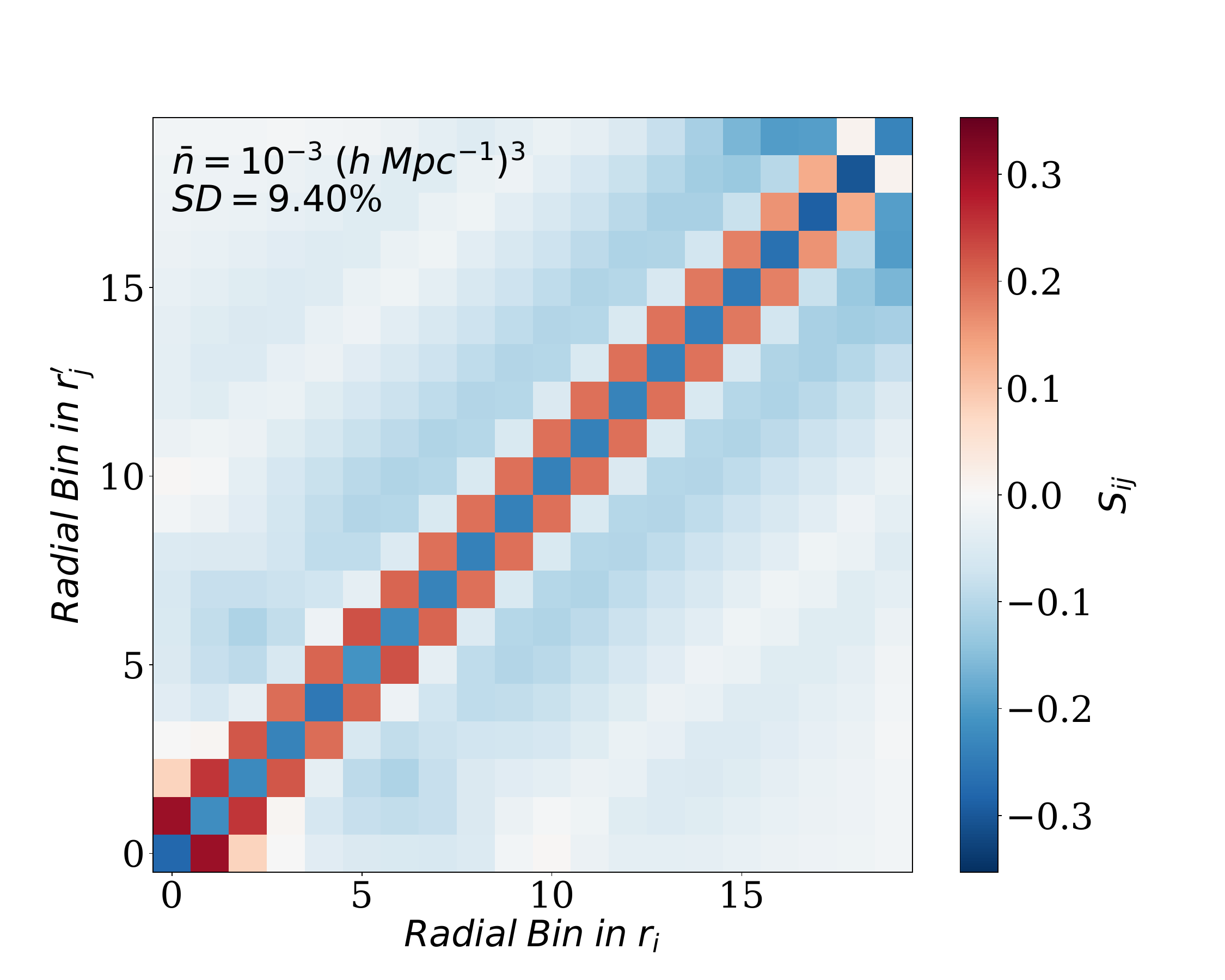}
    \end{subfigure}
    \hspace{-0.25cm}
    \begin{subfigure}[b]{0.45\textwidth}
        \centering
        \includegraphics[width=\textwidth]{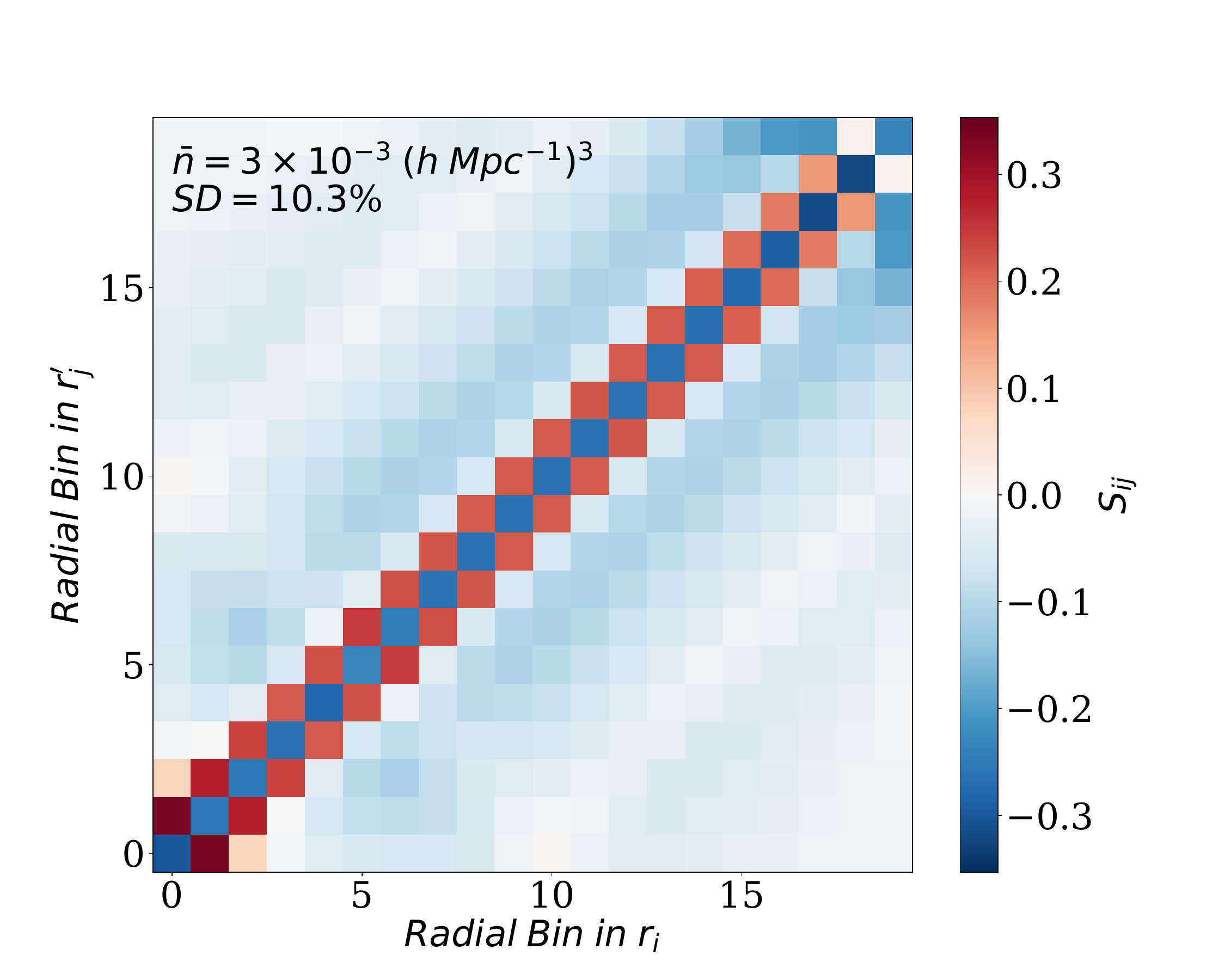}
    \end{subfigure}

    \vspace*{-1.2ex}

    \begin{subfigure}[b]{0.45\textwidth}
        \centering
        \includegraphics[width=\textwidth]{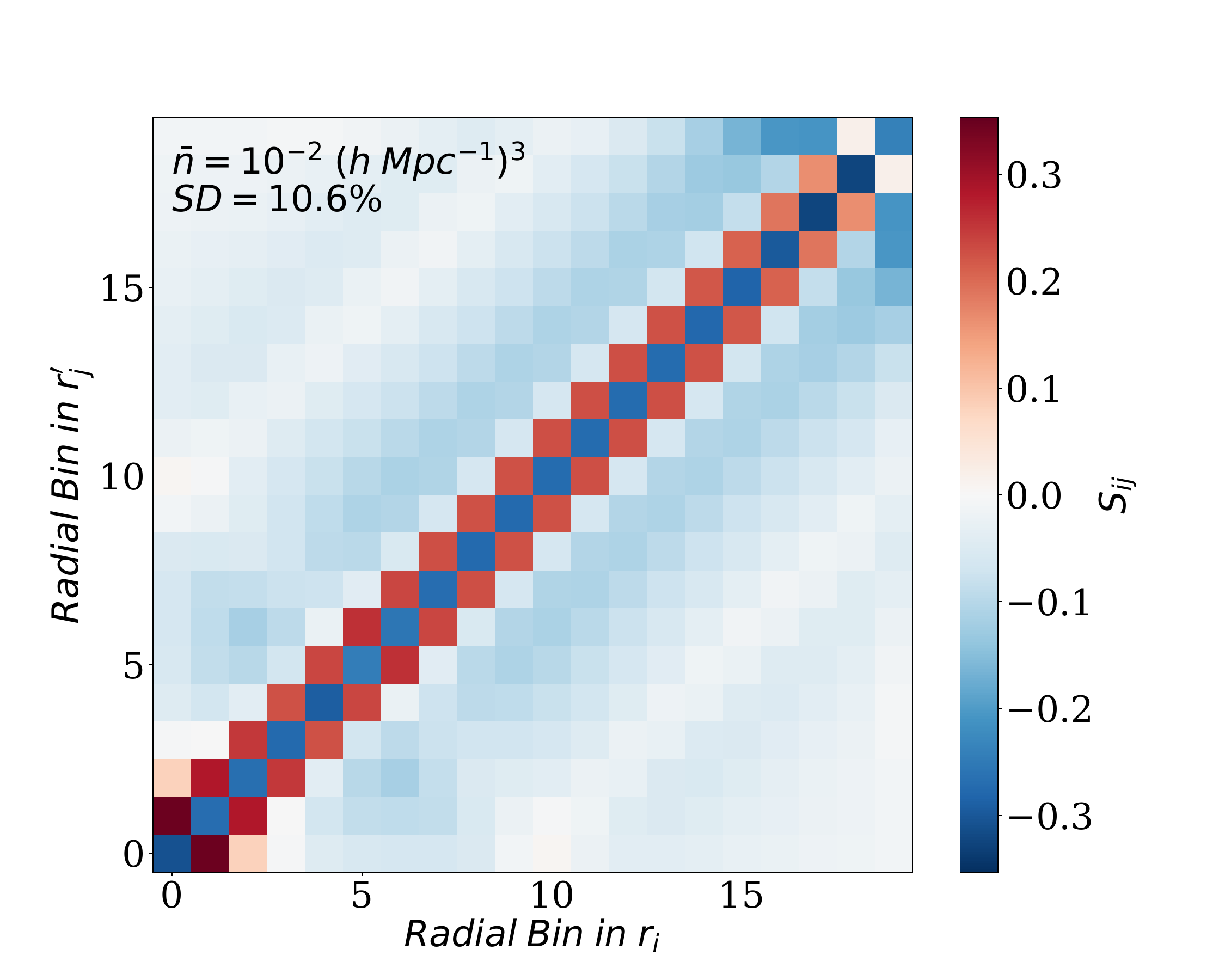}
    \end{subfigure}
    \hspace{-0.25cm}
    \begin{subfigure}[b]{0.45\textwidth}
        \centering
        \includegraphics[width=\textwidth]{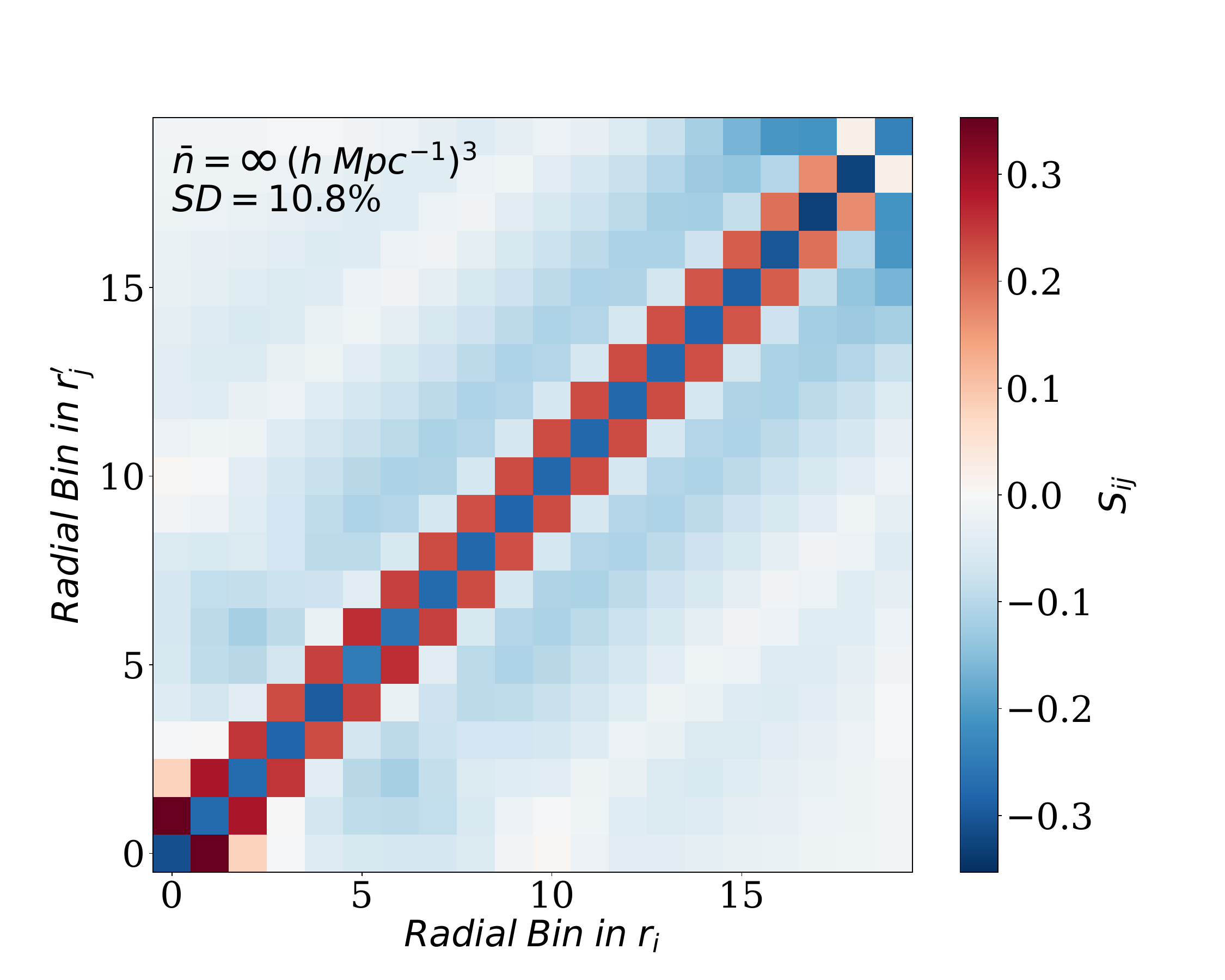}
    \end{subfigure}

    \caption{The half-inverse test ($\mathbf{S} \equiv \mathbf{C}_{\mathrm{model}}^{-1/2}\mathbf{C}_{\mathrm{true}}\mathbf{C}_{\mathrm{model}}^{-1/2} - \mathbf{\mathds{1}}$, \pcref{eqn:half_inv}) comparing the model and true binned covariance matrices of the 2PCF is shown for various realistic number densities as well as the unrealistic case of an infinite number density (\textit{i.\,e.} $1/\bar{n} \rightarrow 0$). If the model and true covariance matrices match, the result of the half-inverse test is the null matrix. The standard deviation (SD), which measures how the half-inverse test deviates from the null matrix on average, is shown for each number density. We report the SD in this way for consistency. $\{r_i,r^{\prime}_j\}$ denotes the radial bins for each galaxy pair. These bins have width 10 $h^{-1}$Mpc and range from 0 $h^{-1}$Mpc to 200 $h^{-1}$Mpc. The model and true covariance matrices match more closely for low number densities, where the shot noise is more dominant than the physical power spectrum and thus our model covariance is closer to the true covariance than at high number densities. Each subplot contains two white ``lines'' parallel to the diagonal and one fainter white ``line'' perpendicular to the diagonal. We hypothesize that these features originate either from the BAO within the true power spectrum, or from the crossing of the model and true power spectra near the BAO scale (as seen in \cref{fig:pk_vs_k}).}
    \label{fig:half_inv}
\end{figure}

The half-inverse test and its SD are shown for various number densities $\bar{n}$ in \cref{fig:half_inv}. With our power spectrum model, we are able to achieve percent-level half-inverse SD values, indicating that our model will indeed allow us to examine the structure of the covariance. 

As the number density increases, the half-inverse test deviates further from the null matrix and the SD increases, showing that the model power spectrum is most accurate for low number densities. This is because for low number densities, the shot noise is more dominant than the physical power spectrum. In the limit as $\bar{n} \rightarrow 0$, any model is exact.

\begin{figure}
    \centering  
    \includegraphics[width=\textwidth]{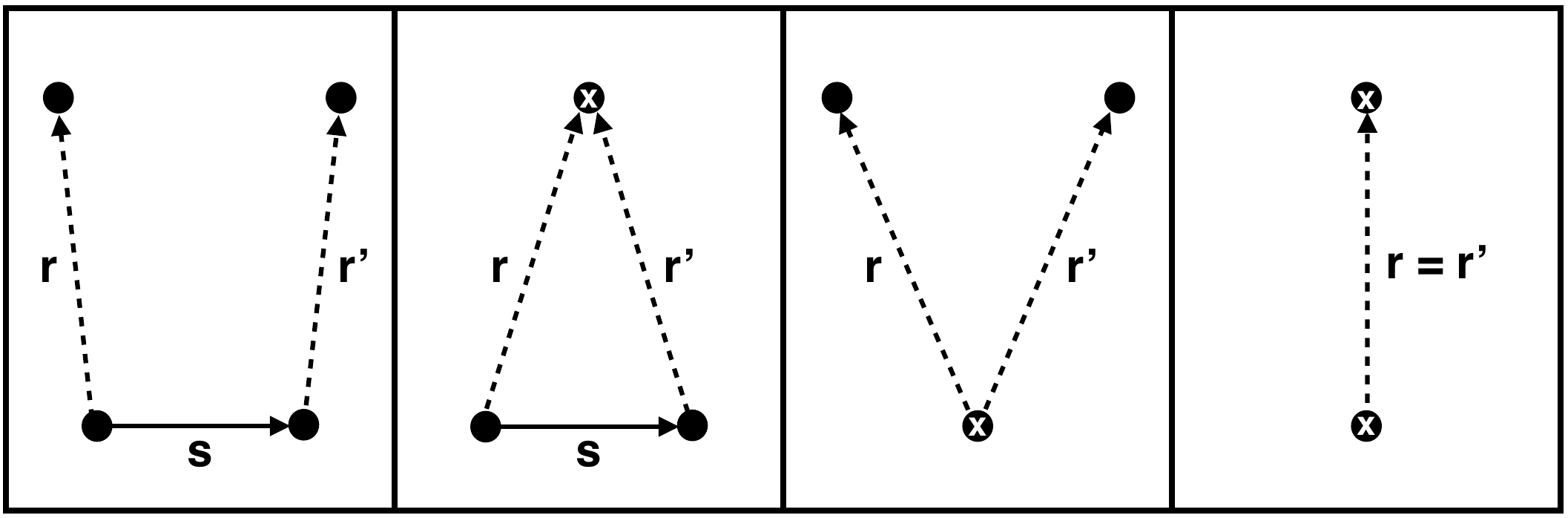}
    \caption{The covariance of the 2PCF has four different configurations of galaxies that impact the shot noise. Here, we display these configurations for equal pair separations $r$ and $r^{\prime}$. Except for the rightmost panel, all configurations are also possible when the side lengths are unequal. The white X represents where two galaxies overlap. \textit{Left:} Neither the primary galaxies nor the secondary galaxies overlap. There is no shot noise contribution to the covariance. \textit{Left center:} The primaries are separated by $s$ and the secondaries overlap. This overlap corresponds to one factor of shot noise $1/\bar{n}$. \textit{Right center:} The primaries overlap while the secondaries do not. Similarly to the left center panel, one factor of shot noise $1/\bar{n}$ is present. \textit{Right:} The primaries overlap and the secondaries overlap. Each overlap contributes one factor of shot noise to the covariance, for a total of $1/\bar{n}^2$. This is only possible when the pair separations $r$ and $r^{\prime}$ are equal, which is represented by the main diagonal of the half-inverse tests in \cref{fig:half_inv}.}
    \label{fig:shotnoise}
\end{figure}

The shot noise piece of the power spectrum, which arises due to sampling a discrete set of galaxies (as the power spectrum is the Fourier transform of the 2PCF), contributes when multiple galaxies are located at the same position. As shown in \cref{fig:shotnoise}, the covariance of the 2PCF has four different configurations of galaxies, three of which give rise to shot noise.

In each panel of \cref{fig:half_inv}, the most nonzero elements of the half-inverse test are on the diagonal, where the binned pair separations are equal. This coincides with the right-most panel of \cref{fig:shotnoise}, where both primaries are located at the same position, and both secondaries are at the same position. Since the shot noise is exact between the model and true power spectra, one may expect the diagonal of the half-inverse test to have the smallest deviation from zero. However, the diagonal of our test matrix may receive contributions from all configurations of the covariance (\cref{fig:shotnoise}), driving the diagonal away from zero.

We recall that in \cref{fig:pk_vs_k}, the model power spectrum deviates from the \textsc{camb} power spectrum at low $k$. To examine the effect this deviation has on the covariance of the 2PCF, we computed the same half-inverse tests shown in \cref{fig:half_inv} while truncating the power spectrum to include only values $k \geq \SI{0.2}{\hHubble\per\Mpc}$, where the two power spectra agree. The truncated half-inverse tests are shown in \cref{fig:half_inv_trunc} of \cref{app_trunc_pk}; their corresponding SD are given in \cref{Tab:rms}. 

We find that excluding the region over which the power spectra deviate (low $k$) does not significantly improve the SD of the half-inverse test. Thus, we conclude that the deviation does not greatly affect the inverse covariance. This is as expected: the Jacobian-weighted power spectrum (lower left panel of \cref{fig:pk_vs_k}), which enters the covariance integrals, is peaked. This quantity is therefore controlled by the region of the peak, over which the model and true power spectra match more closely than at low $k$.

As noted in \cref{Tab:rms}, the improvement made to the half-inverse test SD by truncating the power spectrum decreases as number density increases. This is because at high $\bar{n}$, the shot noise in our model power spectrum (which exactly matches the true power spectrum) contributes less than the physical power spectrum, which we have modeled with $A/k$. In \cref{fig:square_weight_trunc} of \cref{app_trunc_pk}, we display plots similar to the lower left panel of \cref{fig:pk_vs_k} at various number densities to make clear that there is a greater mismatch\footnote{This mismatch is in part due to the amplitude having been set to $A = \SI{277}{\per\hHubble\squared\Mpc\squared}$, which was chosen so that the deviation between power spectra was minimized for $\bar{n} = \SI{3e-4}{\hHubble\cubed\per\Mpc\cubed}$. In practice, $A$ could be tuned for each different number density.} between the model and true physical power spectra as $\bar{n}$ increases. This mismatch is present at the peaks of these plots, and therefore has a greater contribution to the high half-inverse SD than the deviation of the power spectra at low $k$. Thus, truncating the power spectra to include only $k \geq \SI{0.2}{\hHubble\per\Mpc}$ does not remedy this. 

\begin{table}
    \centering
    \begin{tabular}{S[table-format=1.0e+1] *{2}{S[table-format=2.2, table-space-text-post=\,\%]} S[table-format=+1.2, table-space-text-post=\,\%]}
        \toprule
        {\textbf{\boldmath$\bar{n}$ / $\si[detect-all]{\hHubble\cubed\per\Mpc\cubed}$}} & {\textbf{\boldmath SD: Full $P(k)$}} & {\textbf{\boldmath SD: Truncated $P(k)$}} & {\textbf{\boldmath Mean: Full $P(k)$}}
        \\
        \midrule
        e-4  & 4.73 \,\% & 4.40 \,\% & -3.95 \,\% \\
        3e-4 & 7.31 \,\% & 7.14 \,\% & -4.11 \,\% \\
        e-3 & 9.40 \,\% & 9.29 \,\% & -4.16 \,\% \\
        3e-3 & 10.3 \,\% & 10.2 \,\% & -4.18 \,\% \\
        e-2 & 10.6 \,\% & 10.5 \,\% & -4.19 \,\% \\
        {$\infty$} & 10.8 \,\% & 10.7 \,\% & -4.19 \,\% \\
        \bottomrule
    \end{tabular}

    \caption{The half-inverse test (\pcref{eqn:half_inv}) of the covariance of the 2PCF has been computed for various number densities, shown in the leftmost column. The left-center column displays the SD of the half-inverse tests for each of the given number densities. These half-inverse tests, computed using the full range (\SIrange{e-3}{10}{\hHubble\per\Mpc}) of the model and \textsc{camb} power spectra, are shown in \cref{fig:half_inv}. We then computed the half-inverse tests again, truncating the power spectra so that only the region over which they agree ($k \geq \SI{0.2}{\hHubble\per\Mpc}$) was included. The SD from these half-inverse tests are given in the right-center column while these half-inverse tests themselves are displayed in \cref{fig:half_inv_trunc}. Truncating the power spectra to only include the region of agreement did not significantly improve the SD of the half-inverse test; the mean improvement is \num{0.146} percentage points. As number density increases, the improvement to the SD decreases. In the rightmost column, we display the mean value of the elements in each half-inverse test over the full range of $k$.} 
    \label{Tab:rms}
\end{table}

\section{Covariance Matrix of the 2PCF}
\label{sec:2pcf_overview}
In \cref{sec:2pcf}, we obtain the covariance matrix of the 2PCF in the Gaussian Random Field (GRF) limit. This limit assumes that the density fluctuation field is a GRF; thus, all correlations can be described by the 2PCF or products thereof. Then, in \cref{sec:2pcf_structure}, we analyze the structure of the analytic covariance we will have obtained in \cref{sec:2pcf}.

\subsection{Analytic Evaluation}
\label{sec:2pcf}
The unbinned covariance matrix of the 2PCF in the GRF limit is \cite{xu_2pcf_cov}
\setlength{\fboxrule}{0.5pt} 
\begin{empheq}[box=\fbox]{align}
\label{eqn:2pcf_cov_orig}
    \mathrm{Cov}(r,r^{\prime}) &= \frac{2}{V}\int_0^{\infty}\frac{k^2\dl k}{2\pi^2}\;j_0(kr)j_0(kr^{\prime})P(k)^2,
\end{empheq} 
with $V$ the survey volume, $j_0(kr)$ the zero-order sBF, and $r$ and $r^{\prime}$ the (nonzero) separations between the constituent galaxies in each galaxy pair. We do not include any zero-valued separations in the covariance since they account for the correlation of a galaxy with itself, which does not contain any cosmological information. 

Replacing the power spectrum in \cref{eqn:2pcf_cov_orig} with the power spectrum model from \cref{eqn:pk} yields
\begin{align}
\label{eqn:2pcf_cov_with_pk}
    \mathrm{Cov}(r,r^{\prime}) &= \frac{1}{\pi^2V}\int_0^{\infty}k^2\dl k\;j_0(kr)j_0(kr^{\prime})\left[\mleft(\frac{A}{k}\mright)^2+\frac{2A}{\bar{n}k}+\frac{1}{\bar{n}^2} \right].
\end{align}
We do not include damping in the covariance because we are evaluating the covariance analytically, and thus do not have the Gibbs phenomenon that would be present in numerical computations without damping.

We define new integrals to simplify \cref{eqn:2pcf_cov_with_pk}:
\begin{align}
\label{eqn:I_2const}
    I^{[2,\mathrm{const}]}_{\ell}(r,r^{\prime}) &\equiv \int_{0}^{\infty}\dl k\;j_{\ell}(kr)j_{\ell}(kr^{\prime}), \\
\label{eqn:I_2lin}
    I^{[2,\mathrm{lin}]}_{\ell}(r,r^{\prime}) &\equiv \int_{0}^{\infty}k\dl k\;j_{\ell}(kr)j_{\ell}(kr^{\prime}), \\
\label{eqn:I_2quad}
    I^{[2,\mathrm{quad}]}_{\ell}(r,r^{\prime}) &\equiv \int_{0}^{\infty}k^2\dl k\;j_{\ell}(kr)j_{\ell}(kr^{\prime}).
\end{align}
The superscript denotes the number of sBFs, and the text abbreviates the power law term in $k$ in the integrand: ``const'' for constant (no $k$), ``lin'' for linear, and ``quad'' for quadratic. $j_{\ell}(kr)$ is an sBF with integer order $\ell.$

\Cref{eqn:2pcf_cov_with_pk} may now be written as
\begin{align}
\label{eqn:2pcf_cov_in_terms_of_int}
    \mathrm{Cov}(r,r^{\prime}) &= \frac{1}{\pi^2V}\left[ A^2I^{[2,\mathrm{const}]}_0(r,r^{\prime}) + \frac{2A}{\bar{n}}I^{[2,\mathrm{lin}]}_0(r,r^{\prime}) + \frac{1}{\bar{n}^2}I^{[2,\mathrm{quad}]}_0(r,r^{\prime}) \right].
\end{align}
Although \cref{eqn:2pcf_cov_in_terms_of_int} only requires the $\ell = 0$ limit of \cref{eqn:I_2const,eqn:I_2lin,eqn:I_2quad}, we will first evaluate these integrals for general $\ell \geq 0$. These general $\ell$ results will be needed for the covariance matrices of the 3PCF and 4PCF, which depend on integer $\ell$, in the following sections. 

We begin with \cref{eqn:I_2const}, the double-sBF integral with no power law in $k$, which may be evaluated using \cite{GR} equation 6.574.1 (when $r \neq r^{\prime}$) and equation 6.574.2 (when $r = r^{\prime}$). These two cases may be combined into a single formula as
\begin{align}
\label{eqn:I_2const_with_hg}
    I^{[2,\mathrm{const}]}_{\ell}(r,r^{\prime}) &= \int_{0}^{\infty}\dl k\;j_{\ell}(kr)j_{\ell}(kr^{\prime}) \nonumber \\
    &= \frac{\pi}{2(2\ell+1)}g^{-1}\Bigg[H(\chi-1)\;\chi^{-(\ell+1/2)}\twoFone\mleft(\ell+\frac{1}{2}, 0; \ell+\frac{3}{2}; \chi^{-2} \mright) \Bigg. \nonumber \\
    &\Bigg. \qquad + H(1-\chi)\;\chi^{\ell+1/2} \twoFone\mleft(\ell+\frac{1}{2}, 0; \ell+\frac{3}{2}; \chi^2 \mright) + 4\theta(\chi-1)\theta(1-\chi) \Bigg],
\end{align}
where we have defined the geometric mean and ratio of the pair separations:\footnote{We note that $g$ is the logarithmic average of the pair separations while $\chi$ is the logarithmic difference of the pair separations: $\ln(g) = \left(\ln(r) + \ln(r^{\prime}) \right)/2 \; \text{and} \;  \ln(\chi) = \ln(r) - \ln(r^{\prime})$.}
\begin{align}
\label{eqn:chi_g_defs}
    g &\equiv \sqrt{rr^{\prime}},
    \nonumber \\
    \chi &\equiv r/r^{\prime}.
\end{align}

In \cref{eqn:I_2const_with_hg}, $\twoFone$ is the Gauss hypergeometric function, $H(x)$ is the left-continuous Heaviside function (\textit{i.\,e.}, $H(0)=0$), and $\theta(x)$ is the Heaviside function with the half-maximum convention (\textit{i.\,e.}, $\theta(0) = 1/2$). Here, the product $4\theta(\chi-1)\theta(1-\chi)$ indicates that the value of the integral is $(\pi/2)\left(2\ell+1\right)^{-1}g^{-1}$ when $r = r^{\prime}$. In practice, the covariance matrix is computed in radial bins; the $r=r^{\prime}$ contribution in \cref{eqn:I_2const_with_hg} will therefore not contribute to the covariance, since the integral over the point $\chi = 1$ vanishes.

When one of the parameters (one of the first three arguments) of the Gauss hypergeometric function is zero, the function reduces to unity. \Cref{eqn:I_2const_with_hg} thus becomes
\begin{align}
\label{eqn:I_2const_eval}
    I^{[2,\mathrm{const}]}_{\ell}(r,r^{\prime}) &= \frac{\pi}{2g(2\ell+1)}\left[H(\chi-1)\;\chi^{-(\ell+1/2)} + H(1-\chi)\;\chi^{\ell+1/2} + 4\theta(\chi-1)\theta(1-\chi) \right].
\end{align}
\Cref{eqn:I_2const_eval} is symmetric under interchange of $r$ and $r^{\prime}$. 

When $r$ and $r^{\prime}$ are unequal, we choose $r^{\prime}$ to be the larger of the two without loss of generality. We therefore may retain only the $r^{\prime} > r$ and $r = r^{\prime}$ terms of \cref{eqn:I_2const_eval}. In the $\ell=0$ limit, which is needed for the covariance matrix of the 2PCF (\pcref{eqn:2pcf_cov_in_terms_of_int}), \cref{eqn:I_2const_eval} becomes
\setlength{\fboxrule}{0.5pt} 
\begin{empheq}[box=\fbox]{align}
\label{eqn:I_2const_eval_ell0}
    I^{[2,\mathrm{const}]}_0(r,r^{\prime}) &= \frac{\pi}{2g} \left[H(1-\chi)\sqrt{\chi} + 4\theta(\chi-1)\theta(1-\chi)\right].
\end{empheq} 

We now evaluate (without assuming $r^{\prime} > r$ and $\ell=0$) the double-sBF integral with a linear power law in $k$, \cref{eqn:I_2lin}, using \cite{GR} equation 6.512.1:
\begin{align}
\label{eqn:I_2lin_eval}
    I^{[2,\mathrm{lin}]}_{\ell}(r,r^{\prime}) &= \int_{0}^{\infty}k\dl k\;j_{\ell}(kr)j_{\ell}(kr^{\prime})
    \nonumber \\
    &= \frac{\sqrt{\pi} \; \Gamma(\ell+1)}{2\;\Gamma(\ell+3/2)}g^{-2}\Bigg[H(\chi-1)\;\chi^{-(\ell+1)}\twoFone\mleft(\ell+1,\frac{1}{2};\ell+\frac{3}{2};\chi^{-2}\mright) + H(1-\chi) \Bigg.
    \nonumber \\
    &\Bigg. \qquad \times\chi^{\ell+1}\twoFone\Bigg(\ell+1,\frac{1}{2};\ell+\frac{3}{2};\chi^2\Bigg)\Bigg] + 4\theta(\chi-1)\theta(1-\chi)I^{[2,\mathrm{lin}]}_{\ell}(r=r^{\prime}).
\end{align}
The final term in \cref{eqn:I_2lin_eval} (where $r = r^{\prime}$) is divergent; we display an explicit formula for this in a footnote.\footnote{With \cite{NIST} equation 10.53.1, we may rewrite the sBFs in \cref{eqn:I_2lin_eval} as power series to show the ultraviolet divergence of the integral when $r=r^{\prime}$: \setlength{\abovedisplayskip}{0pt} \setlength{\belowdisplayskip}{0pt} \begin{align*} I^{[2,\mathrm{lin}]}_{\ell}(r=r^{\prime}) &= \sum_{\alpha=0}^{\infty}\sum_{\beta=0}^{\infty}\left\{ (-1/2)^{\alpha+\beta}r^{2(\ell+\alpha+\beta)} / [2\alpha!\beta!(2\ell+2\alpha+1)!!(2\ell+2\beta+1)!!(\ell+\alpha+\beta+1)] \right. \\ &\left. \qquad \times \lim_{k\rightarrow\infty}k^{2(\ell+\alpha+\beta+1)} \right\}. \end{align*} } Any integral with the form of the first line of \cref{eqn:I_2lin_eval} will show an ultraviolet divergence when the two sBFs have the same order and the same argument. In practice, the covariance matrix is computed using spherical shell bins for the galaxy pair separations, which will eliminate the divergence at $r=r^{\prime}$ as it will be integrated over and is on a set of measure zero.\footnote{By using the leading-order asymptotic expansion for the sBF at high $k$ ($j_{\ell}(kr) \to \sin\mleft(kr-\ell\pi/2\mright)/(kr)$ \cite{NIST}), then binning the separations $r$ and $r^{\prime}$, and finally taking the two separations to be equal, it can be shown that the ultraviolet divergence will be eliminated from \cref{eqn:I_2lin_eval} for two sBFs with equal orders and arguments.}

We now take the $\ell=0$ limit of \cref{eqn:I_2lin_eval} and retain only the $r^{\prime} > r$ term as is required for \cref{eqn:2pcf_cov_in_terms_of_int}:
\begin{align}
\label{eqn:I_2lin_eval_ell0_notanh}
    I^{[2,\mathrm{lin}]}_0(r,r^{\prime}) &= H(1-\chi)\;g^{-2}\chi\twoFone\mleft(1,\frac{1}{2};\frac{3}{2};\chi^2\mright).
\end{align}
The hypergeometric function in \cref{eqn:I_2lin_eval_ell0_notanh} reduces to an inverse hyperbolic tangent (\cite{GR} equation 9.121.7):
\begin{align}
\label{eqn:2f1_to_tanh_inv}
    \twoFone\mleft(1,\frac{1}{2};\frac{3}{2};\chi^2\mright) &= (2\chi)^{-1}\ln\mleft(\frac{1+\chi}{1-\chi}\mright)
    \nonumber \\
    &= \chi^{-1} \tanh^{-1}(\chi).
\end{align}
Inserting \cref{eqn:2f1_to_tanh_inv} in \cref{eqn:I_2lin_eval_ell0_notanh}, we obtain
\setlength{\fboxrule}{0.5pt} 
\begin{empheq}[box=\fbox]{align}
\label{eqn:I_2lin_eval_ell0}
    I^{[2,\mathrm{lin}]}_0(r,r^{\prime}) &= H(1-\chi) \, g^{-2} \tanh^{-1}(\chi).
\end{empheq} 

Finally, the double-sBF integral with a quadratic power law in $k$, \cref{eqn:I_2quad}, can be evaluated using the sBF closure relation (\cite{GR} equation 6.512.8):
\setlength{\fboxrule}{0.5pt} 
\begin{empheq}[box=\fbox]{align}
\label{eqn:I_2quad_eval}
    I^{[2,\mathrm{quad}]}_{\ell}(r,r^{\prime}) &= \int_{0}^{\infty}k^2\dl k\;j_{\ell}(kr)j_{\ell}(kr^{\prime})
    \nonumber \\
    &= \frac{\pi}{2g^3}\dD(\chi-1).
\end{empheq} 
$\dD(\chi-1)$ is the one-dimensional Dirac $\delta$. \Cref{eqn:I_2quad_eval} thus only contributes to the covariance matrix when $r=r^{\prime}$; this is because the $I^{[2,\mathrm{quad}]}_{\ell}(r,r^{\prime})$ term stems purely from the shot noise in the integrand of \cref{eqn:2pcf_cov_with_pk}. Shot noise is only present when galaxies overlap, which was depicted in \cref{fig:shotnoise}.

The analytic 2PCF covariance matrix can now be obtained by inserting \cref{eqn:I_2const_eval_ell0,eqn:I_2lin_eval_ell0,eqn:I_2quad_eval} into \cref{eqn:2pcf_cov_in_terms_of_int} (recalling that $r^{\prime} \geq r$):
\setlength{\fboxrule}{1.5pt}
\begin{empheq}[box=\fbox]{align}
\label{eqn:2PCF_FINAL}
    \mathrm{Cov}(r,r^{\prime}) &= \frac{1}{2\pi V}\left(\frac{A}{g}\right)^2\left\{H(1-\chi)\left[g\sqrt{\chi} + 4\left(\pi A \bar{n} \right)^{-1} \tanh^{-1}(\chi)\right]\right.
    \nonumber \\
    &\left. \qquad + 4g\;\theta(\chi-1)\theta(1-\chi) + \left(A^2\bar{n}^2g \right)^{-1}\dD(\chi-1) \right\} .
\end{empheq} 
This is our final result for the unbinned covariance matrix of the 2PCF. The first two terms of \cref{eqn:2PCF_FINAL} (proportional to $H(1-\chi)$) contribute when $r^{\prime} > r$ and the final two terms (proportional to $\theta(\chi-1)\theta(1-\chi)$ and $\dD(\chi-1)$) contribute when $r = r^{\prime}$. When the covariance is computed with radial bins in $r$ and $r^{\prime}$, the term proportional to the Heavisides $\theta(\chi-1)\theta(1-\chi)$ will vanish, and the apparent singularity in the Dirac $\delta$ will produce a finite contribution to the covariance.

\subsection{Structure of the Covariance}
\label{sec:2pcf_structure}
We define the following quantities to explore the behavior of \cref{eqn:2PCF_FINAL}:
\begin{align}
\label{eqn:dimensionless_variables}
    N \equiv \bar{n}V, \;\;\;\; N_{\mathrm{g}} \equiv \frac{4}{3}\pi \bar{n}g^3, \;\;\;\; \eta &\equiv k_*^{-1} = (A\bar{n})^{-1}.
\end{align}
$N$ is the total number of galaxies in the survey, determined by the survey number density $\bar{n}$ and the volume of the survey $V$. $N_{\mathrm{g}}$ is the number of galaxies inside a sphere with radius set by the geometric mean of the two separations ($r$ and $r^{\prime}$) entering the covariance. $\eta$, in units of \si{\per\hHubble\Mpc}, is the length scale set by $k_*$, which is the scale of cosmic variance–shot noise equality.

Using the quantities defined in \cref{eqn:dimensionless_variables}, the covariance of the 2PCF (\pcref{eqn:2PCF_FINAL}) becomes
\setlength{\fboxrule}{1.5pt}
\begin{empheq}[box=\fbox]{align}
\label{eqn:2pcf_cov_structure}
    \mathrm{Cov}(r,r^{\prime}) &= \frac{2}{3NN_{\mathrm{g}}}\left\{\left(\frac{g}{\eta}\right)^2 \left[H(1-\chi)\sqrt{\chi} + 4\theta(\chi-1)\theta(1-\chi) \right] \right. \nonumber \\
    &\left. \qquad + H(1-\chi)\frac{4}{\pi }\left(\frac{g}{\eta}\right) \tanh^{-1}(\chi) + \dD(\chi-1) \right\}.
\end{empheq} 
Here, the covariance is split into three terms by the power of $g/\eta$. We recall that $g$ is the geometric mean of the pair separations $r$ and $r^{\prime}$, and $\eta$ shows whether cosmic variance (which is captured by the physical power spectrum) is more dominant than shot noise.

When $g/\eta>1$, the physical power spectrum dominates over shot noise. Conversely, when $g/\eta<1$, shot noise contributes more to the covariance than the physical power spectrum. When $g/\eta=1$, the physical power spectrum and shot noise contribute equally.

\section{Covariance Matrices for Higher-Order Correlation Functions}
\label{sec:higher_order_matrices}
We now examine the covariance matrices of the 3- and 4PCF. We first display (\cref{sec:3pcf}) the covariance of the 3PCF in the isotropic basis, where the covariance depends on the 2PCF and triple-sBF integrals. Then, in \cref{sec:2pcf_from_pk}, we use our power spectrum model to analytically evaluate the 2PCF. Finally, in \cref{sec:4pcf}, we show the covariance of the 4PCF in the isotropic basis, detailing the two coupling cases that contribute to the covariance. The covariance of the 4PCF depends on triple-sBF integrals, similar to the covariance of the 3PCF.

\subsection{Covariance Matrix of the 3PCF}
\label{sec:3pcf}

\begin{figure}
    \centering  
    \includegraphics[width=6cm]{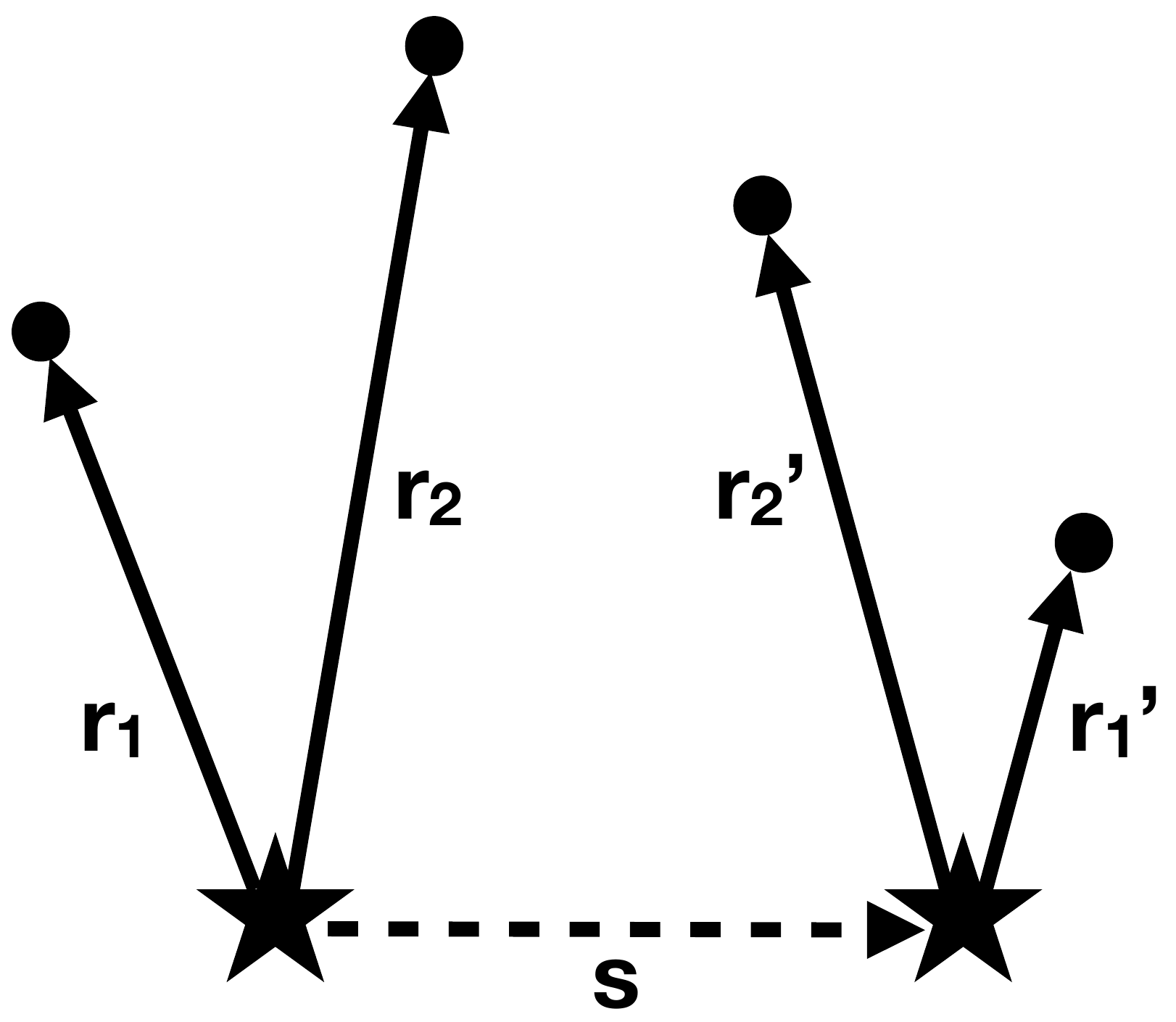}

    \caption{The covariance matrix of the 3PCF is formed by correlating all possible pairs of galaxy triplets, where each vertex of the triplet is a galaxy. Each triplet has a primary vertex (starred) from which two side lengths (solid arrows) extend, connecting the primary vertex to endpoint vertices (indicated with dots). The primaries in each triplet are separated by a distance $s$ (dashed arrow). When computing the 3PCF and its covariance matrix, one must cyclically sum over the galaxies such that each has a chance to serve as the primary vertex.}
    \label{fig:3pcf_diagram}
\end{figure}

The 3PCF measures the excess clustering of galaxy triplets above that expected from a spatially uniform random distribution. The covariance matrix of the 3PCF measures the correlations between these galaxy triplets, as shown schematically in \cref{fig:3pcf_diagram}. $\{r_1,r_2\}$ are the side lengths measured from the galaxy at a primary vertex in one triplet to the other two galaxies in the same triplet. $\{r^{\prime}_1,r^{\prime}_2\}$ are similarly measured for a second triplet. All side lengths are nonzero, since including zero-valued side lengths does not add any cosmological information. $s$ is the separation between the primary vertices of each triplet. $\{\ell,\ell^{\prime},\ell^{\prime\prime}\}$ represent the spherical harmonics that are used to write the covariance matrix in terms of isotropic basis functions \cite{iso}.

The 3PCF covariance matrix is written in the Legendre basis in \cite{ZS_3pcf_N2}; in the isotropic basis \cite{iso} this becomes\footnote{Equation 3 of \cite{iso} may be used to convert from the Legendre basis to the isotropic basis.}
\begin{align}
\label{eqn:3pcf_cov_orig}
    \mathrm{Cov}_{\ell,\ell^{\prime}}(r_1,r_2;r^{\prime}_1,r^{\prime}_2) &= (4\pi)^3\sum_{\ell^{\prime\prime}}\sqrt{(2\ell+1)(2\ell^{\prime}+1)}(2\ell^{\prime\prime}+1)\tj{\ell}{\ell^{\prime}}{\ell^{\prime\prime}}^2
    \nonumber \\ 
    & \qquad \times \int_0^{\infty}\frac{s^2\dl s}{V}\; \Bigl\{ (-1)^{\ell^{\prime\prime}}\Big[\xi(s)f_{\ell,\ell^{\prime},\ell^{\prime\prime}}(r_1,r^{\prime}_1,s)f_{\ell,\ell^{\prime},\ell^{\prime\prime}}(r_2,r^{\prime}_2,s) \Big.\Bigl.
    \nonumber \\
    &\Bigl.\Big. \qquad\qquad + \xi(s)f_{\ell,\ell^{\prime},\ell^{\prime\prime}}(r_1,r^{\prime}_2,s)f_{\ell,\ell^{\prime},\ell^{\prime\prime}}(r_2,r^{\prime}_1,s) \Big]
    \nonumber \\
    & \qquad + (-1)^{(\ell+\ell^{\prime}+\ell^{\prime\prime})/2} \Bigl. \Bigl.\Big[f_{0,\ell^{\prime},\ell^{\prime}}(0,r^{\prime}_1,s)f_{\ell,0,\ell}(r_1,0,s)f_{\ell,\ell^{\prime},\ell^{\prime\prime}}(r_2,r^{\prime}_2,s) \Big.\Bigl.
    \nonumber \\
    &\Bigl.\Big. \qquad\qquad + f_{0,\ell^{\prime},\ell^{\prime}}(0,r^{\prime}_1,s)f_{\ell,\ell^{\prime},\ell^{\prime\prime}}(r_1,r^{\prime}_2,s)f_{\ell,0,\ell}(r_2,0,s) \Big.\Bigl.\nonumber \\
    &\Bigl.\Big. \qquad\qquad + f_{0,\ell^{\prime},\ell^{\prime}}(0,r^{\prime}_2,s)f_{\ell,0,\ell}(r_1,0,s)f_{\ell,\ell^{\prime},\ell^{\prime\prime}}(r_2,r^{\prime}_1,s) \Big.\Bigl. \nonumber \\
    & \qquad\qquad + \Big.\Bigl. f_{0,\ell^{\prime},\ell^{\prime}}(0,r^{\prime}_2,s)f_{\ell,\ell^{\prime},\ell^{\prime\prime}}(r_1,r^{\prime}_1,s)f_{\ell,0,\ell}(r_2,0,s) \Big]\Bigl\},
\end{align}
where $V$ is the survey volume, $\xi(s)$ is the 2PCF, and $f_{\ell,\ell^{\prime},\ell^{\prime\prime}}$ is defined in \cref{eqn:f_orig_def}. The ``matrix'' in the first line of \cref{eqn:3pcf_cov_orig} is a Wigner 3-$j$ symbol. Since the entries in the bottom row are all zero, the Wigner 3-$j$ symbol is only nonzero if the parity of $\ell^{\prime\prime}$ matches that of $\ell+\ell^{\prime}$. $\ell^{\prime\prime}$ is controlled by the triangle inequalities
\begin{align}
\label{eqn:ell_triangle_3pcf}
    |\ell-\ell^{\prime}| \leq \ell^{\prime\prime} \leq \ell+\ell^{\prime}
\end{align}
and thus for fixed $\ell$ and $\ell^{\prime}$, $\ell^{\prime\prime}$ is always finite in range.

The $f$-integrals $f_{\ell,\ell^{\prime},\ell^{\prime\prime}}(r_i,r^{\prime}_j,s)$ \cite{cov} of \cref{eqn:3pcf_cov_orig} are defined as
\setlength{\fboxrule}{0.5pt} 			
\begin{empheq}[box=\fbox]{align}
\label{eqn:f_orig_def}
    f_{\ell,\ell^{\prime},\ell^{\prime\prime}}(r_i,r^{\prime}_j,s) &\equiv \int_{0}^{\infty}\frac{k^2\dl k}{2\pi^2}\;P(k)j_{\ell}(kr_i)j_{\ell^{\prime}}(kr^{\prime}_j)j_{\ell^{\prime\prime}}(ks);
\end{empheq} 
we will further explain this definition and analytically evaluate the $f$-integrals in \cref{sec:f_int}. Before turning to the covariance matrix of the 4PCF, we will now evaluate $\xi(s)$, which appears in \cref{eqn:3pcf_cov_orig}.

\subsection{2PCF from Model Power Spectrum}
\label{sec:2pcf_from_pk}
The 2PCF is represented as $\xi(s)$. In general, the 2PCF is obtained by taking the inverse Fourier transform of the power spectrum. Since the power spectrum is isotropic (in this work, we neglect redshift space distortions, \textit{e.\,g.}, \cite{kaiser_rsd}), this inverse Fourier transform becomes a transform with the order zero sBF via the plane wave expansion. We then have
\setlength{\fboxrule}{0.5pt} 			
\begin{empheq}[box=\fbox]{align}
\label{eqn:2pcf_from_pk_general}
    \xi(s) &= \int_{0}^{\infty}\frac{k^2\dl k}{2\pi^2}\;P(k)j_0(ks).
\end{empheq} 
The 2PCF from the power spectrum model in \cref{eqn:pk} is then
\begin{align}
\label{eqn:2pcf_from_pk_model}
    \xi(s) &= \frac{1}{2\pi^2}\int_{0}^{\infty}k^2\dl k\;\left[\frac{A}{k}+\frac{1}{\bar{n}}\right]j_0(ks)
    \nonumber \\
    &= \frac{A}{2\pi^2}\int_{0}^{\infty}k\dl k\;j_0(ks) + \frac{1}{\bar{n}}\delta_{\mathrm{D}}^{[3]}(\vs)
    \nonumber \\
    &= \frac{A}{2\pi^2s}\int_{0}^{\infty}\dl k\; \sin(ks) + \frac{1}{4\pi \bar{n}s^2}\dD(s)
    \nonumber \\
    &= \frac{A}{2\pi^2s}\int_{0}^{\infty}\dl k\; \mathrm{Im}[e^{iks}] + \frac{1}{4\pi \bar{n}s^2}\dD(s).
\end{align}
In the second line of \cref{eqn:2pcf_from_pk_model}, the first term is from the physical part of the power spectrum ($A/k$), while the second term is the inverse Fourier transform of a constant ($1/\bar{n}$), resulting in a three-dimensional Dirac $\delta$ with a vector argument, $\mathbf{s}$. We used $j_0(ks) = \sin(ks)/(ks)$ and the spherical symmetry of the 3D Dirac $\delta$ to obtain the third line above. The fourth line of \cref{eqn:2pcf_from_pk_model} stems from applying Euler's formula, $e^{ix} = \cos(x)  + i \sin(x)$, to rewrite $\sin(ks)$ as the imaginary part of an exponential. 

To avoid an exponential divergence from performing the integration over $k$, we now insert a factor $\exp(-\alpha k)$ in the integrand, where we will eventually take the $\alpha \rightarrow 0$ limit. \Cref{eqn:2pcf_from_pk_model} then becomes
\begin{align}
\label{eqn:2pcf_with_alpha}
    \xi(s) &= \frac{A}{2\pi^2s}\mathrm{Im}\mleft[\int_{0}^{\infty}\dl k\;e^{-k(\alpha-is)}\mright] + \frac{1}{4\pi \bar{n}s^2}\dD(s)
    \nonumber \\
    &= \frac{A}{2\pi^2s}\mathrm{Im}[(\alpha - is)^{-1}] + \frac{1}{4\pi \bar{n}s^2}\dD(s)
    \nonumber \\
    &= \frac{A}{2\pi^2s}\mathrm{Im}\mleft[\frac{\alpha + is}{\alpha^2 + s^2} \mright] + \frac{1}{4\pi \bar{n}s^2}\dD(s).
\end{align}
The second line of \cref{eqn:2pcf_with_alpha} results from performing the integration over $k$ in the first term. The first term of the third line is obtained after multiplying by unity.

After taking the imaginary part of the second line in the limit $\alpha \rightarrow 0$, we obtain our final result for the 2PCF using the power spectrum model of \cref{eqn:pk}:
\setlength{\fboxrule}{1.5pt} 			
\begin{empheq}[box=\fbox]{align}
\label{eqn:2pcf}
    \xi(s) &= \frac{A}{2\pi^2s^2} + \frac{1}{4\pi \bar{n}s^2}\;\dD(s),
\end{empheq} 
where we recall that the second term could be written in terms of the three-dimensional Dirac $\delta$ as $\delta_{\mathrm{D}}^{[3]}(\vs)/\bar{n}$. \Cref{eqn:2pcf} will tend to infinity at $s=0$ due to the inverse square power of $s$ as well as the Dirac $\delta$.

\subsection{Covariance Matrix of the 4PCF}
\label{sec:4pcf} 

\begin{figure}
    \centering
    %\vspace{-2cm}
    \begin{subfigure}[b]{0.45\textwidth}
        \centering
        \includegraphics[width=\textwidth]{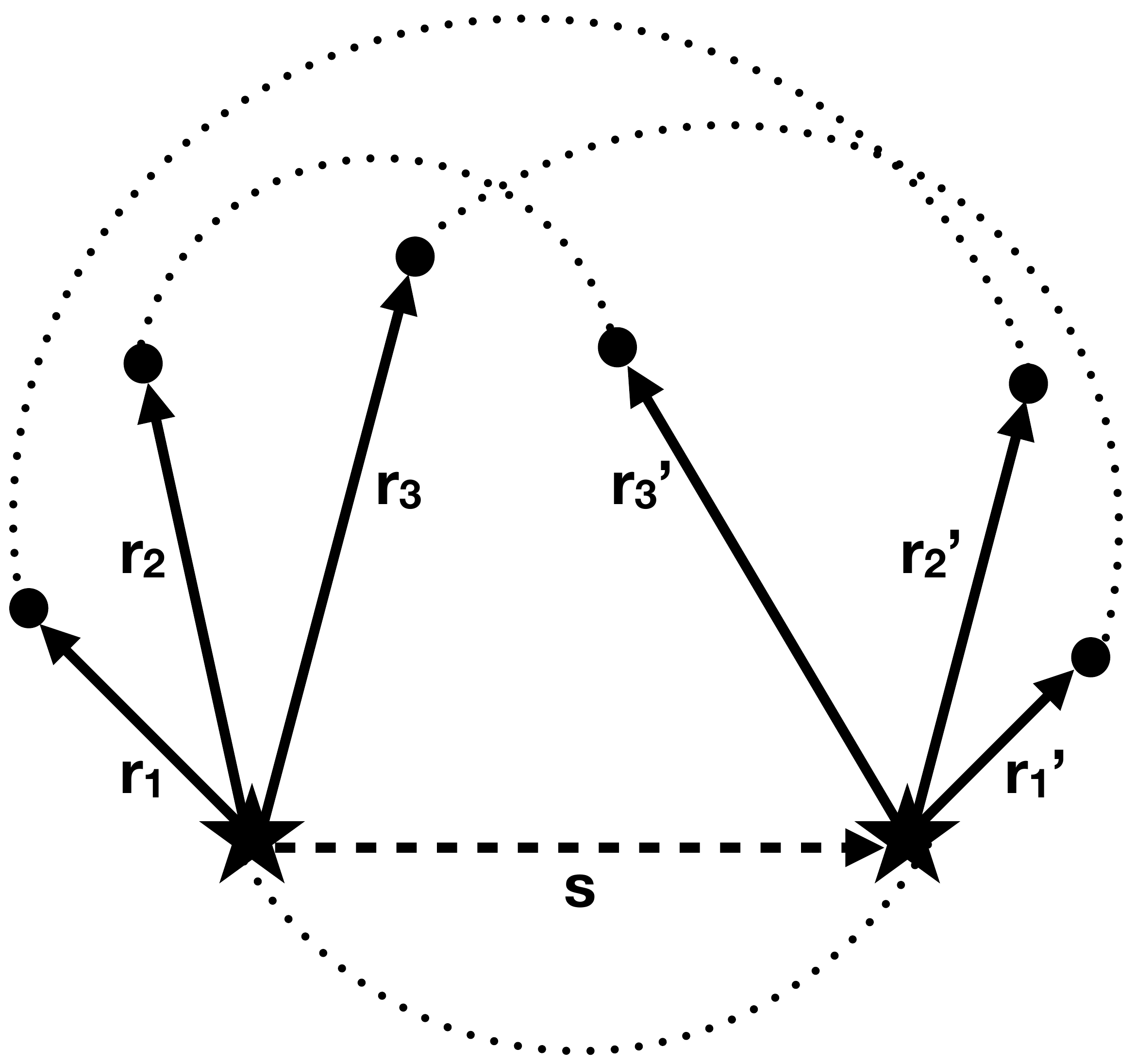}
    \end{subfigure}
    \hspace{0.6cm}
    \begin{subfigure}[b]{0.45\textwidth}
        \centering
        \includegraphics[width=\textwidth]{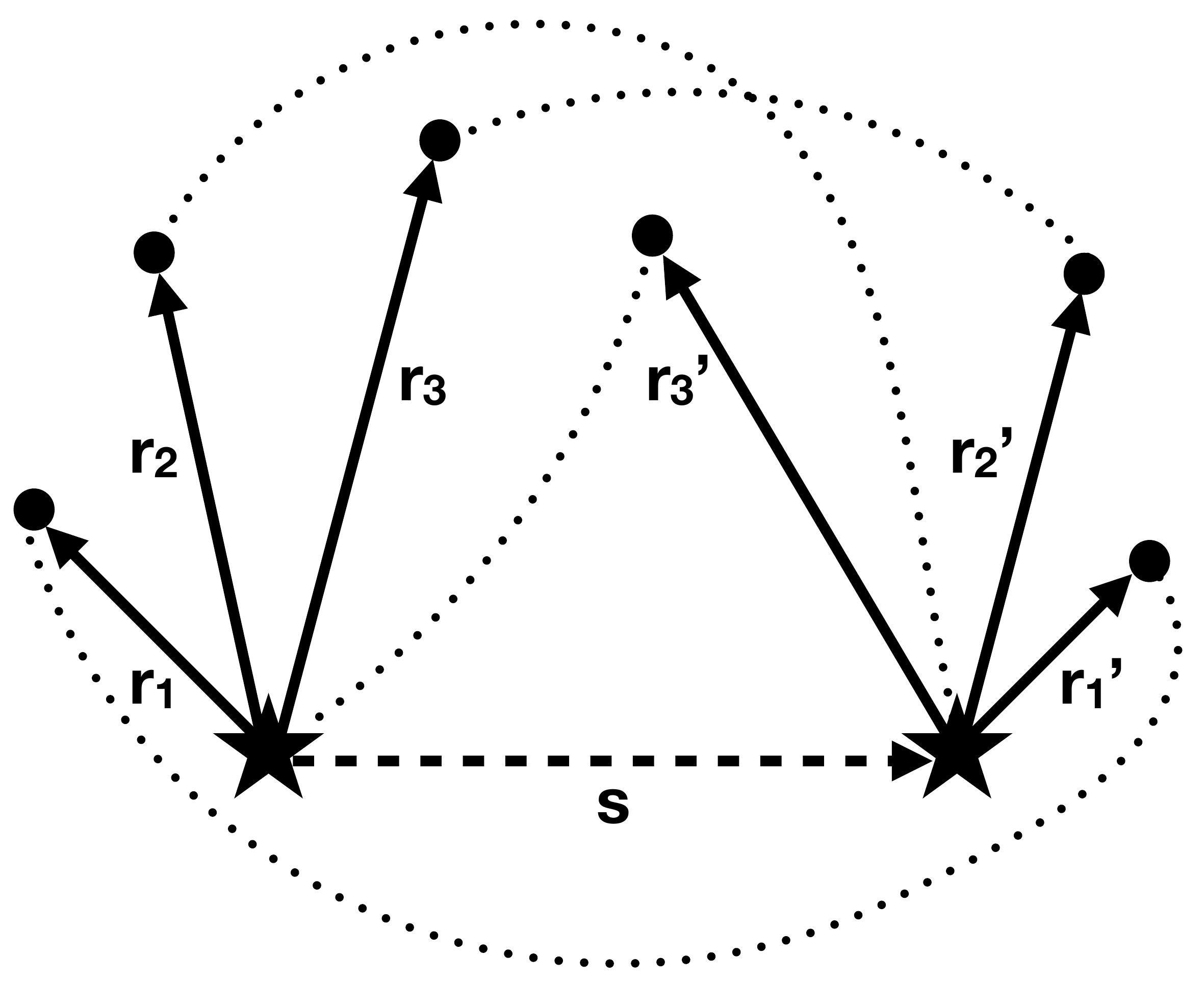}
    \end{subfigure}

    \caption{The covariance of the 4PCF requires correlating each possible combination of two tetrahedra. Each tetrahedron has a primary vertex (starred). The side lengths (solid arrows) of the tetrahedron are measured from the primary vertex to the endpoint vertices (indicated with dots) within the same tetrahedron. The primary vertices of the two correlated tetrahedra are separated by a distance $s$ (dashed arrow). When computing the 4PCF and covariance of the 4PCF, one must cycle over all choices for the primary vertex. The covariance of the 4PCF has two coupling cases, depicted respectively by the left-hand and right-hand diagrams above. \textit{Left-hand diagram:} The primary vertices of each tetrahedra are coupled and the endpoint vertices within opposite tetrahedra are coupled. Each possible coupling of endpoint vertices must be accounted for. \textit{Right-hand diagram:} Each primary vertex is coupled to an endpoint vertex within the opposite tetrahedron. The remaining endpoint vertices within opposite tetrahedra are also coupled. Each possible coupling of vertices must be taken into account.}
    \label{fig:4pcf_diagram}
\end{figure}

The 4PCF measures the excess clustering of tetrahedra, as a function of tetrahedron shape, beyond that which a spatially random distribution of points would have. The covariance matrix of the 4PCF represents the correlations between pairs of tetrahedral configurations. \Cref{fig:4pcf_diagram} shows the two different possibilities for the covariance in the GRF limit: Case \rom{1}, where the primary vertices are coupled (\cref{fig:4pcf_diagram}, left-hand diagram), and Case \rom{2}, where each primary vertex is coupled to an endpoint (non-primary) vertex in the opposite tetrahedron (\cref{fig:4pcf_diagram}, right-hand diagram). Each of these two cases will be described in the following subsections.

As shown in \cref{fig:4pcf_diagram}, the side lengths (unprimed for the first tetrahedron and primed for the second tetrahedron) are measured with respect to the galaxy at the primary vertex of each tetrahedron. The side lengths are all nonzero, since including the zero-valued side lengths in our calculations does not add cosmological information. The primary vertices of the two tetrahedra are separated by a distance $s$. The various $\{\ell, \ell^{\prime}, L \}$ in the following two subsections represent the spherical harmonics composing the isotropic basis functions used for the covariance.

\subsubsection{Case \rom{1}}
\label{ssec:4pcf_case1}
We begin with Case \rom{1} for the covariance of the 4PCF, where primary vertices are coupled to one another and the endpoints of the primed and unprimed tetrahedra are coupled. The Case \rom{1} covariance matrix is given in equation 33 of \cite{cov}; we duplicate it below:
\begin{align}
\label{eqn:c1_orig}
    \mathrm{Cov}^{[\text{\rom{1}}]}_{\ell_{Gi},\ell^{\prime}_i}(r_{Gi},r^{\prime}_i) &= (4\pi)^4\sum_{G}(-1)^{(\ell_1+\ell_2+\ell_3)(1-\mathcal{E}_G)/2}\sum_{L_1L_2L_3}\sqrt{(2L_1+1)(2L_2+1)(2L_3+1)} \nonumber\\
    &\qquad \times (-1)^{L_1+L_2+L_3}\tj{L_1}{L_2}{L_3} \nj{\ell_{G1}}{\ell_{G2}}{\ell_{G3}}{\ell^{\prime}_1}{\ell^{\prime}_2}{\ell^{\prime}_3}{L_1}{L_2}{L_3} \nonumber \\
    &\qquad \times \prod_{i=1}^3 \left[(-1)^{(\ell_{Gi}+\ell^{\prime}_i+3L_i)/2}\sqrt{(2\ell_{Gi}+1)(2\ell^{\prime}_i+1)(2L_i+1)} \right. \nonumber \\
    &\left.\qquad \times \tj{\ell_{Gi}}{\ell^{\prime}_i}{L_i} \int_{0}^{\infty}\frac{s^2\dl s}{V}\;\xi(s)f_{\ell_{Gi},\ell^{\prime}_i,L_i}(r_{Gi},r^{\prime}_i,s)\right].
\end{align}

In \cref{eqn:c1_orig}, the subscript $G$ acts on the index $i$. This denotes the permutation of the three endpoint galaxies, since all possible permutations must be accounted for. $\mathcal{E}_G$ is $1$ if $\{G1,G2,G3\}$ is an even permutation and $-1$ otherwise. 

The $3\times3$ matrix in braces is a Wigner 9-$j$ symbol coupling the spherical harmonics. The integral over the separation $s$ depends on $\xi(s)$ (the 2PCF) and the $f$-integral (which will be discussed in \cref{sec:f_int}), similarly to the covariance of the 3PCF (\pcref{eqn:3pcf_cov_orig}).

By replacing the 2PCF of \cref{eqn:c1_orig} with \cref{eqn:2pcf}, we obtain
\begin{align}
\label{eqn:c1_with_2pcf}
    \mathrm{Cov}^{[\text{\rom{1}}]}_{\ell_{Gi},\ell^{\prime}_i}(r_{Gi},r^{\prime}_i) &= (4\pi)^4\sum_{G}(-1)^{(\ell_1+\ell_2+\ell_3)(1-\mathcal{E}_G)/2}\sum_{L_1L_2L_3}\sqrt{(2L_1+1)(2L_2+1)(2L_3+1)}
    \nonumber \\
    &\qquad \times (-1)^{L_1+L_2+L_3}\tj{L_1}{L_2}{L_3} \nj{\ell_{G1}}{\ell_{G2}}{\ell_{G3}}{\ell^{\prime}_1}{\ell^{\prime}_2}{\ell^{\prime}_3}{L_1}{L_2}{L_3}
    \nonumber \\
    &\qquad \times \prod_{i=1}^3 \left\{(-1)^{(\ell_{Gi}+\ell^{\prime}_i+3L_i)/2}\sqrt{(2\ell_{Gi}+1)(2\ell^{\prime}_i+1)(2L_i+1)} \right.
    \nonumber \\
    &\left.\qquad \times \tj{\ell_{Gi}}{\ell^{\prime}_i}{L_i}\left[\frac{A}{2\pi^2V}\int_{0}^{\infty}\dl s\;f_{\ell_{Gi},\ell^{\prime}_i,L_i}(r_{Gi},r^{\prime}_i,s) \right.\right.
    \nonumber \\
    &\left.\left. \qquad\qquad+ \frac{1}{4\pi\bar{n}V}\int_{0}^{\infty}\dl s\;\dD(s)f_{\ell_{Gi},\ell^{\prime}_i,L_i}(r_{Gi},r^{\prime}_i,s) \right] \right\}.
\end{align}
The final integral of \cref{eqn:c1_with_2pcf} will only be nonzero when $s=0$ by the sifting property of the Dirac $\delta$. This integral will then reduce to the $f$-integral evaluated at $s=0$: $f_{\ell_{Gi},\ell^{\prime}_i,L_i}(r_{Gi},r^{\prime}_i,0)$. Recalling the definition of the $f$-integral (\pcref{eqn:f_orig_def}), we see that $s=0$  corresponds to $j_{L_i}(ks) = j_{L_i}(0)$. $j_0(0)=1$ is the only sBF that is nonzero at zero argument. Thus, the final integral of \cref{eqn:c1_with_2pcf} becomes $f_{\ell_{Gi},\ell^{\prime}_i,0}(r_{Gi},r^{\prime}_i,0)$.

The 3-$j$ symbol in \cref{eqn:c1_with_2pcf} requires that $\ell_{Gi}+\ell^{\prime}_i+L_i$ is even and also that
\begin{align}
\label{eqn:ell_triangle}
    |\ell_{Gi}-\ell^{\prime}_i| \leq L_i \leq \ell_{Gi}+\ell^{\prime}_i.
\end{align}
When $L_i=0$, $\ell_{Gi}$ must be equal to $\ell^{\prime}_i$. Hence, the final integral in \cref{eqn:c1_with_2pcf} becomes an $f$-integral with one sBF of order zero and argument zero, and with the other two sBF orders equal.

We may now rewrite the Case $\text{\rom{1}}$ covariance as
\begin{align}
\label{eqn:c1_final}
    \mathrm{Cov}^{[\text{\rom{1}}]}_{\ell_{Gi},\ell^{\prime}_i}(r_{Gi},r^{\prime}_i) &= (4\pi)^4\sum_{G}(-1)^{(\ell_1+\ell_2+\ell_3)(1-\mathcal{E}_G)/2}\sum_{L_1L_2L_3}\sqrt{(2L_1+1)(2L_2+1)(2L_3+1)}
    \nonumber\\
    &\qquad \times (-1)^{L_1+L_2+L_3}\tj{L_1}{L_2}{L_3} \nj{\ell_{G1}}{\ell_{G2}}{\ell_{G3}}{\ell^{\prime}_1}{\ell^{\prime}_2}{\ell^{\prime}_3}{L_1}{L_2}{L_3}
    \nonumber \\
    &\qquad \times \prod_{i=1}^3 \left\{(-1)^{(\ell_{Gi}+\ell^{\prime}_i+3L_i)/2}\sqrt{(2\ell_{Gi}+1)(2\ell^{\prime}_i+1)(2L_i+1)} \right.
    \nonumber \\
    &\left.\qquad \times \tj{\ell_{Gi}}{\ell^{\prime}_i}{L_i}\left[\frac{A}{2\pi^2V}\int_{0}^{\infty}\dl s\;f_{\ell_{Gi},\ell^{\prime}_i,L_i}(r_{Gi},r^{\prime}_i,s) \right.\right.
    \nonumber \\
    &\left.\left. \qquad\qquad+ \frac{1}{4\pi\bar{n}V}\;\kD_{\ell_{Gi}\ell^{\prime}_i}\;\kD_{L_i0}\;f_{\ell_{Gi},\ell^{\prime}_i,0}(r_{Gi},r^{\prime}_i,0) \right] \right\}.
\end{align}
We have included Kronecker $\delta$s ($\kD$) in the final term of \cref{eqn:c1_final} to indicate that it is zero unless $L_i = 0$, which causes $\ell_{Gi}$ to equal $\ell^{\prime}_i$. 

\subsubsection{Case \rom{2}}
\label{ssec:4pcf_case2}
Case \rom{2} for the covariance of the 4PCF arises when one primary vertex is coupled to an endpoint vertex within the opposite tetrahedron. The covariance matrix contribution from this coupling is given by equation 41 of \cite{cov}; we duplicate it here:
\begin{align}
\label{eqn:c2_orig}
    \mathrm{Cov}^{[\text{\rom{2}}]}_{\ell_{Gi},\ell^{\prime}_{Hi}}(r_{Gi},r^{\prime}_{Hi}) &= (4\pi)^4\sum_{G,H}(-1)^{(\ell^{\prime}_1+\ell^{\prime}_2+\ell^{\prime}_3)(1-\mathcal{E}_H)/2}\sum_{L_1,L_2,L_3}\sqrt{(2L_1+1)(2L_2+1)} \nonumber \\
    &\qquad \times \sqrt{2L_3+1}(-1)^{L_1+L_2+L_3} \tj{L_1}{L_2}{L_3} \nj{\ell_{G1}}{\ell_{G2}}{\ell_{G3}}{\ell^{\prime}_{H1}}{\ell^{\prime}_{H2}}{\ell^{\prime}_{H3}}{L_1}{L_2}{L_3} \nonumber \\
    &\qquad \times \prod_{i=1}^3\left[ (-1)^{(\ell_{Gi}+\ell^{\prime}_{Hi}+3L_i)/2} \sqrt{(2\ell_{Gi}+1)(2\ell^{\prime}_{Hi}+1)(2L_i+1)} \right. \nonumber \\
    &\left.\qquad \times \tj{\ell_{Gi}}{\ell^{\prime}_{Hi}}{L_i} \right] \int_0^{\infty}\frac{s^2\dl s\;}{V} \left[ f_{\ell_{G1},0,\ell_{G1}}(r_{G1},0,s) \right. \nonumber \\
    &\left.\qquad \times f_{0,\ell^{\prime}_{H1},\ell^{\prime}_{H1}}(0,r^{\prime}_{H1},s)f_{\ell_{G2},\ell^{\prime}_{H2},L_2}(r_{G2},r^{\prime}_{H2},s) \right. \nonumber \\
    &\left.\qquad \times f_{\ell_{G3},\ell^{\prime}_{H3},L_3}(r_{G3},r^{\prime}_{H3},s)\right].
\end{align}

In \cref{eqn:c2_orig}, $G$ represents the permutation of endpoint galaxies in the tetrahedron with unprimed side lengths. $H$ is a similar permutation for the tetrahedron with primed side lengths. As a result of the contraction of a primary vertex with endpoint vertices in the opposite tetrahedron, we may choose $G$ to be a cyclic permutation, as long as $H$ remains a standard permutation. $\mathcal{E}_H$ is $1$ when $\{H_1,H_2,H_3\}$ is a cyclic permutation and $-1$ when it is non-cyclic.

The integral over the separation $s$ in \cref{eqn:c2_orig} depends only on the product of $f$-integrals. We now focus on analytically evaluating these $f$-integrals.

\section{\boldmath Analytic Evaluation of the \texorpdfstring{$f$}{f}-Integrals}
\label{sec:f_int}
The covariance matrices of the 3- and 4PCF (\pcref{eqn:3pcf_cov_orig,eqn:c1_final,eqn:c2_orig}) depend on $f$-integrals \cite{cov}, which we recall were defined as
\begin{align}
\label{eqn:f_int}
    f_{\ell,\ell^{\prime},\ell^{\prime\prime}}(r_i,r^{\prime}_j,s) &\equiv \int_{0}^{\infty}\frac{k^2\dl k}{2\pi^2}\;P(k)j_{\ell}(kr_i)j_{\ell^{\prime}}(kr^{\prime}_j)j_{\ell^{\prime\prime}}(ks).
\end{align}
The $\{\ell,\ell^{\prime},\ell^{\prime\prime}\}$ represent the spherical harmonics used to write the covariance in the isotropic basis, and $\{r_i,r^{\prime}_j\}$ represent the side lengths in the unprimed and primed shapes of which we are computing the covariance. $s$ is the separation between these unprimed and primed shapes.

We now turn to the analytic evaluation of \cref{eqn:f_int}. The covariance of the 3PCF (\pcref{eqn:3pcf_cov_orig}) and both cases of the covariance of the 4PCF (\pcref{eqn:c1_final,eqn:c2_orig}) depend on the same two types of $f$-integrals. The first type has one sBF with both argument zero and order zero (\textit{i.\,e.} $j_0(0)$), and two sBFs with equal orders. The second type has three sBFs with nonzero arguments. We will evaluate each case separately.

Before we evaluate the $f$-integrals analytically, we use the power spectrum model (\pcref{eqn:pk}) to rewrite the $P(k)$ within \cref{eqn:f_int}; we then have
\setlength{\fboxrule}{0.5pt} 			
\begin{empheq}[box=\fbox]{align}
\label{eqn:f_int_pk}
    f_{\ell,\ell^{\prime},\ell^{\prime\prime}}(r_i,r^{\prime}_j,s) &= \int_{0}^{\infty}\frac{k^2\dl k}{2\pi^2}\;\left(\frac{A}{k}+\frac{1}{\bar{n}}\right)j_{\ell}(kr_i)j_{\ell^{\prime}}(kr^{\prime}_j)j_{\ell^{\prime\prime}}(ks).
\end{empheq} 
For each of the two types of $f$-integral previously described, we will use \cref{eqn:f_int_pk} as the starting point.

\subsection{One Zero-Argument sBF with Corresponding Order Zero}
\label{sec:f_one_zero_sbf}
Since the $f$-integrals depend on a product of three sBFs, the result we obtain here will hold for any sBF in that triple with argument zero. If the argument of an sBF is zero, the sBF is only nonzero if the order is also zero. Starting with \cref{eqn:f_int_pk}, then setting the order and argument of the second sBF equal to zero ($j_{\ell^{\prime}}(kr^{\prime}_j)=j_0(0)=1$), the first type of $f$-integral is
\setlength{\fboxrule}{0.5pt} 			
\begin{empheq}[box=\fbox]{align}
\label{eqn:f_int_zero_unequal_orders}
    f_{\ell,0,\ell^{\prime\prime}}(r_i,0,s) &= \int_{0}^{\infty}\frac{k^2\dl k}{2\pi^2}\;\left(\frac{A}{k}+\frac{1}{\bar{n}}\right)j_{\ell}(kr_i)j_{\ell^{\prime\prime}}(ks).
\end{empheq} 

The Wigner 3-$j$ and 9-$j$ symbols within the covariance matrices of the 3- and 4PCF (\pcref{eqn:3pcf_cov_orig,eqn:c1_final,eqn:c2_orig}) enforce triangle inequalities\footnote{The Wigner 9-$j$ symbol may be defined in terms of Wigner 3-$j$ symbols (\cite{NIST} equation 34.6.1) and thus restricts the sBF orders such that they satisfy the triangle inequalities dictated by the 3-$j$ symbols. There is such a constraint on each row and column of the Wigner 9-$j$ symbol.} on the orders of the sBFs. Therefore, if one sBF in the $f$-integral has order zero, the orders of the other two sBFs must be equal. We use this to rewrite \cref{eqn:f_int_zero_unequal_orders} as
\begin{align}
\label{eqn:f_int_zero}
    f_{\ell,0,\ell}(r_i,0,s) &= \int_{0}^{\infty}\frac{k^2\dl k}{2\pi^2}\;\left(\frac{A}{k}+\frac{1}{\bar{n}}\right)j_{\ell}(kr_i)j_{\ell}(ks).
\end{align}

Recalling the definitions of $I^{[2,\mathrm{lin}]}_{\ell}(r,r^{\prime})$ (\pcref{eqn:I_2lin}) and $I^{[2,\mathrm{quad}]}_{\ell}(r,r^{\prime})$ (\pcref{eqn:I_2quad}), we may rewrite \cref{eqn:f_int_zero}:
\begin{align}
\label{f_int_zero_in_terms_of_I}
    f_{\ell,0,\ell}(r_i,0,s) &= \frac{1}{2\pi^2}\left[AI^{[2,\mathrm{lin}]}_{\ell}(r_i,s)+\frac{1}{\bar{n}}I^{[2,\mathrm{quad}]}_{\ell}(r_i,s) \right].
\end{align}
We will now use the results for $I^{[2,\mathrm{lin}]}_{\ell}(r,r^{\prime})$ and $I^{[2,\mathrm{quad}]}_{\ell}(r,r^{\prime})$ (\pcref{eqn:I_2lin_eval,eqn:I_2quad_eval}, respectively) with the free arguments of the sBFs replaced by $r_i$ and $s$, to evaluate \cref{f_int_zero_in_terms_of_I}. Here, we neglect the $r_i=s$ contribution from \cref{eqn:I_2lin_eval}, since it is on a set of measure zero and will not contribute to the covariance. This is because the covariance matrices of the 3- and 4PCF (\pcref{eqn:3pcf_cov_orig,eqn:c1_final,eqn:c2_orig}) require products of $f$-integrals integrated over the separation $s$, and the integral of a function at one point (such as $r_i=s$) vanishes.

\Cref{f_int_zero_in_terms_of_I} is now evaluated using the results from \cref{eqn:I_2lin_eval,eqn:I_2quad_eval}:
\setlength{\fboxrule}{1.5pt} 			
\begin{empheq}[box=\fbox]{align}
\label{eqn:f_int_zero_eval}
    f_{\ell,0,\ell}(r_i,0,s) &= \frac{1}{4\pi^2r_is}\left\{\frac{A\sqrt{\pi}\;\Gamma(\ell+1)}{\Gamma(\ell+3/2)}\left[H(r_i-s)\left(\frac{s}{r_i}\right)^{\ell+1}  \right.\right. \nonumber \\
    &\left.\left. \qquad \times \twoFone\mleft(\ell+1,\frac{1}{2};\ell+\frac{3}{2}; \left(\frac{s}{r_i}\right)^2 \mright) + H(s-r_i)\left(\frac{r_i}{s}\right)^{\ell+1}  \right. \right. \nonumber \\
    &\left.\left. \qquad \times \twoFone\mleft(\ell+1,\frac{1}{2};\ell+\frac{3}{2}; \left(\frac{r_i}{s}\right)^2 \mright)\right] +\frac{\pi}{\bar{n}}\dD(r_i-s) \right\}.
\end{empheq} 

The hypergeometric functions in \cref{eqn:f_int_zero_eval} will converge when the argument ($(s/r_i)^2$ or the inverse) is less than unity, which is guaranteed by the Heaviside functions. The Dirac $\delta$ term of \cref{eqn:f_int_zero_eval} is only nonzero when $r_i=s$; at this point, it appears that a singularity is present. However, to obtain the covariance, one must integrate products of $f$-integrals. Both the hypergeometric functions and the Dirac $\delta$ are integrable and will result in finite contributions to the covariance.

\subsection{All Nonzero sBF Arguments}
\label{sec:f_nonzero_sbfs}
The second case of $f$-integrals needed for the covariance matrices of the 3- and 4PCF (\pcref{eqn:3pcf_cov_orig,eqn:c1_final,eqn:c2_orig}) requires that all three sBFs have nonzero arguments. Their orders may be zero or positive integers. This case was shown in \cref{eqn:f_int_pk}. 

We define two new integrals to evaluate \cref{eqn:f_int_pk}:
\begin{align}
\label{eqn:I_3lin}
    I^{[3,\mathrm{lin}]}_{\ell,\ell^{\prime},\ell^{\prime\prime}}(r_i,r^{\prime}_j,s) &\equiv \int_{0}^{\infty}k\dl k\;j_{\ell}(kr_i)j_{\ell^{\prime}}(kr^{\prime}_j)j_{\ell^{\prime\prime}}(ks), \\
\label{eqn:I_3quad}
    I^{[3,\mathrm{quad}]}_{\ell,\ell^{\prime},\ell^{\prime\prime}}(r_i,r^{\prime}_j,s) &\equiv \int_{0}^{\infty}k^2\dl k\;j_{\ell}(kr_i)j_{\ell^{\prime}}(kr^{\prime}_j)j_{\ell^{\prime\prime}}(ks).
\end{align}
The number in each superscript represents the number of sBFs, while ``lin'' and ``quad'' denote the power of $k$ in the integrand.
\Cref{eqn:I_3lin,eqn:I_3quad} can be used to rewrite \cref{eqn:f_int_pk}, yielding
\setlength{\fboxrule}{0.5pt} 			
\begin{empheq}[box=\fbox]{align}
\label{eqn:f_in_terms_of_I}
    f_{\ell,\ell^{\prime},\ell^{\prime\prime}}(r_i,r^{\prime}_j,s) = \frac{1}{2\pi^2}\left[AI^{[3,\mathrm{lin}]}_{\ell,\ell^{\prime},\ell^{\prime\prime}}(r_i,r^{\prime}_j,s)+\frac{1}{\bar{n}}I^{[3,\mathrm{quad}]}_{\ell,\ell^{\prime},\ell^{\prime\prime}}(r_i,r^{\prime}_j,s) \right].
\end{empheq} 

$I^{[3,\mathrm{lin}]}_{\ell,\ell^{\prime},\ell^{\prime\prime}}(r_i,r^{\prime}_j,s)$ and $I^{[3,\mathrm{quad}]}_{\ell,\ell^{\prime},\ell^{\prime\prime}}(r_i,r^{\prime}_j,s)$ (\pcref{eqn:I_3lin,eqn:I_3quad}, respectively), have the same general structure:
\setlength{\fboxrule}{0.5pt} 			
\begin{empheq}[box=\fbox]{align}
\label{eqn:I3_orig}
    I^{[3,n]}_{\ell,\ell^{\prime},\ell^{\prime\prime}}(r_i,r^{\prime}_j,s) &\equiv \int_{0}^{\infty}k^n\dl k\;j_{\ell}(kr_i)j_{\ell^{\prime}}(kr^{\prime}_j)j_{\ell^{\prime\prime}}(ks),
\end{empheq} 
with $n=1$ for $I^{[3,\mathrm{lin}]}_{\ell,\ell^{\prime},\ell^{\prime\prime}}(r_i,r^{\prime}_j,s)$ and $n=2$ for $I^{[3,\mathrm{quad}]}_{\ell,\ell^{\prime},\ell^{\prime\prime}}(r_i,r^{\prime}_j,s)$.

Although we may directly use equation 3.21 of \cite{k2_3sbf} to evaluate $I^{[3,\mathrm{quad}]}_{\ell,\ell^{\prime},\ell^{\prime\prime}}(r_i,r^{\prime}_j,s)$, we will instead simplify \cref{eqn:I3_orig} as much as possible before substituting in the specific values for $n$. We do this so that the final results we obtain for $I^{[3,\mathrm{lin}]}_{\ell,\ell^{\prime},\ell^{\prime\prime}}(r_i,r^{\prime}_j,s)$ and $I^{[3,\mathrm{quad}]}_{\ell,\ell^{\prime},\ell^{\prime\prime}}(r_i,r^{\prime}_j,s)$ will have the same structure, allowing for easier analysis of the structure of the $f$-integral we are seeking (\pcref{eqn:f_in_terms_of_I}), and thus of the covariance.

As shown in equation 37 of \cite{meigs}, $k^nj_{\ell^{\prime\prime}}(ks)$ can be rewritten as
\begin{align}
\label{eqn:k_times_jL}
    k^nj_{\ell^{\prime\prime}}(ks) &= \int_{0}^{\infty}k^{\prime n}\dl k^{\prime}\;j_{\ell^{\prime\prime}}(k^{\prime}s)\;\dD(k-k^{\prime}) \nonumber \\
    &= \frac{2}{\pi}\int_{0}^{\infty}k^{\prime n+2}\dl k^{\prime}\;j_{\ell^{\prime\prime}}(k^{\prime}s)\int_{0}^{\infty}u^2\dl u\;j_{\lambda}(ku)j_{\lambda}(k^{\prime}u).
\end{align}
The first line of \cref{eqn:k_times_jL} is obtained as a result of the sifting property of the Dirac $\delta$. The second line of \cref{eqn:k_times_jL} then comes from the first line by using the orthogonality relation for sBFs to rewrite the Dirac $\delta$. Since this orthogonality relation holds for any two sBFs of the same order, $\lambda$ is free. Using the second line of \cref{eqn:k_times_jL} to rewrite $k^nj_{\ell^{\prime\prime}}(ks)$ in \cref{eqn:I3_orig} gives
\setlength{\fboxrule}{0.5pt} 			
\begin{empheq}[box=\fbox]{align}
\label{eqn:I3_threeint}
    I^{[3,n]}_{\ell,\ell^{\prime},\ell^{\prime\prime}}(r_i,r^{\prime}_j,s) = \frac{2}{\pi}\int_{0}^{\infty}u^2\dl u\;\int_{0}^{\infty}k^{\prime n}\dl k^{\prime}j_{\ell^{\prime\prime}}(k^{\prime}s)j_{\lambda}(k^{\prime}u)\int_{0}^{\infty}k^2\dl k\;j_{\ell}(kr_i)j_{\ell^{\prime}}(kr^{\prime}_j)j_{\lambda}(ku),
\end{empheq} 
which is equation 38 in \cite{meigs}. 

The innermost $k$-integral of \cref{eqn:I3_threeint} can be evaluated with \cite{k2_3sbf} if the sum of the orders of the sBFs in the integrand, $\ell+\ell^{\prime}+\lambda$, is even and the orders satisfy the triangular inequalities:
\begin{align}
\label{eqn:triangle_mehrem}
    |\ell-\ell^{\prime}| \leq \lambda \leq \ell+\ell^{\prime}\;.
\end{align}
Since $\lambda$ is free, we may choose it to satisfy these conditions. Using equation 3.21 of \cite{k2_3sbf}, the innermost $k$-integral of \cref{eqn:I3_threeint} then becomes
\begin{align}
\label{eqn:mehrem_with_M}
    \int_{0}^{\infty}k^2\dl k\;j_{\ell}(kr_i)j_{\ell^{\prime}}(kr^{\prime}_j)j_{\lambda}(ku) &= \frac{\pi \beta(\Delta)}{4r_ir^{\prime}_ju}\;i^{\ell+\ell^{\prime}-\lambda}\sqrt{2\lambda+1}\left(\frac{r_i}{u} \right)^{\lambda}\tj{\ell}{\ell^{\prime}}{\lambda}^{-1} \nonumber \\
    & \qquad \times \sum_{\mathcal{L}=0}^\lambda\binom{2\lambda}{2\mathcal{L}}^{1/2}\left(\frac{r^{\prime}_j}{r_i}\right)^{\mathcal{L}}\sum_{J}(2J+1)\tj{\ell}{\lambda-\mathcal{L}}{J} \nonumber \\
    & \qquad \times \tj{\ell^{\prime}}{\mathcal{L}}{J}\sj{\ell}{\ell^{\prime}}{\lambda}{\mathcal{L}}{\lambda-\mathcal{L}}{J}P_J(\Delta).
\end{align}
$\binom{2\lambda}{2\mathcal{L}}$ is a binomial coefficient, the matrix in curly braces is a Wigner 6-$j$ symbol, and $P_J(\Delta)$ is a Legendre polynomial of order $J$. $\beta(\Delta)$ and $\Delta$ are defined below, as in \cite{k2_3sbf}:
\begin{align}
\label{eqn:beta_delta_mehrem}    
    \beta(\Delta) &\equiv \theta(1-\Delta)\theta(1+\Delta), \nonumber \\
    \Delta &\equiv \frac{r_i^2+r^{\prime 2}_j-u^2}{2r_ir^{\prime}_j}.
\end{align}
$\theta$ is the Heaviside function using the half-maximum convention ($\theta(0) = 1/2$). $\Delta$ is the cosine of the angle between $r_i$ and $r^{\prime}_j$ in the triangle with sides $r_i,r^{\prime}_j$ and $u$. The sum over $J$ in \cref{eqn:mehrem_with_M} is bounded by the two Wigner 3-$j$ symbols containing $J$ such that 
\begin{align}
\label{eqn:J_def}
    |\ell - (\lambda-\mathcal{L})| \leq &\;J \leq \ell + (\lambda-\mathcal{L}) \qquad \text{and}
    \nonumber \\
    |\ell^{\prime}-\mathcal{L}| \leq &\;J \leq \ell^{\prime}+\mathcal{L}.
\end{align}

The first Wigner 3-$j$ symbol in \cref{eqn:mehrem_with_M} constrains the free index $\lambda$: $\lambda$ must satisfy the triangular inequalities (\pcref{eqn:triangle_mehrem}) and $\ell+\ell'+\lambda$ must be even. We recall that from the Wigner 3-$j$ and 9-$j$ symbols within the covariance matrices of the 3- and 4PCF (\pcref{eqn:3pcf_cov_orig,eqn:c1_final,eqn:c2_orig}), the orders of the sBFs in the $f$-integral also satisfy these same two constraints. Thus, for our starting \cref{eqn:f_int_pk}, we see that $\ell^{\prime\prime}$ is constrained in the same way as $\lambda$, so that we may choose $\lambda=\ell^{\prime\prime}$. 

The Legendre polynomial in \cref{eqn:mehrem_with_M} can be expanded as a finite series in $u$ (\cite{GR} equation 8.911.1):
\begin{align}
\label{eqn:binom_exp}
    P_J(\Delta) &= 2^{-J}\sum_{a=0}^{\lfloor J/2 \rfloor}(-1)^a\binom{J}{a}\binom{2(J-a)}{J}\Delta^{J-2a}
    \nonumber \\
    &= 2^{-J}\sum_{a=0}^{\lfloor J/2 \rfloor}(-1)^a\binom{J}{a}\binom{2(J-a)}{J}(2r_ir^{\prime}_j)^{2a-J}\sum_{b=0}^{J-2a}\binom{J-2a}{b}(r_i^2+r^{\prime 2}_j)^{J-2a-b}
    \nonumber \\
    & \qquad \times(-1)^bu^{2b}.
\end{align}
$\lfloor J/2 \rfloor$ is the floor function, which gives the greatest integer value less than or equal to its argument. The floor function gives a falling power series in $\Delta$ in the first line of \cref{eqn:binom_exp}. We have used a binomial expansion on $\Delta^{J-2a}$ in the first line of \cref{eqn:binom_exp} to obtain the second line. 

Choosing $\lambda = \ell^{\prime\prime}$, as previously discussed, and using the Legendre polynomial expansion \cref{eqn:binom_exp} to rewrite $P_J(\Delta)$ in \cref{eqn:mehrem_with_M} yields
\setlength{\fboxrule}{0.5pt}
\begin{empheq}[box=\fbox]{align}
\label{eqn:mehrem_with_ell''}
    \int_{0}^{\infty}k^2\dl k\;j_{\ell}(kr_i)j_{\ell^{\prime}}(kr^{\prime}_j)j_{\ell^{\prime\prime}}(ku) &= \frac{\pi \beta(\Delta)}{4r_ir^{\prime}_ju}\;i^{\ell+\ell^{\prime}-\ell^{\prime\prime}}\sqrt{2\ell^{\prime\prime}+1}\left(\frac{r_i}{u} \right)^{\ell^{\prime\prime}}\tj{\ell}{\ell^{\prime}}{\ell^{\prime\prime}}^{-1}
    \nonumber \\
    & \qquad \times \sum_{\mathcal{L}=0}^{\ell^{\prime\prime}}\binom{2\ell^{\prime\prime}}{2\mathcal{L}}^{1/2} \left(\frac{r^{\prime}_j}{r_i}\right)^{\mathcal{L}}
    \nonumber \\
    & \qquad \times \sum_{J}(2J+1)\tj{\ell}{\ell^{\prime\prime}-\mathcal{L}}{J}
    \nonumber \\
    & \qquad \times \tj{\ell^{\prime}}{\mathcal{L}}{J} \sj{\ell}{\ell^{\prime}}{\ell^{\prime\prime}}{\mathcal{L}}{\ell^{\prime\prime}-\mathcal{L}}{J}2^{-J}
    \nonumber \\
    &\qquad \times \sum_{a=0}^{\lfloor J/2 \rfloor}(-1)^a\binom{J}{a}
    \binom{2(J-a)}{J}(2r_ir^{\prime}_j)^{2a-J}
    \nonumber \\
    & \qquad \times \sum_{b=0}^{J-2a}\binom{J-2a}{b}
    (r_i^2+r^{\prime 2}_j)^{J-2a-b}(-u^2)^b.
\end{empheq} 

We may now insert \cref{eqn:mehrem_with_ell''} to replace the innermost $k$-integral of \cref{eqn:I3_threeint}, yielding
\begin{align}
\label{eqn:I3_twoint}
    I^{[3,n]}_{\ell,\ell^{\prime},\ell^{\prime\prime}}(r_i,r^{\prime}_j,s) &= \frac{1}{2r_ir^{\prime}_j}\;i^{\ell+\ell^{\prime}-\ell^{\prime\prime}}\sqrt{2\ell^{\prime\prime}+1}r_i^{\ell^{\prime\prime}}\tj{\ell}{\ell^{\prime}}{\ell^{\prime\prime}}^{-1}\sum_{\mathcal{L}=0}^{\ell^{\prime\prime}}\binom{2\ell^{\prime\prime}}{2\mathcal{L}}^{1/2}\left(\frac{r^{\prime}_j}{r_i}\right)^{\mathcal{L}}\sum_{J}(2J+1) \nonumber \\
    & \qquad \times \tj{\ell}{\ell^{\prime\prime}-\mathcal{L}}{J}\tj{\ell^{\prime}}{\mathcal{L}}{J}\sj{\ell}{\ell^{\prime}}{\ell^{\prime\prime}}{\mathcal{L}}{\ell^{\prime\prime}-\mathcal{L}}{J}2^{-J}\sum_{a=0}^{\lfloor J/2 \rfloor}(-1)^a\binom{J}{a} \nonumber \\
    & \qquad \times \binom{2(J-a)}{J}(2r_ir^{\prime}_j)^{2a-J}\sum_{b=0}^{J-2a}\binom{J-2a}{b}(r_i^2+r^{\prime 2}_j)^{J-2a-b}(-1)^b \nonumber \\
    & \qquad \times \int_{|r_i-r^{\prime}_j|}^{r_i+r^{\prime}_j}u^{1+2b-\ell^{\prime\prime}}\dl u\; \int_{0}^{\infty}k^{\prime n}\dl k^{\prime}\;j_{\ell^{\prime\prime}}(k^{\prime}s)j_{\ell^{\prime\prime}}(k^{\prime}u).
\end{align}
$\beta(\Delta)$ (\pcref{eqn:beta_delta_mehrem}), which controls the bounds of the outermost $u$-integral, has been applied. We define
\begin{align}
\label{eqn:C_def}
    \mathcal{C}_{\ell,\ell^{\prime},\ell^{\prime\prime}}(r_i,r^{\prime}_j) &\equiv \frac{1}{2r_ir^{\prime}_j}\;i^{\ell+\ell^{\prime}-\ell^{\prime\prime}}\sqrt{2\ell^{\prime\prime}+1}\;r_i^{\ell^{\prime\prime}}\tj{\ell}{\ell^{\prime}}{\ell^{\prime\prime}}^{-1}
\end{align}
and
\begin{align}
\label{eqn:W_def}
    W_{\ell,\ell^{\prime},\ell^{\prime\prime}}^{\mathcal{L},J,a,b}(r_i,r^{\prime}_j) &\equiv \binom{2\ell^{\prime\prime}}{2\mathcal{L}}^{1/2}\left(\frac{r^{\prime}_j}{r_i}\right)^{\mathcal{L}}(2J+1)\tj{\ell}{\ell^{\prime\prime}-\mathcal{L}}{J}\tj{\ell^{\prime}}{\mathcal{L}}{J}\sj{\ell}{\ell^{\prime}}{\ell^{\prime\prime}}{\mathcal{L}}{\ell^{\prime\prime}-\mathcal{L}}{J} \nonumber \\
    & \qquad \times 2^{-J}\binom{J}{a}\binom{2(J-a)}{J}(2r_ir^{\prime}_j)^{2a-J}\binom{J-2a}{b}(r_i^2+r^{\prime 2}_j)^{J-2a-b}(-1)^{a+b}
\end{align}
so that \cref{eqn:I3_twoint} can be written as
\setlength{\fboxrule}{0.5pt} 			
\begin{empheq}[box=\fbox]{align}
\label{eqn:I3_twoint_W}
    I^{[3,n]}_{\ell,\ell^{\prime},\ell^{\prime\prime}}(r_i,r^{\prime}_j,s) &= \mathcal{C}_{\ell,\ell^{\prime},\ell^{\prime\prime}}(r_i,r^{\prime}_j)\sum_{\mathcal{L}=0}^{\ell^{\prime\prime}}\sum_{J}\sum_{a=0}^{\lfloor J/2 \rfloor}\sum_{b=0}^{J-2a} W_{\ell,\ell^{\prime},\ell^{\prime\prime}}^{\mathcal{L},J,a,b}(r_i,r^{\prime}_j) \nonumber \\
    &\qquad \times \int_{|r_i-r^{\prime}_j|}^{r_i+r^{\prime}_j}\dl u\;u^{1+2b-\ell^{\prime\prime}} \int_{0}^{\infty}k^{\prime n}\dl k^{\prime}\;j_{\ell^{\prime\prime}}(k^{\prime}s)j_{\ell^{\prime\prime}}(k^{\prime}u).
\end{empheq}

Although $\mathcal{C}_{\ell,\ell^{\prime},\ell^{\prime\prime}}(r_i,r^{\prime}_j)$ is symmetric under interchange of $\ell$ and $\ell^{\prime}$, it is not symmetric under interchange of $\ell^{\prime\prime}$ with either of the other two sBF orders. Additionally, neither $\mathcal{C}_{\ell,\ell^{\prime},\ell^{\prime\prime}}(r_i,r^{\prime}_j)$ nor $W_{\ell,\ell^{\prime},\ell^{\prime\prime}}^{\mathcal{L},J,a,b}(r_i,r^{\prime}_j)$ are symmetric under interchange of $r_i$ with $r^{\prime}_j$.

All sums in \cref{eqn:I3_twoint_W} are finite in range. $\ell^{\prime\prime}$ is an integer that stems from writing the covariance in the isotropic basis; therefore the sum over $\mathcal{L}$ is finite. $J$ is controlled by the integer sBF orders, which can be seen by examining \cref{eqn:J_def} and recalling that we have chosen $\lambda = \ell^{\prime\prime}$. Since $J$ is a finite value, the sums over $a$ and $b$ will also be finite.

The innermost $k^{\prime}$-integral of \cref{eqn:I3_twoint_W} will have different results depending on the value of $n$; thus, we must now evaluate the needed $n=2$ and $n=1$ cases individually.

\subsubsection[\texorpdfstring{$n = 2$}{n = 2}]{\boldmath$n = 2$}
\label{sec:n2}
We begin with the $n=2$ case, which will give the result for $I^{[3,\mathrm{quad}]}_{\ell,\ell^{\prime},\ell^{\prime\prime}}(r_i,r^{\prime}_j,s)$ (\cref{eqn:I_3quad}). We first substitute $n=2$ into \cref{eqn:I3_twoint_W}:
\setlength{\fboxrule}{0.5pt} 			
\begin{empheq}[box=\fbox]{align}
\label{eqn:I_3quad_sub_n}
    I^{[3,\mathrm{quad}]}_{\ell,\ell^{\prime},\ell^{\prime\prime}}(r_i,r^{\prime}_j,s) &= \mathcal{C}_{\ell,\ell^{\prime},\ell^{\prime\prime}}(r_i,r^{\prime}_j)\sum_{\mathcal{L}=0}^{\ell^{\prime\prime}}\sum_{J}\sum_{a=0}^{\lfloor J/2 \rfloor}\sum_{b=0}^{J-2a}W_{\ell,\ell^{\prime},\ell^{\prime\prime}}^{\mathcal{L},J,a,b}(r_i,r^{\prime}_j) \nonumber \\
    &\qquad \times \int_{|r_i-r^{\prime}_j|}^{r_i+r^{\prime}_j}\dl u\;u^{1+2b-\ell^{\prime\prime}} \int_{0}^{\infty}k^{\prime 2}\dl k^{\prime}\;j_{\ell^{\prime\prime}}(k^{\prime}s)j_{\ell^{\prime\prime}}(k^{\prime}u).
\end{empheq}
The innermost $k^{\prime}$-integral can be evaluated using \cref{eqn:I_2quad_eval} with the order of each sBF replaced by $\ell^{\prime\prime}$ and the free arguments of the sBFs replaced by $s$ and $u$:
\begin{align}
\label{eqn:I_3quad_with_dirac}
    I^{[3,\mathrm{quad}]}_{\ell,\ell^{\prime},\ell^{\prime\prime}}(r_i,r^{\prime}_j,s) &= \frac{\pi}{2s}\;\mathcal{C}_{\ell,\ell^{\prime},\ell^{\prime\prime}}(r_i,r^{\prime}_j)\sum_{\mathcal{L}=0}^{\ell^{\prime\prime}}\sum_{J}\sum_{a=0}^{\lfloor J/2 \rfloor}\sum_{b=0}^{J-2a}W_{\ell,\ell^{\prime},\ell^{\prime\prime}}^{\mathcal{L},J,a,b}(r_i,r^{\prime}_j) \nonumber \\
    &\qquad \times \int_{|r_i-r^{\prime}_j|}^{r_i+r^{\prime}_j}\dl u\;u^{2b-\ell^{\prime\prime}}\dD(s-u).
\end{align}
From the sifting property of the Dirac $\delta$, the $u$-integral of \cref{eqn:I_3quad_with_dirac} assumes the value of the integrand at $u=s$. The result for $I^{[3,\mathrm{quad}]}_{\ell,\ell^{\prime},\ell^{\prime\prime}}(r_i,r^{\prime}_j,s)$ is then
\setlength{\fboxrule}{1.5pt}
\begin{empheq}[box=\fbox]{align}
\label{eqn:I_3quad_eval}
    I^{[3,\mathrm{quad}]}_{\ell,\ell^{\prime},\ell^{\prime\prime}}(r_i,r^{\prime}_j,s) &= \frac{\pi}{2}\;\mathcal{C}_{\ell,\ell^{\prime},\ell^{\prime\prime}}(r_i,r^{\prime}_j)\sum_{\mathcal{L}=0}^{\ell^{\prime\prime}}\sum_{J}\sum_{a=0}^{\lfloor J/2 \rfloor}\sum_{b=0}^{J-2a}W_{\ell,\ell^{\prime},\ell^{\prime\prime}}^{\mathcal{L},J,a,b}(r_i,r^{\prime}_j)s^{2b-\ell^{\prime\prime}-1}.
\end{empheq} 

This result matches that given in equation 3.21 of \cite{k2_3sbf}. Although \cref{eqn:I_3quad_eval} appears to have a possible divergence when $s=0$, we recall that for our initial integral (\pcref{eqn:f_in_terms_of_I}), it was specified that $r_i$, $r_j^{\prime}$, and $s$ must all be nonzero. Therefore, \cref{eqn:I_3quad_eval} is convergent.

We now recall that to obtain the $f$-integral needed for the covariance (\pcref{eqn:f_in_terms_of_I}), we must evaluate both $I^{[3,\mathrm{quad}]}_{\ell,\ell^{\prime},\ell^{\prime\prime}}(r_i,r^{\prime}_j,s)$ and $I^{[3,\mathrm{lin}]}_{\ell,\ell^{\prime},\ell^{\prime\prime}}(r_i,r^{\prime}_j,s)$. The evaluation of $I^{[3,\mathrm{lin}]}_{\ell,\ell^{\prime},\ell^{\prime\prime}}(r_i,r^{\prime}_j,s)$ will be shown in the following subsection.

\subsubsection[\texorpdfstring{$n = 1$}{n = 1}]{\boldmath$n = 1$}
\label{sec:n1}
We now evaluate the $n=1$ case of \cref{eqn:I3_twoint_W},
\setlength{\fboxrule}{0.5pt} 			
\begin{empheq}[box=\fbox]{align}
\label{eqn:I_3lin_2int}
    I^{[3,\mathrm{lin}]}_{\ell,\ell^{\prime},\ell^{\prime\prime}}(r_i,r^{\prime}_j,s) &= \mathcal{C}_{\ell,\ell^{\prime},\ell^{\prime\prime}}(r_i,r^{\prime}_j)\sum_{\mathcal{L}=0}^{\ell^{\prime\prime}}\sum_{J}\sum_{a=0}^{\lfloor J/2 \rfloor}\sum_{b=0}^{J-2a} W_{\ell,\ell^{\prime},\ell^{\prime\prime}}^{\mathcal{L},J,a,b}(r_i,r^{\prime}_j)\int_{|r_i-r^{\prime}_j|}^{r_i+r^{\prime}_j}\dl u\;u^{1+2b-\ell^{\prime\prime}}
    \nonumber \\
    &\qquad \times \int_{0}^{\infty}k^{\prime}\dl k^{\prime}\;j_{\ell^{\prime\prime}}(k^{\prime}s)j_{\ell^{\prime\prime}}(k^{\prime}u),
\end{empheq} 
to obtain the result for $I^{[3,\mathrm{lin}]}_{\ell,\ell^{\prime},\ell^{\prime\prime}}(r_i,r^{\prime}_j,s)$ (\pcref{eqn:I_3lin}). The innermost $k^{\prime}$-integral of \cref{eqn:I_3lin_2int} can be evaluated using \cref{eqn:I_2lin_eval} with the orders of the sBFs replaced with $\ell^{\prime\prime}$ and the free arguments replaced with $s$ and $u$:
\begin{align}
\label{eqn:I_3lin_with_heavisides}
    I^{[3,\mathrm{lin}]}_{\ell,\ell^{\prime},\ell^{\prime\prime}}(r_i,r^{\prime}_j,s) &= \frac{\sqrt{\pi}\;\Gamma(\ell^{\prime\prime}+1)}{2\;\Gamma(\ell^{\prime\prime}+3/2)}\mathcal{C}_{\ell,\ell^{\prime},\ell^{\prime\prime}}(r_i,r^{\prime}_j)\sum_{\mathcal{L}=0}^{\ell^{\prime\prime}}\sum_{J}\sum_{a=0}^{\lfloor J/2 \rfloor}\sum_{b=0}^{J-2a} W_{\ell,\ell^{\prime},\ell^{\prime\prime}}^{\mathcal{L},J,a,b}(r_i,r^{\prime}_j)
    \nonumber \\
    & \qquad \times \left[s^{-\ell^{\prime\prime}-2}\int_{|r_i-r^{\prime}_j|}^{r_i+r^{\prime}_j}\dl u\; H(s-u)u^{1+2b}\twoFone\mleft(\ell^{\prime\prime}+1,\frac{1}{2};\ell^{\prime\prime}+\frac{3}{2};\left(\frac{u}{s}\right)^2\mright) \right.
    \nonumber \\
    &\left. \qquad + s^{\ell^{\prime\prime}}\int_{|r_i-r^{\prime}_j|}^{r_i+r^{\prime}_j}\dl u\;H(u-s)u^{-1+2b-2\ell^{\prime\prime}}\twoFone\mleft(\ell^{\prime\prime}+1,\frac{1}{2};\ell^{\prime\prime}+\frac{3}{2};\left(\frac{s}{u}\right)^2\mright) \right].
\end{align}

We do not include the $s=u$ term from \cref{eqn:I_2lin_eval}. The covariance requires an integration of a product of $f$-integrals (shown in \pcref{eqn:3pcf_cov_orig,eqn:c1_final,eqn:c2_orig}), and thus a pointwise contribution (such as at $s=u$) to $I^{[3,\mathrm{lin}]}_{\ell,\ell^{\prime},\ell^{\prime\prime}}(r_i,r^{\prime}_j,s)$ will vanish in the computation of the covariance.

\Cref{eqn:I_3lin_with_heavisides} has two distinct terms from the Heavisides: the first integral, with $H(s-u)$, is nonzero when $s>u$ while the second integral, with $H(u-s)$, is nonzero when $u>s$. To move these Heaviside functions outside of the integrals, we will now use the arguments of the Heavisides to change the bounds of the $u$-integrals.

%\level{4}[s>u]{$\boldsymbol{s > u}$}
\paragraph{\boldmath$s > u$}
\label{sec:s_greater_u}
~\\
We begin with the first integral of \cref{eqn:I_3lin_with_heavisides}, which is only nonzero when $s>u$:
\begin{align}
\label{eqn:H_s_greater_u}
    I_{\ell^{\prime\prime}}^{\mathrm{S}}(s,u;b) &\equiv s^{-\ell^{\prime\prime}-2}\int_{|r_i-r^{\prime}_j|}^{r_i+r^{\prime}_j}\dl u\;H(s-u)u^{1+2b}\twoFone\mleft(\ell^{\prime\prime}+1,\frac{1}{2};\ell^{\prime\prime}+\frac{3}{2};\left(\frac{u}{s}\right)^2\mright).
\end{align}
The $\mathrm{S}$ in the superscript of \cref{eqn:H_s_greater_u} denotes that $s$ is greater than $u$.

The Heaviside in \cref{eqn:H_s_greater_u} shows that $s$ must always be greater than $u$; therefore, $s$ can never be less than the minimum value of $u$, which is $|r_i-r^{\prime}_j|$ from the lower integration bound. This is the same as writing $H(s-|r_i-r^{\prime}_j|)$ and keeping $|r_i-r^{\prime}_j|$ as the lower integration bound.

There are now two cases to consider: $s > r_i+r^{\prime}_j$ and $r_i+r^{\prime}_j > s$, where the upper bound on the $u$-integral will be set to the minimum value of $\{s,r_i+r^{\prime}_j\}$. However, the Heaviside in \cref{eqn:H_s_greater_u} shows that we cannot have $s=u$. Therefore, the upper bound on the $u$-integral will be set to the minimum value of $\{s-\epsilon,r_i+r^{\prime}_j\}$, where $\epsilon$ is positive and small. We now can represent these two cases as two separate integrals with $H(s-(r_i+r^{\prime}_j))$ when the upper bound is $r_i+r^{\prime}_j$ and $H(r_i+r^{\prime}_j - s)$ when the upper bound is $s-\epsilon$. 

\Cref{eqn:H_s_greater_u} may now be expressed as
\begin{align}
\label{eqn:H_s_greater_u_rewrite}
    I_{\ell^{\prime\prime}}^{\mathrm{S}}(s,u;b) &= H(s-(r_i+r^{\prime}_j))H(s-|r_i-r^{\prime}_j|)s^{-\ell^{\prime\prime}-2} \int_{|r_i-r^{\prime}_j|}^{r_i+r^{\prime}_j}\dl u\;u^{1+2b}
    \nonumber \\
    &\qquad\qquad\qquad\qquad\qquad\qquad\qquad\qquad\qquad
    \times \twoFone\mleft(\ell^{\prime\prime}+1,\frac{1}{2};\ell^{\prime\prime}+\frac{3}{2};\left(\frac{u}{s}\right)^2\mright) &
    \displaybreak[1]
    \nonumber \\
    &\qquad +H(r_i+r^{\prime}_j-s)H(s-|r_i-r^{\prime}_j|)s^{-\ell^{\prime\prime}-2}\int_{|r_i-r^{\prime}_j|}^{s-\epsilon}\dl u\;u^{1+2b}
    \nonumber \\
    &\qquad\qquad\qquad\qquad\qquad\qquad\qquad\qquad\qquad
    \times \twoFone\mleft(\ell^{\prime\prime}+1,\frac{1}{2};\ell^{\prime\prime}+\frac{3}{2};\left(\frac{u}{s}\right)^2\mright). &
\end{align}
From the triangle inequality, 
\begin{align}
    r_i+r^{\prime}_j \geq |r_i-r^{\prime}_j|.
\end{align}
The first Heaviside in the first line of \cref{eqn:H_s_greater_u_rewrite}, $H(s-(r_i+r^{\prime}_j))$, shows that $s>r_i+r^{\prime}_j$, which implies that $s>|r_i-r^{\prime}_j|$. The second Heaviside in the first line of \cref{eqn:H_s_greater_u_rewrite}, $H(s-|r_i-r^{\prime}_j|)$, is less restrictive than the first Heaviside and can thus be omitted. 

We define
\begin{align}
\label{eqn:R_def}
    R_{\pm,ij} &\equiv \frac{|r_i \pm r^{\prime}_j|}{s} 
\end{align}
and rewrite \cref{eqn:H_s_greater_u_rewrite} after dividing the arguments of the Heaviside functions in it by $s$:
\setlength{\fboxrule}{0.5pt} 			
\begin{empheq}[box=\fbox]{align}
\label{eqn:H_s_greater_u_div_s}
    I_{\ell^{\prime\prime}}^{\mathrm{S}}(s,u;b) &= H(1-R_{+,ij})s^{-\ell^{\prime\prime}-2}\int_{|r_i-r^{\prime}_j|}^{r_i+r^{\prime}_j}\dl u\;u^{1+2b}\twoFone\mleft(\ell^{\prime\prime}+1,\frac{1}{2};\ell^{\prime\prime}+\frac{3}{2};\left(\frac{u}{s}\right)^2\mright)
    \nonumber \\
    &\qquad + H(R_{+,ij}-1) H(1-R_{-,ij}) s^{-\ell^{\prime\prime}-2}\int_{|r_i-r^{\prime}_j|}^{s-\epsilon}\dl u\;u^{1+2b}
    \nonumber \\
    &\qquad\qquad\qquad\qquad\qquad\qquad\qquad\qquad \times \twoFone\mleft(\ell^{\prime\prime}+1,\frac{1}{2};\ell^{\prime\prime}+\frac{3}{2};\left(\frac{u}{s}\right)^2\mright).
\end{empheq} 
$R_{\pm,ij}$ is symmetric under interchange of $r_i$ and $r^{\prime}_j$.

The argument of the hypergeometric functions in \cref{eqn:H_s_greater_u_div_s}, $(u/s)^2$, will reach its maximum value when evaluated at the maximum value of $u$ and minimum value of $s$. We now use these arguments to examine the convergence of the hypergeometric functions.

The upper bound of the first integral of \cref{eqn:H_s_greater_u_div_s} shows that the maximum value of $u$ is $r_i+r^{\prime}_j$, meaning that the maximum value of $(u/s)^2$ is $((r_i+r^{\prime}_j)/s)^2$. The Heaviside in front of the first integral of \cref{eqn:H_s_greater_u_div_s} ensures that $s>(r_i+r^{\prime}_j)$; therefore, the argument of the hypergeometric function in the first integral of \cref{eqn:H_s_greater_u_div_s} will always be less than unity. This shows that the hypergeometric function will converge.

The second integral of \cref{eqn:H_s_greater_u_div_s} has a maximum value of $u$ equal to $s-\epsilon$. The maximum value of $(u/s)^2$ is thus $((s-\epsilon)/s)^2$. Since this quantity is always less than unity, the hypergeometric function converges.

%\level{4}[u>s]{$\boldsymbol{u > s}$}
\paragraph{\boldmath$u > s$}
\label{sec:u_greater_s}
~\\
We now repeat the above process for the second integral of \cref{eqn:I_3lin_with_heavisides},
\begin{align}
\label{eqn:H_u_greater_s}
    I_{\ell^{\prime{\prime}}}^{\mathrm{U}}(s,u;b) &\equiv  s^{\ell^{\prime\prime}}\int_{|r_i-r^{\prime}_j|}^{r_i+r^{\prime}_j}\dl u\;H(u-s)u^{-1+2b-2\ell^{\prime\prime}}\twoFone\mleft(\ell^{\prime\prime}+1,\frac{1}{2};\ell^{\prime\prime}+\frac{3}{2};\left(\frac{s}{u}\right)^2\mright),
\end{align}
rewriting it so that the Heaviside is no longer part of the integrand. \Cref{eqn:H_u_greater_s} is nonzero when $u>s$; this is represented by the superscript $\mathrm{U}$.

The Heaviside in \cref{eqn:H_u_greater_s}, $H(u-s)$, shows that $u$ must always be greater than $s$. In other words, $s$ cannot exceed the maximum value of $u$, which is $r_i+r^{\prime}_j$ from the upper bound of the $u$-integral. This condition can be represented by the Heaviside $H(r_i+r^{\prime}_j-s)$. 

The upper bound of the $u$-integral will thus remain $r_i+r^{\prime}_j$ while the lower bound will be set by the maximum of $\{s,|r_i-r^{\prime}_j|\}$. However, the Heaviside in \cref{eqn:H_u_greater_s} dictates that the integral vanishes when $s=u$; therefore, the lower bound is set by the maximum of $\{s+\epsilon,|r_i-r^{\prime}_j|\}$, where $\epsilon$ is again positive and small. The integral bounded below by $|r_i-r^{\prime}_j|$ will require $H(|r_i-r^{\prime}_j|-s)$ while the integral bounded below by $s+\epsilon$ will require $H(s-|r_i-r^{\prime}_j|)$. 

\Cref{eqn:H_u_greater_s} can now be split into two integrals:
\begin{align}
\label{eqn:H_u_greater_s_rewrite}
    I_{\ell^{\prime{\prime}}}^{\mathrm{U}}(s,u;b) &= H(r_i+r^{\prime}_j-s)H(|r_i-r^{\prime}_j|-s)s^{\ell^{\prime\prime}}\int_{|r_i-r^{\prime}_j|}^{r_i+r^{\prime}_j}\dl u\;u^{-1+2b-2\ell^{\prime\prime}}
    \nonumber \\
    &\qquad\qquad\qquad\qquad\qquad\qquad\qquad\qquad\qquad
    \times\twoFone\mleft(\ell^{\prime\prime}+1,\frac{1}{2};\ell^{\prime\prime}+\frac{3}{2};\left(\frac{s}{u}\right)^2\mright)
    \nonumber \\
    &\qquad+H(r_i+r^{\prime}_j-s)H(s-|r_i-r^{\prime}_j|)s^{\ell^{\prime\prime}}\int_{s+\epsilon}^{r_i+r^{\prime}_j}\dl u\;u^{-1+2b-2\ell^{\prime\prime}}
    \nonumber \\
    &\qquad\qquad\qquad\qquad\qquad\qquad\qquad\qquad\qquad
    \times\twoFone\mleft(\ell^{\prime\prime}+1,\frac{1}{2};\ell^{\prime\prime}+\frac{3}{2};\left(\frac{s}{u}\right)^2\mright).
\end{align}

The second Heaviside in front of the first integral of \cref{eqn:H_u_greater_s_rewrite}, $H(|r_i-r^{\prime}_j|-s)$, shows that $|r_i-r^{\prime}_j|>s$, which implies that $r_i+r^{\prime}_j>s$. The first Heaviside before the first integral of \cref{eqn:H_u_greater_s_rewrite}, $H(r_i+r^{\prime}_j-s)$, can be omitted since it is less restrictive than $H(|r_i-r^{\prime}_j|-s)$. 

After dividing the arguments of the Heavisides by $s$, \cref{eqn:H_u_greater_s_rewrite} may be rewritten as
\setlength{\fboxrule}{0.5pt} 			
\begin{empheq}[box=\fbox]{align}
\label{eqn:H_u_greater_s_div_s}
    I_{\ell^{\prime{\prime}}}^{\mathrm{U}}(s,u;b) &= H(R_{-,ij}-1) s^{\ell^{\prime\prime}}\int_{|r_i-r^{\prime}_j|}^{r_i+r^{\prime}_j}\dl u\;u^{-1+2b-2\ell^{\prime\prime}}\twoFone\mleft(\ell^{\prime\prime}+1,\frac{1}{2};\ell^{\prime\prime}+\frac{3}{2};\left(\frac{s}{u}\right)^2\mright)
    \nonumber \\
    &\qquad + H(R_{+,ij}-1) H(1-R_{-,ij}) s^{\ell^{\prime\prime}}\int_{s+\epsilon}^{r_i+r^{\prime}_j}\dl u\;u^{-1+2b-2\ell^{\prime\prime}}
    \nonumber \\
    &\qquad\qquad\qquad\qquad\qquad\qquad\qquad\qquad
    \times \twoFone\mleft(\ell^{\prime\prime}+1,\frac{1}{2};\ell^{\prime\prime}+\frac{3}{2};\left(\frac{s}{u}\right)^2\mright).
\end{empheq} 

We now show that the hypergeometric functions in \cref{eqn:H_u_greater_s_div_s} converge. The argument of the hypergeometric functions, $(s/u)^2$, is at a maximum when $s$ is maximized and $u$ is minimized. 

In the first integral of \cref{eqn:H_u_greater_s_div_s}, the minimum value of $u$ is $|r_i-r^{\prime}_j|$ from the lower integration bound; the maximum value of $(s/u)^2$ is thus $(s/|r_i-r^{\prime}_j|)^2$. From the Heaviside preceding the first integral of \cref{eqn:H_u_greater_s_div_s}, $|r_i-r^{\prime}_j| > s$. Therefore, the argument of the hypergeometric function in the first integral of \cref{eqn:H_u_greater_s_div_s} will always be less than unity and the hypergeometric function will not diverge. 

From the lower bound of the second integral in \cref{eqn:H_u_greater_s_div_s}, the minimum value of $u$ is $s+\epsilon$. The maximum value of the argument of the hypergeometric function in the final line of \cref{eqn:H_u_greater_s_div_s} is therefore $(s/s+\epsilon)^2$. Since the argument will never exceed unity, the hypergeometric function is convergent.

%\level{4}{Evaluation of the Hypergeometric Integrals}
%\vspace{-1.349cm}
\paragraph{Evaluation of the Hypergeometric Integrals}
\label{sec:hypergeometric_integrals}
~\\
We may now rewrite \cref{eqn:I_3lin_with_heavisides} with no Heaviside functions in the integrands. To do this, we replace the $H(s-u)$ term of \cref{eqn:I_3lin_with_heavisides} with \cref{eqn:H_s_greater_u_div_s} and the $H(u-s)$ term with \cref{eqn:H_u_greater_s_div_s}:
%\vspace{-1.34cm}
\begin{align}
\label{eqn:I_3lin_with_bounds}
    I^{[3,\mathrm{lin}]}_{\ell,\ell^{\prime},\ell^{\prime\prime}}(r_i,r^{\prime}_j,s) &= \frac{\sqrt{\pi}\;\Gamma(\ell^{\prime\prime}+1)}{2\;\Gamma(\ell^{\prime\prime}+3/2)}\mathcal{C}_{\ell,\ell^{\prime},\ell^{\prime\prime}}(r_i,r^{\prime}_j)\sum_{\mathcal{L}=0}^{\ell^{\prime\prime}}\sum_{J}\sum_{a=0}^{\lfloor J/2 \rfloor}\sum_{b=0}^{J-2a} W_{\ell,\ell^{\prime},\ell^{\prime\prime}}^{\mathcal{L},J,a,b}(r_i,r^{\prime}_j)
    \nonumber \\
    \allowbreak
    &\qquad \times \left[H(1-R_{+,ij})s^{-\ell^{\prime\prime}-2}\int_{|r_i-r^{\prime}_j|}^{r_i+r^{\prime}_j}\dl u\;u^{1+2b}\twoFone\mleft(\ell^{\prime\prime}+1,\frac{1}{2};\ell^{\prime\prime}+\frac{3}{2};\left(\frac{u}{s}\right)^2\mright) \right.
    \nonumber \\
    &\left. \qquad + H(R_{+,ij}-1) H(1-R_{-,ij})s^{-\ell^{\prime\prime}-2}\int_{|r_i-r^{\prime}_j|}^{s-\epsilon}\dl u\;u^{1+2b} \right.
    \nonumber \\
    &\left. \qquad\qquad\qquad\qquad\qquad\qquad\qquad\qquad\qquad\times \twoFone\mleft(\ell^{\prime\prime}+1,\frac{1}{2};\ell^{\prime\prime}+\frac{3}{2};\left(\frac{u}{s}\right)^2\mright) \right.
    \nonumber \\
    &\left. \qquad + H(R_{-,ij}-1)s^{\ell^{\prime\prime}}\int_{|r_i-r^{\prime}_j|}^{r_i+r^{\prime}_j}\dl u\;u^{-1+2b-2\ell^{\prime\prime}}\twoFone\mleft(\ell^{\prime\prime}+1,\frac{1}{2};\ell^{\prime\prime}+\frac{3}{2};\left(\frac{s}{u}\right)^2\mright) \right.
    \nonumber \\
    &\left. \qquad + H(R_{+,ij}-1) H(1-R_{-,ij})s^{\ell^{\prime\prime}}\int_{s+\epsilon}^{r_i+r^{\prime}_j}\dl u\;u^{-1+2b-2\ell^{\prime\prime}} \right.
    \nonumber \\
    &\left. \qquad\qquad\qquad\qquad\qquad\qquad\qquad\qquad \times \twoFone\mleft(\ell^{\prime\prime}+1,\frac{1}{2};\ell^{\prime\prime}+\frac{3}{2};\left(\frac{s}{u}\right)^2\mright) \right].
\end{align}

We use the changes of variable $x=(u/s)^2$ and $y=(s/u)^2$ to make the arguments of the hypergeometric functions in \cref{eqn:I_3lin_with_bounds} depend on a single variable:
\begin{align}
\label{eqn:I_3lin_change_var}
    I^{[3,\mathrm{lin}]}_{\ell,\ell^{\prime},\ell^{\prime\prime}}(r_i,r^{\prime}_j,s) &= \frac{\sqrt{\pi}\;\Gamma(\ell^{\prime\prime}+1)}{4\;\Gamma(\ell^{\prime\prime}+3/2)}\mathcal{C}_{\ell,\ell^{\prime},\ell^{\prime\prime}}(r_i,r^{\prime}_j)\sum_{\mathcal{L}=0}^{\ell^{\prime\prime}}\sum_{J}\sum_{a=0}^{\lfloor J/2 \rfloor}\sum_{b=0}^{J-2a} W_{\ell,\ell^{\prime},\ell^{\prime\prime}}^{\mathcal{L},J,a,b}(r_i,r^{\prime}_j)s^{2b-\ell^{\prime\prime}}
    \nonumber \\
    \allowbreak
    & \qquad \times \left[H(1-R_{+,ij})
    \int_{R_{-,ij}^2}^{R_{+,ij}^2}\dl x\;x^b\twoFone\mleft(\ell^{\prime\prime}+1,\frac{1}{2};\ell^{\prime\prime}+\frac{3}{2};x\mright) \right.
    \nonumber \\
    &\left. \qquad + H(R_{+,ij}-1) H(1-R_{-,ij}) \int_{R_{-,ij}^2}^{[(s-\epsilon)/s]^2}\dl x\;x^b\twoFone\mleft(\ell^{\prime\prime}+1,\frac{1}{2};\ell^{\prime\prime}+\frac{3}{2};x\mright) \right.
    \nonumber \\
    &\left. \qquad + H(R_{-,ij}-1) \int_{R_{+,ij}^{-2}}^{R_{-,ij}^{-2}}\dl y\;y^{\ell^{\prime\prime}-b-1}\twoFone\mleft(\ell^{\prime\prime}+1,\frac{1}{2};\ell^{\prime\prime}+\frac{3}{2};y\mright) \right.
    \nonumber \\
    &\left. \qquad + H(R_{+,ij}-1) H(1-R_{-,ij}) \int_{R_{+,ij}^{-2}}^{[s/(s+\epsilon)]^2}\dl y\;y^{\ell^{\prime\prime}-b-1} \right.
    \nonumber \\
    &\left. \qquad\qquad\qquad\qquad\qquad\qquad\qquad\qquad\qquad \times \twoFone\mleft(\ell^{\prime\prime}+1,\frac{1}{2};\ell^{\prime\prime}+\frac{3}{2};y\mright) \right].
\end{align}
Since $b$ is always an integer greater than or equal to zero (as shown in the fourth summation of \pcref{eqn:I_3lin_change_var}), the first two integrals (\textit{i.\,e.} those over $x$) of \cref{eqn:I_3lin_change_var} can be evaluated as\footnote{\url{https://functions.wolfram.com/HypergeometricFunctions/Hypergeometric2F1/21/01/02/01/0002/}}
\begin{align}
\label{eqn:int_2f1_x}
    I_{\ell^{\prime\prime}}(x;b) &\equiv \int_{x_{\mathrm{min}}}^{x_{\mathrm{max}}} \dl x\;x^b\twoFone\mleft(\ell^{\prime\prime}+1,\frac{1}{2};\ell^{\prime\prime}+\frac{3}{2};x\mright)
    \nonumber \\
    &= \frac{x^{b+1}}{b+1}\;\threeFtwo\mleft(\ell^{\prime\prime}+1,\frac{1}{2},b+1;\ell^{\prime\prime}+\frac{3}{2},b+2;x\mright)\Bigg|_{x_{\mathrm{min}}}^{x_{\mathrm{max}}}.
\end{align}

The final two integrals (\textit{i.\,e.} those over $y$) of \cref{eqn:I_3lin_change_var} involve $y$ raised to the power $\ell^{\prime\prime}-b-1$. This power can be positive, zero, or negative, but will always be an integer. The $y$-integrals of \cref{eqn:I_3lin_change_var} may be evaluated as 
\begin{align}
\label{eqn:int_2f1_y}
    I_{\ell^{\prime\prime}}(y;b) &\equiv \int_{y_{\mathrm{min}}}^{y_{\mathrm{max}}} \dl y\;y^{\ell^{\prime\prime}-b-1}\twoFone\mleft(\ell^{\prime\prime}+1,\frac{1}{2};\ell^{\prime\prime}+\frac{3}{2};y\mright)
    \nonumber \\
    &= (-1)^{\ell^{\prime\prime}-b-1}\frac{\Gamma(\ell^{\prime\prime}+3/2)}{\sqrt{\pi}\;\Gamma(\ell^{\prime\prime}+1)}\;\MeijerGTwoTwoThreeThree{y}{1}{-\ell^{\prime\prime}+b+1}{b+3/2}{b+1}{-\ell^{\prime\prime}+b+1/2}{0}\Bigg|_{-1/y_{\mathrm{min}}}^{-1/y_{\mathrm{max}}}.
\end{align}
$G^{2,2}_{3,3}$ is a Meijer $G$-function, which is a contour integral defined with the Mellin-Barnes representation in equation 9.301 of \cite{GR}. \Cref{eqn:int_2f1_y} holds for all the values we may have for $\ell^{\prime\prime}-b-1$. For further details on the derivation of \cref{eqn:int_2f1_y}, see \cref{integral_hyp_to_meijer}.

We now use \cref{eqn:int_2f1_x,eqn:int_2f1_y} to evaluate the integrals within \cref{eqn:I_3lin_change_var}:
\setlength{\fboxrule}{1.5pt} 			
\begin{empheq}[box=\fbox]{align}
    \label{eqn:I_3lin_eval_int}
    I^{[3,\mathrm{lin}]}_{\ell,\ell^{\prime},\ell^{\prime\prime}}(r_i,r^{\prime}_j,s) &= \frac{1}{4}\mathcal{C}_{\ell,\ell^{\prime},\ell^{\prime\prime}}(r_i,r^{\prime}_j)\sum_{\mathcal{L}=0}^{\ell^{\prime\prime}}\sum_{J}\sum_{a=0}^{\lfloor J/2 \rfloor}\sum_{b=0}^{J-2a} W_{\ell,\ell^{\prime},\ell^{\prime\prime}}^{\mathcal{L},J,a,b}(r_i,r^{\prime}_j)s^{2b-\ell^{\prime\prime}} \Bigg\{ \Bigg.
    \nonumber \\
    & \Bigg. \qquad \textcolor{neworange}{ H(1-R_{+,ij}) \frac{\sqrt{\pi}\;\Gamma(\ell^{\prime\prime}+1)}{(b+1)\Gamma(\ell^{\prime\prime}+3/2)}\left(R_{+,ij}^{2(b+1)}F_+ - R_{-,ij}^{2(b+1)}F_-\right) }  \Bigg. \nonumber \\
    &\Bigg. \qquad + \textcolor{newpurple}{H(R_{+,ij}-1)} \textcolor{newpurple}{H(1-R_{-,ij}) \Bigg[\frac{\sqrt{\pi}\;\Gamma(\ell^{\prime\prime}+1)}{(b+1)\Gamma(\ell^{\prime\prime}+3/2)} } \Bigg. \Bigg.
    \nonumber \\
    &\Bigg.\Bigg. \qquad\qquad \textcolor{newpurple}{\times \left( \omega_-^{2(b+1)} F_\omega - R_{-,ij}^{2(b+1)}F_-\right) + (-1)^{\ell^{\prime\prime}-b-1}\left(G_{\omega} - G_+ \right) \Bigg]}
    \nonumber \\
    &\qquad + \textcolor{midgreen}{H(R_{-,ij}-1)(-1)^{\ell^{\prime\prime}-b-1} \left(G_- - G_+ \right)} \Bigg\}.
\end{empheq}
We have defined the following quantities:
\begingroup\allowdisplaybreaks[1]%
\begin{align}
\label{eqn:R_and_F_and_G_defs}
    \omega_{\pm} &\equiv 1 \pm \frac{\epsilon}{s}
    \nonumber \\
    F_{\pm} &\equiv \;\threeFtwo\mleft(\ell^{\prime\prime}+1,\frac{1}{2},b+1;\ell^{\prime\prime}+\frac{3}{2},b+2;R_{\pm,ij}^2 \mright),
    \nonumber \\ 
    F_{\omega} &\equiv \;\threeFtwo\mleft(\ell^{\prime\prime}+1,\frac{1}{2},b+1;\ell^{\prime\prime}+\frac{3}{2},b+2; \omega_-^2 \mright),
    \nonumber \\
    G_{\pm} &\equiv \;\MeijerGTwoTwoThreeThree{-R_{\pm,ij}^2}{1}{-\ell^{\prime\prime}+b+1}{b+3/2}{b+1}{-\ell^{\prime\prime}+b+1/2}{0},
    \nonumber \\ 
    G_{\omega} &\equiv \;\MeijerGTwoTwoThreeThree{-\omega_+^2}{1}{-\ell^{\prime\prime}+b+1}{b+3/2}{b+1}{-\ell^{\prime\prime}+b+1/2}{0}.
\end{align}%
\endgroup

\begin{figure}
    \centering
    \includegraphics[width=\textwidth]{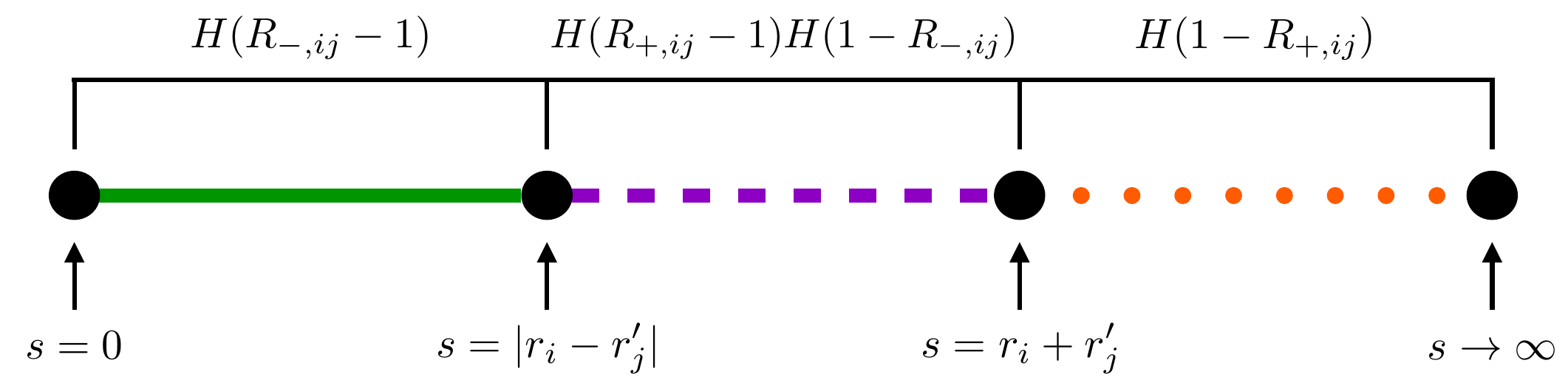}

    \caption{The Heaviside functions in \cref{eqn:I_3lin_eval_int} correspond to regions with distinct $s$ values. The first term of \cref{eqn:I_3lin_eval_int}, with Heaviside $H(1-R_{+,ij})$, is only nonzero when $s>r_i+r^{\prime}_j$ (dotted orange line). The second set of Heavisides, $H(R_{+,ij}-1) H(1-R_{-,ij})$, ensures that $|r_i-r^{\prime}_j|<s<r_i+r^{\prime}_j$ (dashed purple line). The final Heaviside, $H(R_{-,ij} -1)$, shows that the term associated with it contributes to $I^{[3,\mathrm{lin}]}_{\ell,\ell^{\prime},\ell^{\prime\prime}}(r_i,r^{\prime}_j,s)$ only when $|r_i-r^{\prime}_j|>s$ (solid green line). We note that the dashed purple (center) region is the only one in which closed triangles may be formed between $r_i$, $r^{\prime}_j$, and $s$. As will be shown in \cref{sec:sparse_matrices}, the greatest contributions to the $f$-integrals stem from this region.}
    \label{fig:heavisides}
\end{figure}

There are three distinct terms within \cref{eqn:I_3lin_eval_int} corresponding to different configurations of $r_i$, $r^{\prime}_j$, and $s$. Therefore, only one piece of \cref{eqn:I_3lin_eval_int} needs to be evaluated at a time. This is shown schematically in \cref{fig:heavisides}.

% \clearpage
%\level{4}{Discussion of Convergence}
\paragraph{Discussion of Convergence}
~\\
First, we explore the convergence of the original $I^{[3,\mathrm{lin}]}_{\ell,\ell^{\prime},\ell^{\prime\prime}}(r_i,r^{\prime}_j,s)$-integral (\pcref{eqn:I_3lin}). As $k \rightarrow 0$, the small-argument behavior of the sBFs leads to the integrand becoming $k^{\ell+\ell^{\prime}+\ell^{\prime\prime}+1}$, showing that there is no infrared divergence. As $k \rightarrow \infty$, the sBFs may be approximated as $j_{\ell}(x) \rightarrow \sin(x-\ell\pi/2)/x$. The integrand then becomes the product of $1/k^2$ and three sinusoids, which have arguments $k(r_i \pm r^{\prime}_j \pm s)$. In this ultraviolet limit, there is no divergence because convergence is assured by the $1/k^2$ envelope. 

We now turn to a discussion of our final result for $I^{[3,\mathrm{lin}]}_{\ell,\ell^{\prime},\ell^{\prime\prime}}(r_i,r^{\prime}_j,s)$, given in \cref{eqn:I_3lin_eval_int}, for which we had previously specified that $r_i$, $r_j^{\prime}$, and $s$ are all nonzero. Any possible divergences in \cref{eqn:I_3lin_eval_int} would stem from the special functions within, which we will now show are all convergent.

The hypergeometric functions within \cref{eqn:I_3lin_eval_int} can be expanded as power series of their arguments. As demonstrated in \cite{NIST} equation 16.2.1, a hypergeometric function is thus convergent when the magnitude of its argument is less than unity. 

The first distinct term in \cref{eqn:I_3lin_eval_int} is proportional to $H(1-R_{+,ij})$, or, in other words, this term is nonzero only when $1 > R_{+,ij}$. If $1 > R_{+,ij}$, then it also must be true that $1 > R_{-,ij}$. The arguments of the hypergeometric functions $F_+$ and $F_-$ in this first term are $R_{+,ij}^2$ and $R_{-,ij}^2$, respectively, which must have magnitudes less than unity. Thus, these two hypergeometric functions converge.

The second term in \cref{eqn:I_3lin_eval_int} is only nonzero when $R_{+,ij} > 1 > R_{-,ij}$, by the product of Heavisides $H(R_{+,ij}-1)H(1-R_{-,ij})$. We first focus on the two hypergeometric functions in this term, $F_-$ and $F_{\omega}$. Since $1 > R_{-,ij}$, the argument of $F_-$ is less than unity, meaning that $F_-$ is convergent. $F_{\omega}$ has argument $\omega_-^2 = (1-\epsilon/s)^2$. As we take $\epsilon \rightarrow 0$, we must have $\epsilon/s \rightarrow 0$, showing that $\omega_-^2$ approaches but is not equal to unity. The hypergeometric function $F_{\omega}$ thus is also convergent.

Equation 16.21.1 of \cite{NIST} shows that the Meijer $G$-function $G^{m,n}_{p,q}$ has a singularity when the argument is equal to $0, (-1)^{p-m-n}$, or $\infty$. For the Meijer $G$-functions in \cref{eqn:I_3lin_eval_int}, an infinite argument corresponds to complex side lengths, which we do not have. We therefore only need to examine whether the Meijer $G$-functions can have arguments equal to $0$ or $(-1)^{p-m-n} = -1$.

The second distinct term in \cref{eqn:I_3lin_eval_int} has Heavisides $H(R_{+,ij}-1)H(1-R_{-,ij})$, which show that $R_{+,ij} > 1 > R_{-,ij}$. This term has two Meijer $G$-functions in it: $G_{+}$ and $G_{\omega}$. Since $R_{+,ij} > 1$, we know that the argument of $G_+$, which is $-R_{+,ij}^2$, must be less than $-1$. Since the argument $-R_{+,ij}^2$ cannot equal $0$ or $-1$, there is no divergence in $G_+$. The argument of $G_{\omega}$ is $-\omega_+^2 = -(1+\epsilon/s)^2$. As $\epsilon \rightarrow 0$, $-\omega_+^2$ will approach but not be equal to negative unity. $G_{\omega}$ is therefore convergent.

The third term of \cref{eqn:I_3lin_eval_int} contains the Heaviside $H(R_{-,ij} -1)$ and Meijer $G$-functions $G_-$ and $G_+$. The Heaviside dictates that $R_{-,ij} > 1$, which implies that $R_{+,ij} > 1$. By these inequalities, the arguments of $G_-$ and $G_+$, which are $-R_{-,ij}^2$ and $-R_{+,ij}^2$, respectively, must both be less than $-1$. These two Meijer $G$-functions thus cannot diverge.

\subsubsection[Analytic Result for \texorpdfstring{$f$}{f}-Integral with Nonzero sBF Orders and Arguments]{\boldmath Analytic Result for \texorpdfstring{$f$}{f}-Integral with Nonzero sBF Orders and Arguments}
\label{sec:f_full_result}
We are now able to evaluate the $f$-integral with nonzero orders and arguments for all three sBFs within it (\pcref{eqn:f_in_terms_of_I}, duplicated below):
\begin{align}
\label{eqn:f_in_terms_of_I_duplicate}
    f_{\ell,\ell^{\prime},\ell^{\prime\prime}}(r_i,r^{\prime}_j,s) &= \frac{1}{2\pi^2}\left[ AI^{[3,\mathrm{lin}]}_{\ell,\ell^{\prime},\ell^{\prime\prime}}(r_i,r^{\prime}_j,s) + \frac{1}{\bar{n}}I^{[3,\mathrm{quad}]}_{\ell,\ell^{\prime},\ell^{\prime\prime}}(r_i,r^{\prime}_j,s)\right].
\end{align}

After replacing $I^{[3,\mathrm{lin}]}_{\ell,\ell^{\prime},\ell^{\prime\prime}}(r_i,r^{\prime}_j,s)$ and $I^{[3,\mathrm{quad}]}_{\ell,\ell^{\prime},\ell^{\prime\prime}}(r_i,r^{\prime}_j,s)$ appearing in \cref{eqn:f_in_terms_of_I_duplicate} with \cref{eqn:I_3lin_eval_int,eqn:I_3quad_eval}, we obtain
\setlength{\fboxrule}{1.5pt} 			
\begin{empheq}[box=\fbox]{align}
\label{eqn:f_FINAL}
    f_{\ell,\ell^{\prime},\ell^{\prime\prime}}(r_i,r^{\prime}_j,s) &= \frac{A}{8\pi^2}\;\mathcal{C}_{\ell,\ell^{\prime},\ell^{\prime\prime}}(r_i,r^{\prime}_j)\sum_{\mathcal{L}=0}^{\ell^{\prime\prime}}\sum_{J}\sum_{a=0}^{\lfloor J/2 \rfloor}\sum_{b=0}^{J-2a} W_{\ell,\ell^{\prime},\ell^{\prime\prime}}^{\mathcal{L},J,a,b}(r_i,r^{\prime}_j)s^{2b-\ell^{\prime\prime}} \Bigg\{
    \nonumber \\
    &\qquad \textcolor{neworange}{H(1-R_{+,ij}) \frac{\sqrt{\pi}\;\Gamma(\ell^{\prime\prime}+1)}{(b+1)\Gamma(\ell^{\prime\prime}+3/2)}\left(R_{+,ij}^{2(b+1)}F_+ - R_{-,ij}^{2(b+1)}F_- \right)} \Bigg.
    \nonumber \\
    &\Bigg. \qquad+ \textcolor{newpurple}{H(R_{+,ij}-1) H(1-R_{-,ij}) \bigg[\frac{\sqrt{\pi}\;\Gamma(\ell^{\prime\prime}+1)}{(b+1)\Gamma(\ell^{\prime\prime}+3/2)} \bigg.\Bigg.}
    \nonumber \\
    &\qquad\qquad \textcolor{newpurple}{\Bigg.\bigg. \times \left(\omega_-^{2(b+1)}F_{\omega} - R_{-,ij}^{2(b+1)}F_- \right) + (-1)^{\ell^{\prime\prime}-b-1}\left(G_{\omega}-G_+\right) \bigg]} \Bigg. \nonumber \\
    &\qquad +\textcolor{midgreen}{H(R_{-,ij}-1)(-1)^{\ell^{\prime\prime}-b-1}} \textcolor{midgreen}{\Bigg. \left(G_--G_+\right)} +\frac{2\pi}{A\bar{n}s} \Bigg\}.
\end{empheq}

We now have all the necessary pieces to compute the covariance of the 3- and 4PCF. We recall that the covariance of the 3PCF (\pcref{eqn:3pcf_cov_orig}) depends on products of the 2PCF with two $f$-integrals, and the product of three $f$-integrals. The covariance of the 4PCF depends on single $f$-integrals (Case \rom{1}, \pcref{eqn:c1_final}) and the product of four $f$-integrals (Case \rom{2}, \pcref{eqn:c2_orig}). With our analytic results for the 2PCF (\pcref{eqn:2pcf}), the $f$-integral with one sBF of order and corresponding argument zero (\pcref{eqn:f_int_zero_eval}), and the $f$-integral for all nonzero sBF orders and arguments (\pcref{eqn:f_FINAL}), the covariance is now tractable.

\section{Discussion}
\label{sec:discussion}
We first examine the structure of the $f$-integrals, which leads to near-zero contributions to the covariance outside of the region where the various $r_i$, $r^{\prime}_j$, and $s$ can form a closed triangle (\cref{sec:sparse_matrices}). In \cref{sec:inverse_cov}, we explore a matrix inversion lemma used to efficiently invert the covariance. We also determine the correction term. Finally, in \cref{sec:computational_complexity} we discuss the computational complexity of the analytic covariance formulae found in this work.

\subsection{Sparse Matrices}
\label{sec:sparse_matrices}
Here, we compare the $f$-integrals (\pcref{eqn:f_int}) from our model power spectrum to the $f$-integrals computed with the \textsc{camb} power spectrum. In \cref{subsec:f_plots_fixed_s}, we show these $f$-integrals at a fixed separation $s = \SI{50}{\per\hHubble\Mpc}$, while in \cref{subsec:f_plots_diff_s}, they are displayed at various separations ($s = \SIlist{20; 70; 130}{\per\hHubble\Mpc}$). In \cref{subsec:Difference_Plots}, we find the deviation between the model and true $f$-integrals. In \cref{subsubsec:extrapolation_cov}, we examine how the trends present in the $f$-integrals may give insight into sparse covariance matrices. Finally, in \cref{subsubsec:comparison_of_cov}, we compare the covariance of the 3PCF and 4PCF using the \textsc{camb} power spectrum to those computed with our model power spectrum.

In each of the figures in \cref{subsec:f_plots_fixed_s,subsec:f_plots_diff_s}, the $f$-integral is weighted by $(r_i/10)^2(r^{\prime}_j/10)^2$, which is a scaled version of the weighting that would enter the covariance integrals when binning in radial separation, as done in a real analysis. The left-hand columns display the weighted $f$-integrals from our model while the right-hand columns show those from \textsc{camb}. We display various $\{\ell,\ell^{\prime},\ell^{\prime\prime}\}$ channels, where each combination of these sBF orders sums to an even number and satisfies the triangle inequality. The black dashed lines trace the Heaviside functions of \cref{eqn:f_FINAL}. The greatest contributions to the $f$-integrals are within the rectangular boxes traced by these black dashed lines. These boxes correspond to the $H(R_{+,ij}-1) H(1-R_{-,ij})$ term in \cref{eqn:f_FINAL} and the purple dashed region in \cref{fig:heavisides} (where closed triangles may be formed). The $f$-integrals are near zero outside of this region; this corresponds to the $H(1-R_{+,ij})$ and $H(R_{-,ij} -1)$ terms of  \cref{eqn:f_FINAL} (dotted orange and solid green lines in \cref{fig:heavisides}, respectively). This structure within the $f$-integrals leads to sparse covariance matrices, and thus, sparse precision matrices.

\subsubsection[\texorpdfstring{$f$}{f}-Integrals at Fixed Separation]{\boldmath \texorpdfstring{$f$}{f}-Integrals at Fixed Separation}
\label{subsec:f_plots_fixed_s}

We first investigate the weighted $f$-integrals at a fixed separation $s = \SI{50}{\per\hHubble\Mpc}$. This separation was chosen as it is mildly nonlinear, avoids the BAO feature, and corresponds to the region at which our model power spectrum begins to agree with the true power spectrum (\cref{fig:pk_vs_k}). 

\begin{figure}
    \centering
    %\vspace{-2.8cm}
    \begin{subfigure}[b]{0.65\textwidth}
        \centering
        \includegraphics[width=\textwidth]{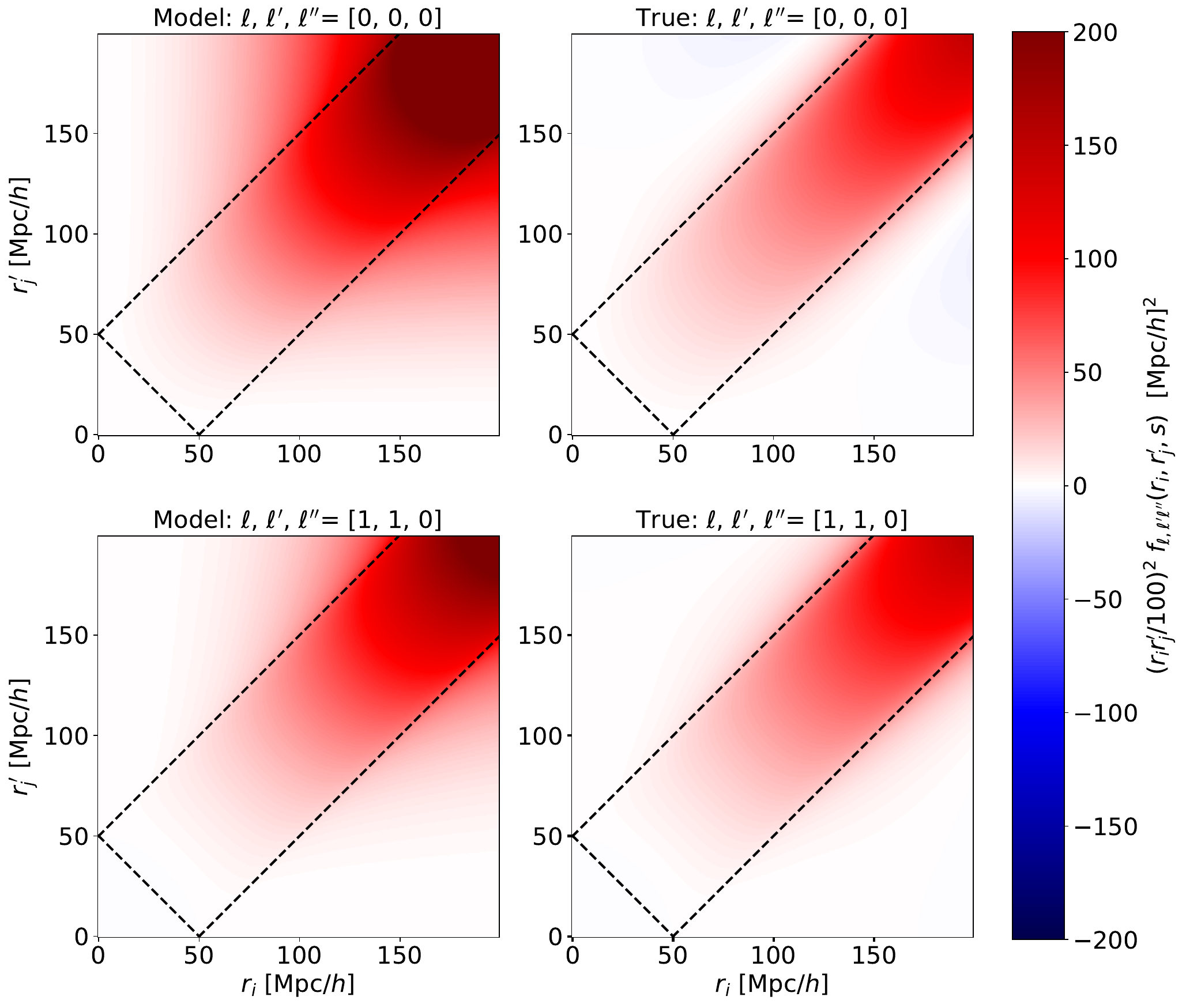}
    \end{subfigure}

    \vspace{0.2cm}

    \begin{subfigure}[b]{0.65\textwidth}
        \centering
        \includegraphics[width=\textwidth]{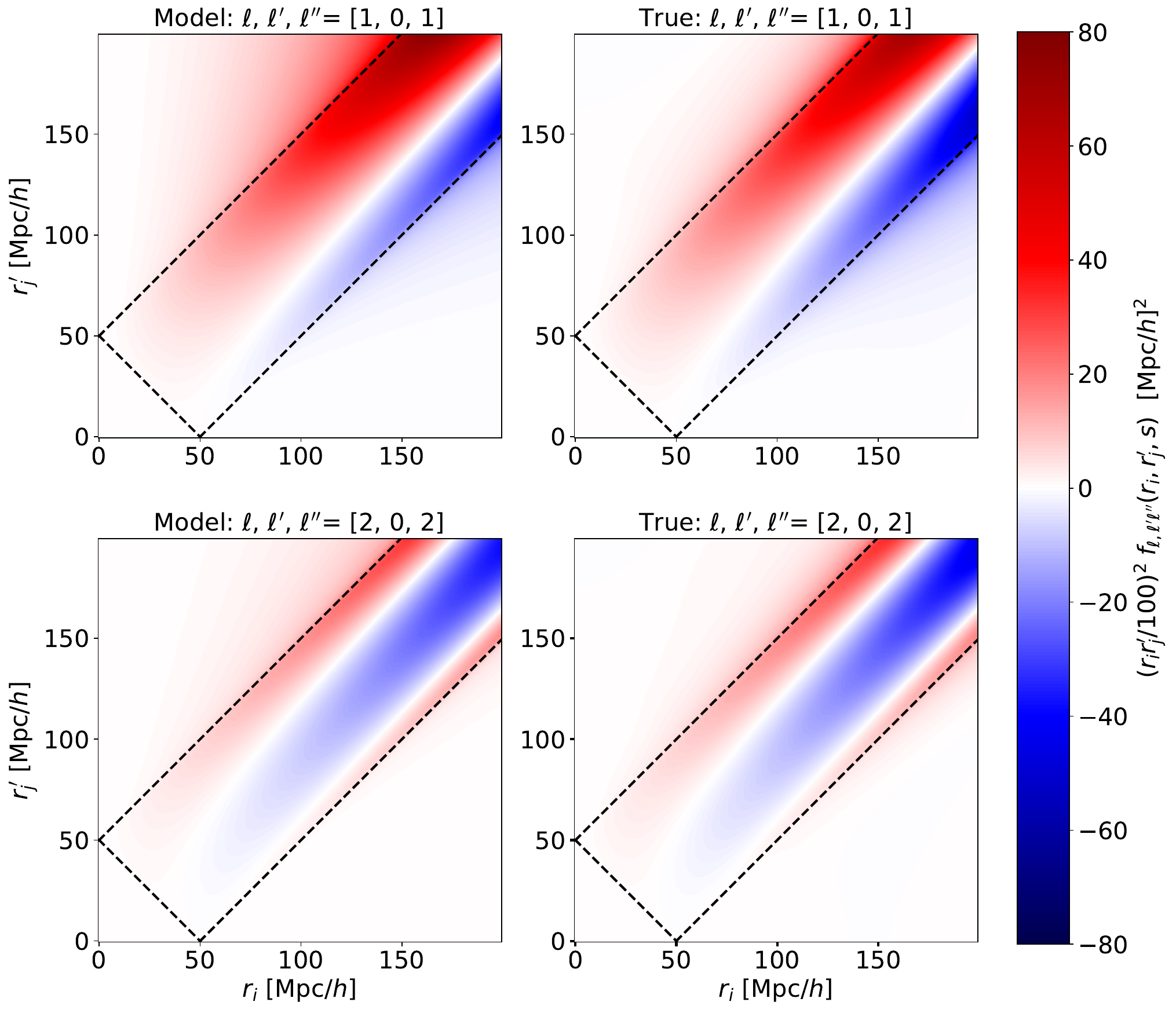}
    \end{subfigure}

    %\vspace{-0.2cm}

    \caption{The $f$-integral (\pcref{eqn:f_int}), weighted by $(r_i/10)^2(r^{\prime}_j/10)^2$, is plotted at $s = \SI{50}{\per\hHubble\Mpc}$ for various $\{\ell,\ell^{\prime},\ell^{\prime\prime}\}$ channels where one of these sBF orders is zero. The \textit{left-hand column} was computed from our model power spectrum (\pcref{eqn:pk}) while the \textit{right-hand column} was determined using the \textsc{camb} power spectrum. The colorbar of the first two rows (where there is symmetry because $\ell = \ell^{\prime}$) differs from that of the last two rows (where the symmetry is broken). The black dashed lines trace the Heaviside functions of \cref{eqn:f_FINAL}; inside this region, $r_i$, $r^{\prime}_j$, and $s$ can form a closed triangle. There is a contribution to the $f$-integrals outside of this region, which we do not see as notably in the following figures.}
    \label{fig:f_integral_order_zero}
\end{figure}

We begin with the weighted $f$-integrals with at least one zero-order sBF. In \cref{fig:f_integral_order_zero}, we display four different $\{\ell,\ell^{\prime},\ell^{\prime\prime}\}$ combinations, where at least one of these sBF orders is zero. We caution that the colorbar for the first two rows of \cref{fig:f_integral_order_zero} does not match the colorbar of the final two rows.

The first two rows (where one $\ell=0$ sBF has argument $s$) have greater contributions to the $f$-integral than do the final two rows (where the $\ell=0$ sBF has argument $r^{\prime}_j$). We note that when one $\ell=0$ sBF has argument $s$, our model slightly overpredicts the values of the $f$-integrals at large $r$ and $r^{\prime}$, especially for the $\{0,0,0\}$ case. Additionally, the contributions to the $f$-integral are greater for lower values of $\ell^{\prime\prime}.$ We also see that in each subplot (both model and \textsc{camb}), the nonzero contributions to the $f$-integrals are not all centralized within the black dashed box traced by the Heaviside functions of \cref{eqn:f_FINAL}, where $r_i$, $r^{\prime}_j$, and $s$ can form a closed triangle. Outside of this box, $r_i$, $r^{\prime}_j$, and $s$ do not form a closed triangle.

\begin{figure}
    \centering  
    \includegraphics[width=0.7\textwidth]{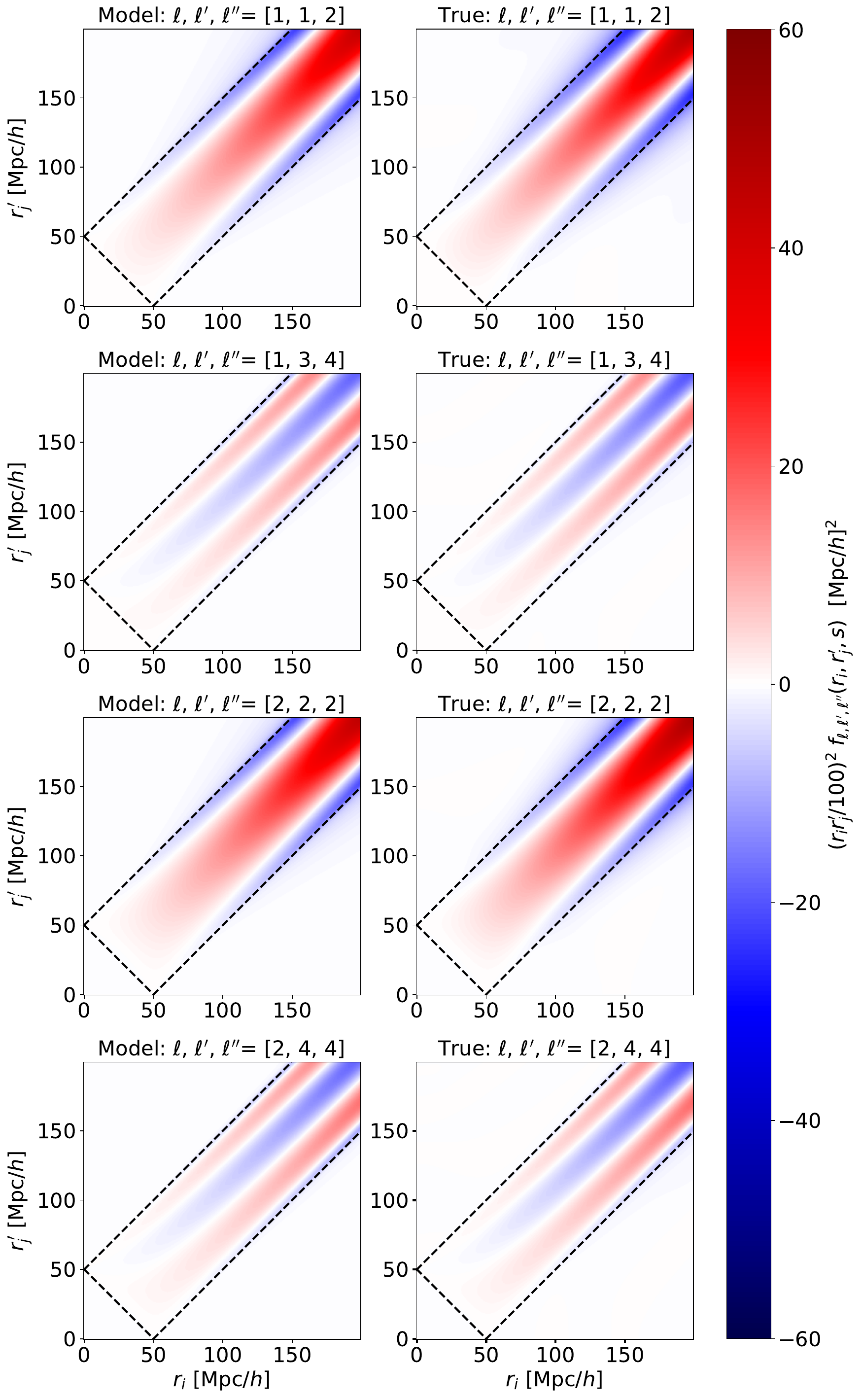}
    \caption{The weighted $f$-integral is plotted at $s = \SI{50}{\per\hHubble\Mpc}$ for various $\{\ell,\ell^{\prime},\ell^{\prime\prime}\}$ channels, where the parity of $\ell$ \emph{matches} that of $\ell^{\prime}$. The \textit{left-hand column} was computed from our model power spectrum (\pcref{eqn:pk}) while the \textit{right-hand column} uses the \textsc{camb} power spectrum. The black dashed lines trace the Heaviside functions of \cref{eqn:f_FINAL}. The first and third rows (or second and fourth rows) have similar patterns due to having the same value of $\ell^{\prime\prime}$. We note that there are greater contributions to the $f$-integral for smaller values of $\ell^{\prime\prime}$. Additionally, there is more oscillation between positive and negative values than in \cref{fig:f_integral_order_zero} (where at least one sBF was of order zero).}
    \label{fig:f_integral_same_parity}
\end{figure}

In \cref{fig:f_integral_same_parity}, we display the weighted $f$-integrals for various $\{\ell,\ell^{\prime},\ell^{\prime\prime}\}$ channels where $\ell$ and $\ell^{\prime}$ have the same parity. The $f$-integrals are near zero outside of the rectangular regions given by the Heaviside functions of \cref{eqn:f_FINAL} (traced by the black dashed lines). The first and third rows (those with $\ell^{\prime\prime}=2$) have similar interference patterns; so do the second and fourth rows (those with $\ell^{\prime\prime}=4$). Similarly to \cref{fig:f_integral_order_zero} (where at least one sBF had order zero), there is a decrease in the values of the $f$-integrals as $\ell^{\prime\prime}$ increases. However, the $f$-integrals here are more oscillatory and have lower values than those in \cref{fig:f_integral_order_zero}. Overall, when $\ell$ and $\ell^{\prime}$ have the same parity, our model power spectrum accurately reproduces the $f$-integrals we expect from the true power spectrum. In other words, the left column faithfully captures the right column.

\begin{figure}
    \centering  
    \includegraphics[width=0.7\textwidth]{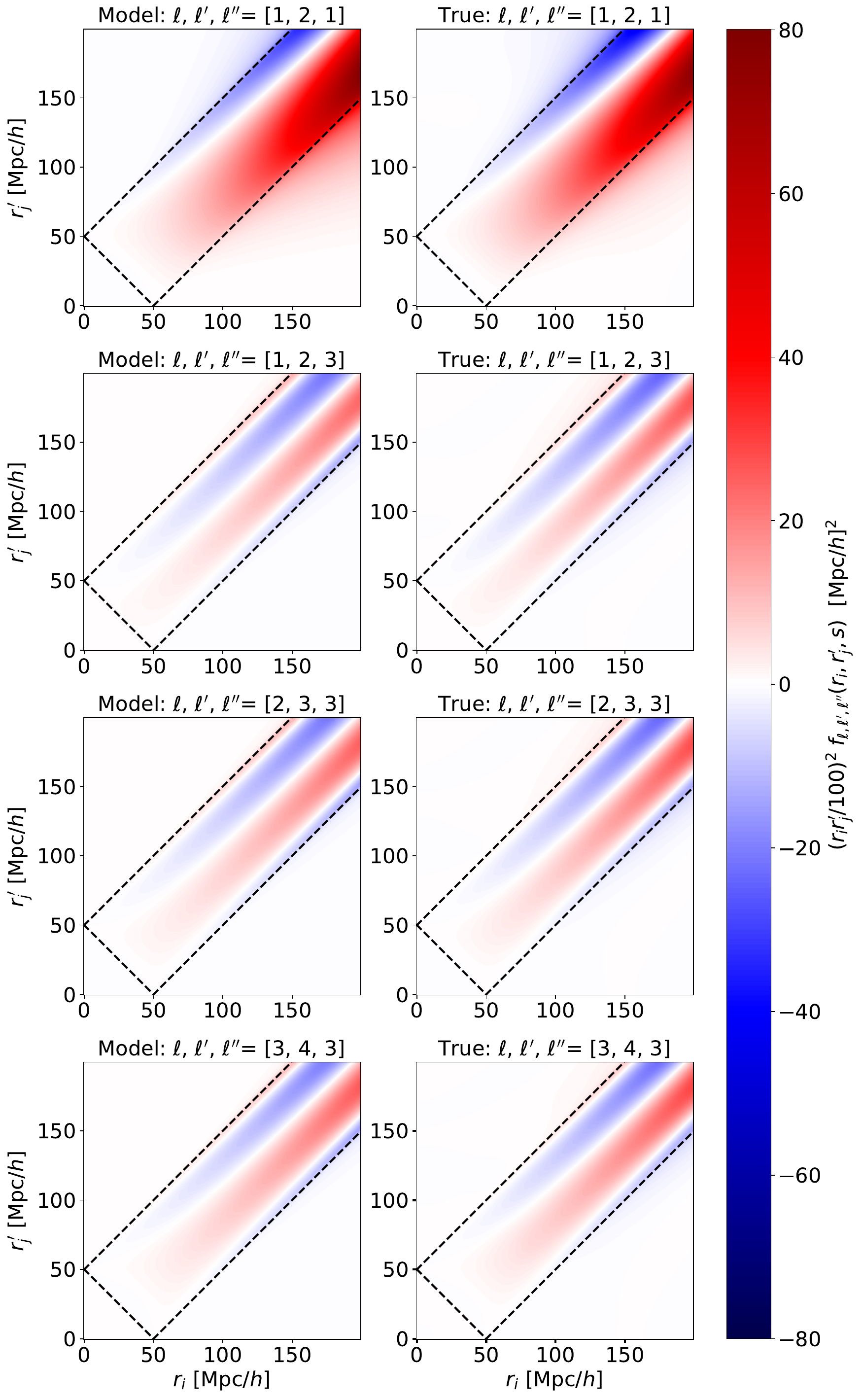}
    \caption{The weighted $f$-integral is plotted at $s = \SI{50}{\per\hHubble\Mpc}$ for various $\{\ell,\ell^{\prime},\ell^{\prime\prime}\}$ channels, where the parity of $\ell$ is \emph{opposite} that of $\ell^{\prime}$. The \textit{left-hand column} was computed from our model power spectrum (\pcref{eqn:pk}) while the \textit{right-hand column} uses the \textsc{camb} power spectrum. The black dashed lines trace the Heaviside functions of \cref{eqn:f_FINAL}. Similarly to \namecrefs{fig:f_integral_order_zero} \labelcref{fig:f_integral_order_zero} (where there is at least one zero-order sBF) and \labelcref{fig:f_integral_same_parity} (where $\ell$ and $\ell^{\prime}$ have the same parity), the $f$-integrals here have lesser values for higher $\ell^{\prime\prime}$.}
    \label{fig:f_integral_opp_parity}
\end{figure}

Next, we show the weighted $f$-integrals where $\ell$ and $\ell^{\prime}$ have opposite parity (\cref{fig:f_integral_opp_parity}). In the first row of \cref{fig:f_integral_opp_parity}, the contributions to the $f$-integrals are not fully contained within the rectangular box traced by the Heaviside functions, although they fall off quickly outside of it. In the other subplots of \cref{fig:f_integral_opp_parity}, the $f$-integrals are near zero outside of these boxed regions. The interference patterns of the final three rows resemble one another; this is due to the sBFs associated with the separation $s$ having the same order. We once again see that the $f$-integral contributions decrease as $\ell^{\prime\prime}$ increases, which was also seen in \cref{fig:f_integral_order_zero,fig:f_integral_same_parity}. Between \cref{fig:f_integral_same_parity,fig:f_integral_opp_parity} we do not see a noticeable difference in the $f$-integrals, regardless of whether $\ell$ and $\ell^{\prime}$ have the same parity or not. Again, our model power spectrum accurately reproduces the $f$-integrals from the true power spectrum.  

\begin{figure}
    \centering  
    \includegraphics[width=0.7\textwidth]{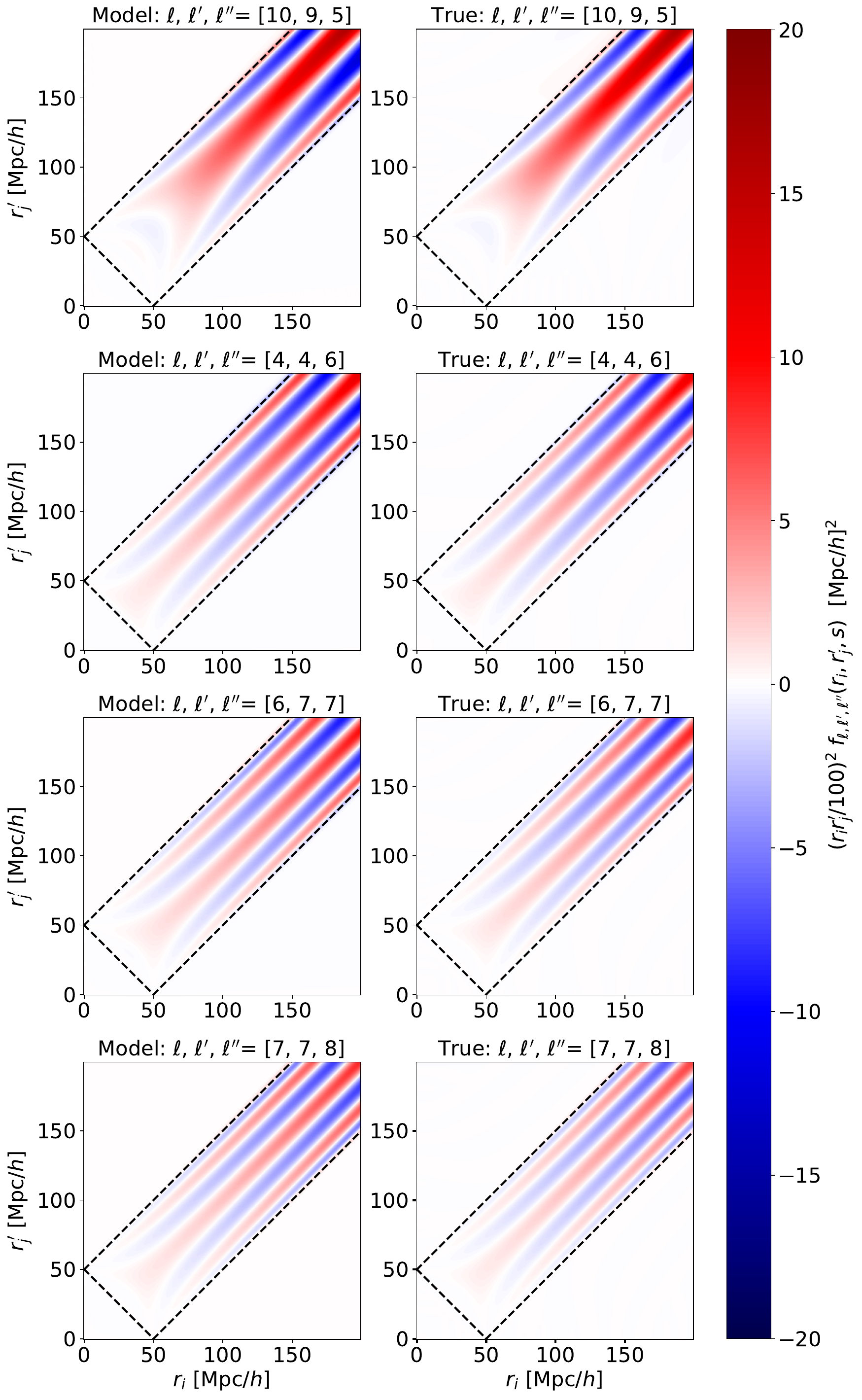}
    \caption{The weighted $f$-integral is plotted at $s = \SI{50}{\per\hHubble\Mpc}$ for various $\{\ell,\ell^{\prime},\ell^{\prime\prime}\}$ channels, with higher values than those in the previous plots. The \textit{left-hand column} was computed from our model power spectrum (\pcref{eqn:pk}) while the \textit{right-hand column} uses the \textsc{camb} power spectrum. The black dashed lines trace the Heaviside functions of \cref{eqn:f_FINAL}. Here, we see more oscillation between positive and negative values than in \crefrange{fig:f_integral_order_zero}{fig:f_integral_opp_parity}, since the sBFs in these $f$-integrals are of higher orders than the previous figures. The sines and cosines of which they are composed are thus modified by an increased amount of polynomials, which weight the trigonometric functions by greater inverse powers of their arguments.}
    \label{fig:f_integral_high_ell}
\end{figure}

In \cref{fig:f_integral_high_ell}, we show the weighted $f$-integrals at higher values of $\{\ell,\ell^{\prime},\ell^{\prime\prime}\}$ than those of the previous figures. Again, we see that these $f$-integrals are near zero outside of the rectangular region bounded by the Heaviside functions from \cref{eqn:f_FINAL}. As the value of $\ell^{\prime\prime}$ increases, the $f$-integrals become more oscillatory. Overall, we see that our model power spectrum faithfully captures the oscillatory nature of the $f$-integrals.

\subsubsection[\texorpdfstring{$f$}{f}-Integrals at Different Separations]{\boldmath \texorpdfstring{$f$}{f}-Integrals at Different Separations}
\label{subsec:f_plots_diff_s}

\begin{figure}
    \centering  
    \includegraphics[width=0.7\textwidth]{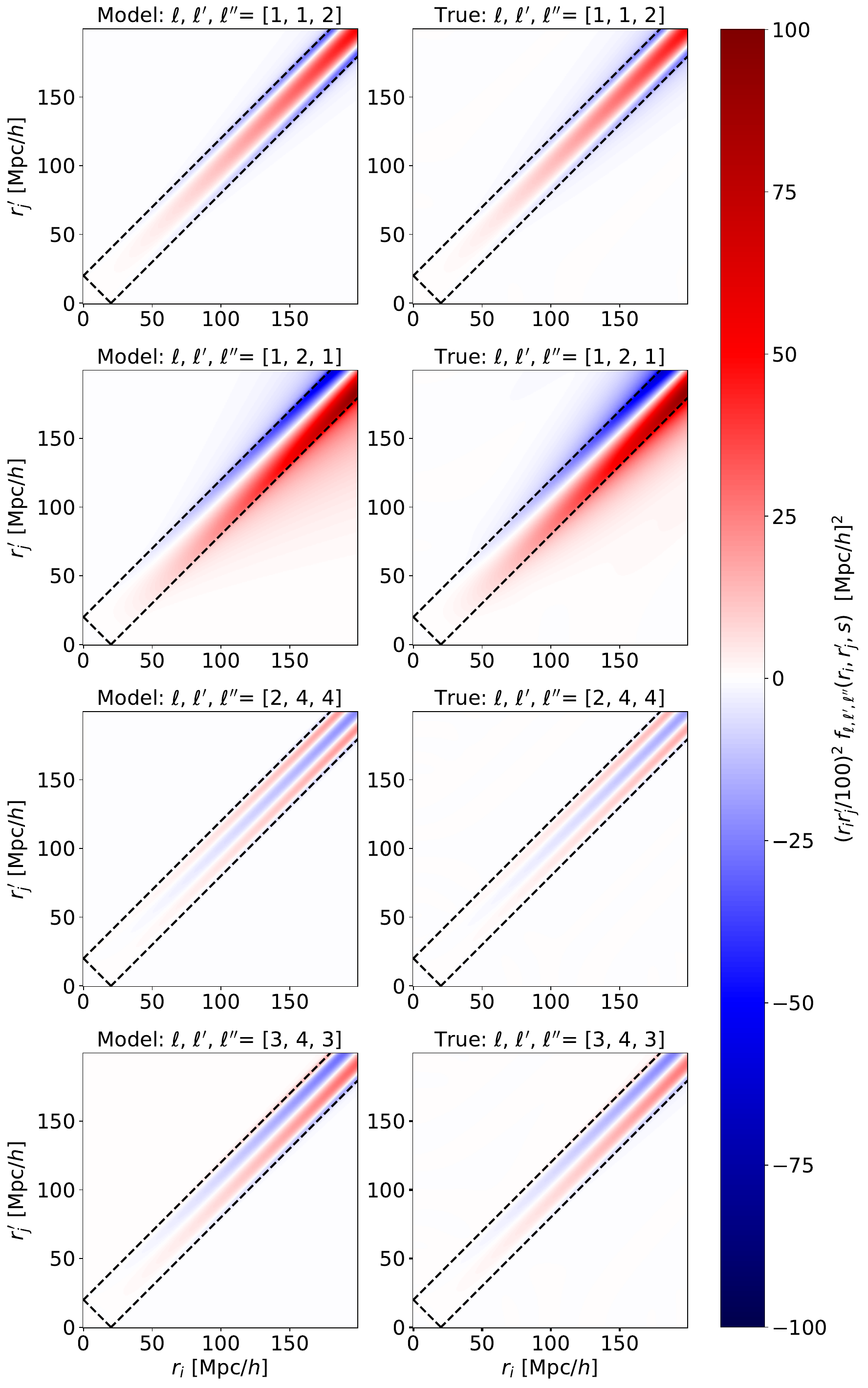}
    \caption{The $f$-integral (\pcref{eqn:f_int}), weighted by $(r_i/10)^2(r^{\prime}_j/10)^2$, is plotted at $s = \SI{20}{\per\hHubble\Mpc}$ for various $\{\ell,\ell^{\prime},\ell^{\prime\prime}\}$ channels. The \textit{left-hand column} was computed from our model power spectrum (\pcref{eqn:pk}) while the \textit{right-hand column} uses the \textsc{camb} power spectrum. The black dashed lines trace the Heaviside functions of \cref{eqn:f_FINAL}. We see that for small separation $s$, the triangular region between $r_i$, $r^{\prime}_j$, and $s$ shrinks to a narrow box. The $f$-integrals here have a compact region of support.}
    \label{fig:f_integral_s=20}
\end{figure}

\begin{figure}
    \centering  
    \includegraphics[width=0.7\textwidth]{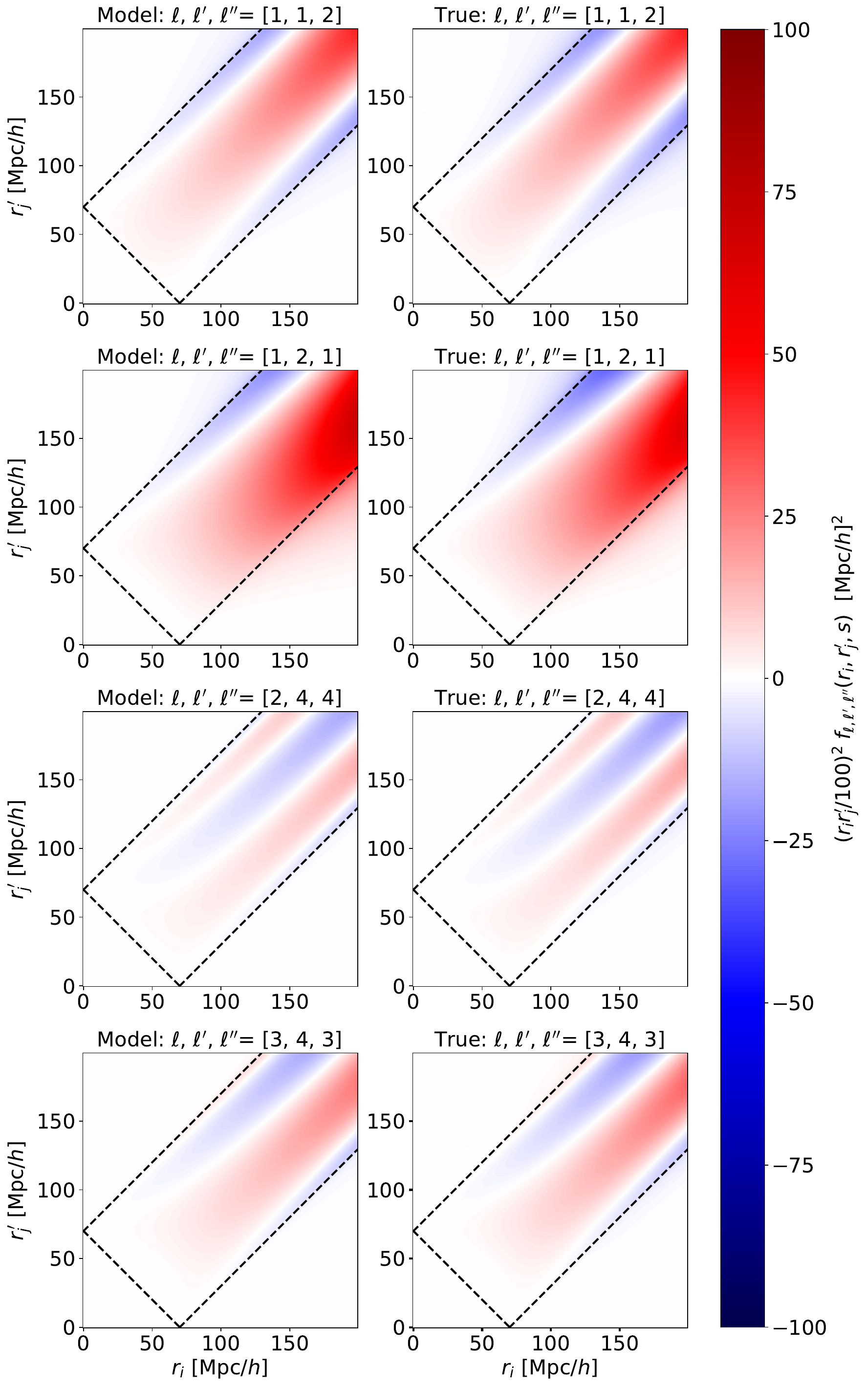}
    \caption{The weighted $f$-integral is plotted at $s = \SI{70}{\per\hHubble\Mpc}$ for various $\{\ell,\ell^{\prime},\ell^{\prime\prime}\}$ channels. The \textit{left-hand column} was computed from our model power spectrum (\pcref{eqn:pk}) while the \textit{right-hand column} uses the \textsc{camb} power spectrum. The black dashed lines trace the Heaviside functions of \cref{eqn:f_FINAL}. Here, we see a larger triangular region than in \cref{fig:f_integral_s=20}, where the separation was smaller. The contributions within these triangular regions have lesser values than the corresponding plots in \cref{fig:f_integral_s=20}.}
    \label{fig:f_integral_s=70}
\end{figure}

\begin{figure}
    \centering  
    \includegraphics[width=0.7\textwidth]{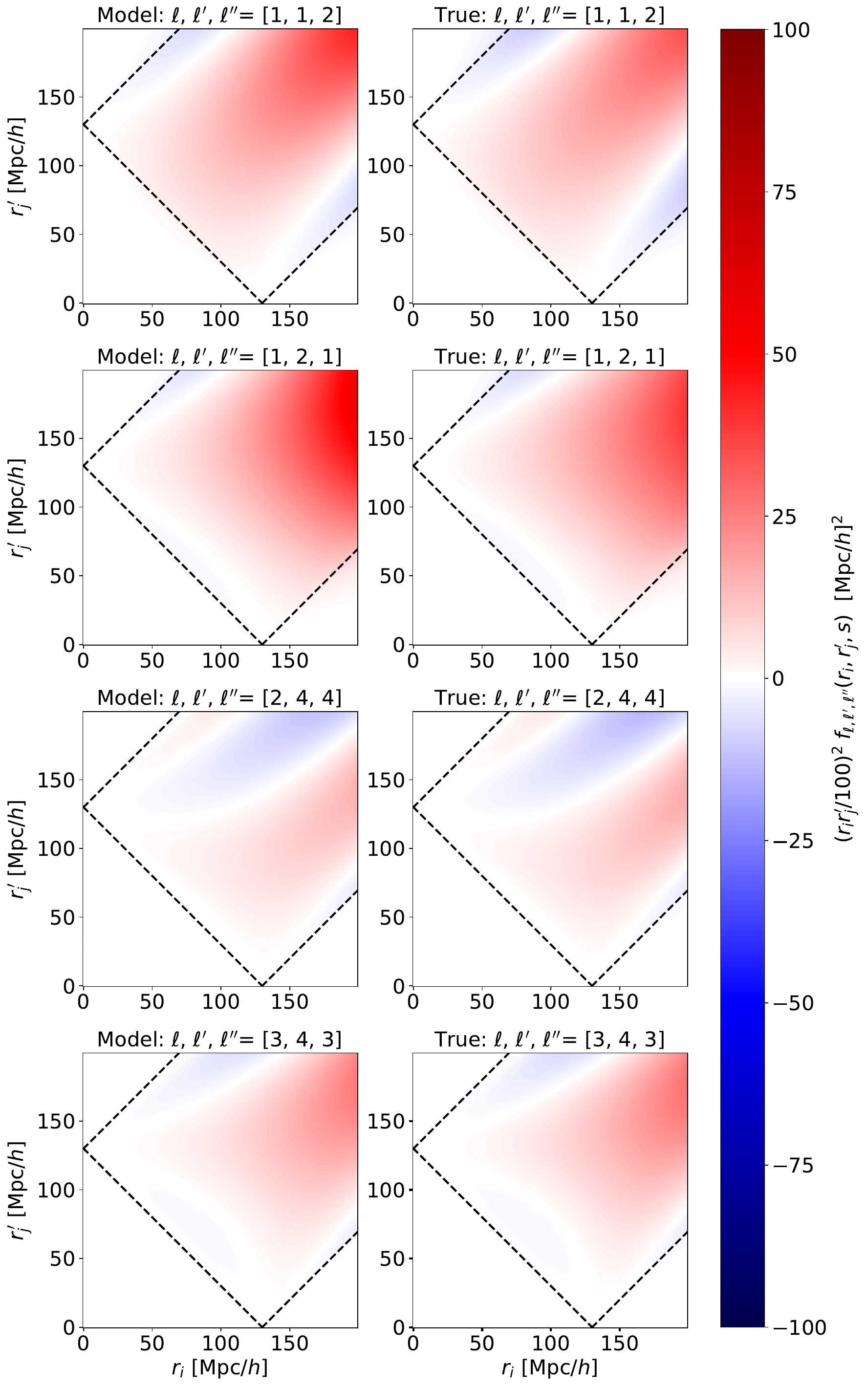}
    \caption{The weighted $f$-integral is plotted at $s = \SI{130}{\per\hHubble\Mpc}$ for various $\{\ell,\ell^{\prime},\ell^{\prime\prime}\}$ channels. The \textit{left-hand column} was computed from our model power spectrum (\pcref{eqn:pk}) while the \textit{right-hand column} uses the \textsc{camb} power spectrum. The black dashed lines trace the Heaviside functions of \cref{eqn:f_FINAL}. We again see a larger triangular region compared to \cref{fig:f_integral_s=20,fig:f_integral_s=70}, which each had smaller separations. As the separation $s$ increases, the contributions within the triangular regions decrease in value; the subplots shown here have lesser values than those displayed in \cref{fig:f_integral_s=20,fig:f_integral_s=70}.}
    \label{fig:f_integral_s=130}
\end{figure}

We now examine how increasing the separation $s$ affects the $f$-integrals. Here, we display four $\{\ell,\ell^{\prime},\ell^{\prime\prime}\}$ combinations, each at three different separations. We choose $s = \SI{20}{\per\hHubble\Mpc}$, which is in the nonlinear regime (\cref{fig:f_integral_s=20}); $s = \SI{70}{\per\hHubble\Mpc}$, which is in the linear regime (\cref{fig:f_integral_s=70}); and $s = \SI{130}{\per\hHubble\Mpc}$, which is above the scale of the BAO feature (\cref{fig:f_integral_s=130}). \Crefrange{fig:f_integral_s=20}{fig:f_integral_s=130} use the same colorbar. Again, we see that our model power spectrum accurately reproduces the $f$-integrals from the true power spectrum. 

The subplots for $\{\ell,\ell^{\prime},\ell^{\prime\prime}\} = \{1,2,1\}$ in \crefrange{fig:f_integral_s=20}{fig:f_integral_s=130} have nonzero contributions extending outside of the rectangular box tracing the Heaviside functions from \cref{eqn:f_FINAL}. In other words, there are nonzero contributions in the region where $r_i$, $r^{\prime}_j$, and $s$ do not form a closed triangle, in addition to the contributions where $r_i$, $r^{\prime}_j$, and $s$ may form a closed triangle. The non-triangular contributions quickly fall off away from the boxed region. This behavior is similar to the subplot of $\{\ell,\ell^{\prime},\ell^{\prime\prime}\} = \{1,2,1\}$ displayed in \cref{fig:f_integral_opp_parity}. We see similar behavior in the subplots for $\{\ell,\ell^{\prime},\ell^{\prime\prime}\} = \{1,1,2\}$ in \crefrange{fig:f_integral_s=20}{fig:f_integral_s=130}, which is consistent with the  $\{\ell,\ell^{\prime},\ell^{\prime\prime}\} = \{1,1,2\}$ subplot in \cref{fig:f_integral_same_parity}.

As the separation $s$ increases, the rectangular box bounded by the Heaviside functions (where $r_i$, $r^{\prime}_j$, and $s$ may form a closed triangle) becomes wider and shifts upward in the $\{r_i,r^{\prime}_j\}$ plane. We display this in \crefrange{fig:f_integral_s=20}{fig:f_integral_s=130}. In particular, this widening of the triangular region leads to a greater number of $\{r_i,r^{\prime}_j\}$ combinations that may produce nonzero contributions to the $f$-integrals. As $s$ increases, the contributions within the triangular region decrease in value.

\subsubsection[Comparison of Model and True \texorpdfstring{$f$}{f}-Integrals]{\boldmath Comparison of Model and True \texorpdfstring{$f$}{f}-Integrals}
\label{subsec:Difference_Plots}

To compare the $f$-integrals from our model power spectrum to those from the \textsc{camb} power spectrum, we find the difference between the two for each set of subplots in \crefrange{fig:f_integral_order_zero}{fig:f_integral_s=130}. We define the difference between the unweighted $f$-integrals as
\begin{equation}
    \mathcal{D} \equiv f_{\text{Model}} - f_{\textsc{camb}}.
    \label{eqn:D}
\end{equation}
Here, $f_{\text{Model}}$ is the $f$-integral defined in \cref{eqn:f_int} computed with the model power spectrum, while $f_{\textsc{camb}}$ is that computed with the \textsc{camb} power spectrum.

\begin{figure}
    \centering  
    \includegraphics[width=0.85\textwidth]{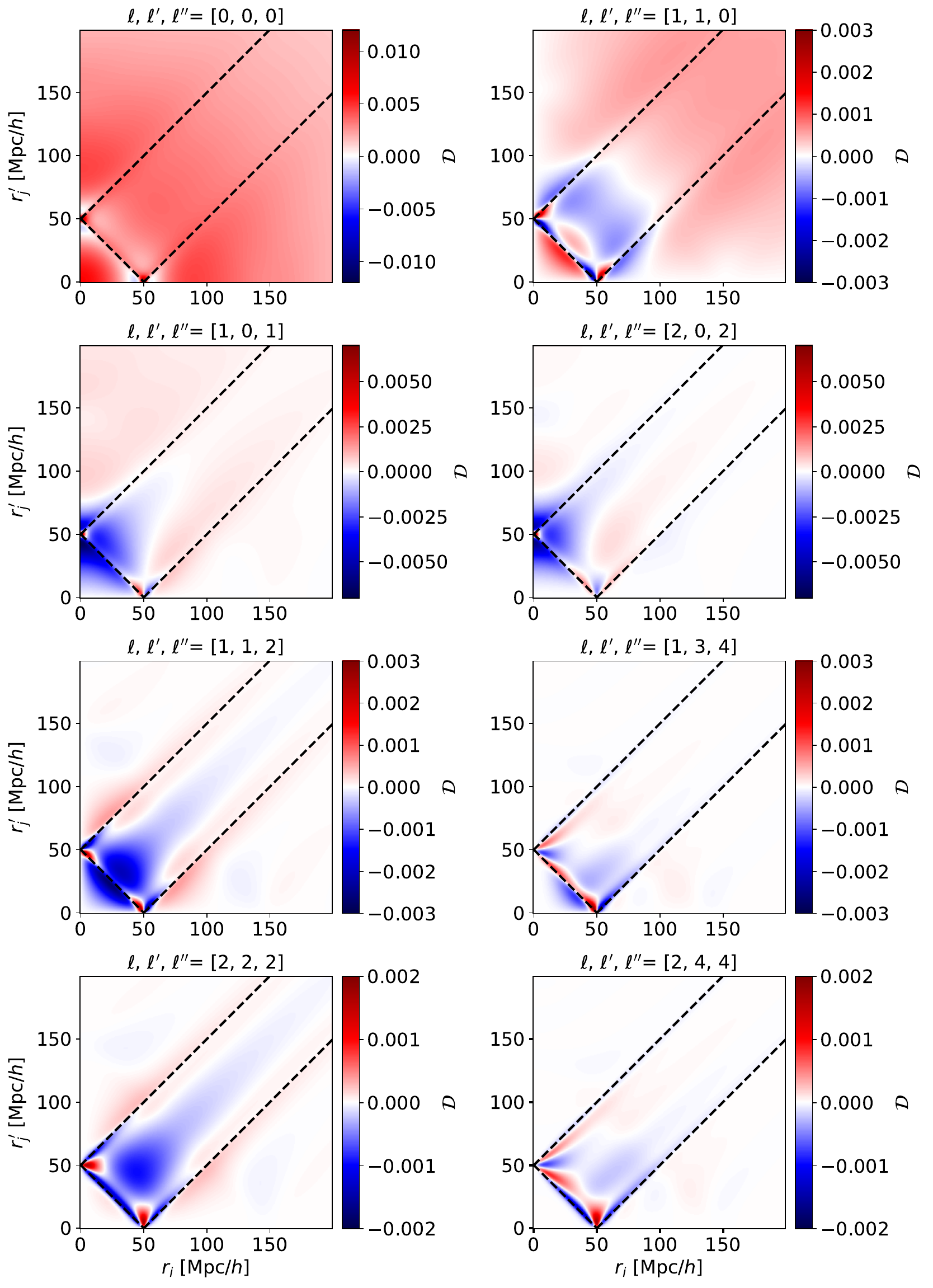}
    \caption{Here, we display the differences ($\mathcal{D}$, \pcref{eqn:D}) between the unweighted model and true $f$-integrals for the $\{\ell,\ell^{\prime},\ell^{\prime\prime}\}$ combinations displayed in \cref{fig:f_integral_order_zero,fig:f_integral_same_parity}, at the fixed separation $s = \SI{50}{\per\hHubble\Mpc}$. The red regions show where the model overpredicts the true $f$-integrals; the blue regions show where the model underpredicts the true $f$-integrals. We see that our model $f$-integrals accurately reproduce the true $f$-integrals, with the greatest deviations between the two at small values of $r_i$ and $r^{\prime}_j$.}
    \label{fig:Difference_f_integral}
\end{figure}

\begin{figure}
    \centering  
    \includegraphics[width=0.85\textwidth]{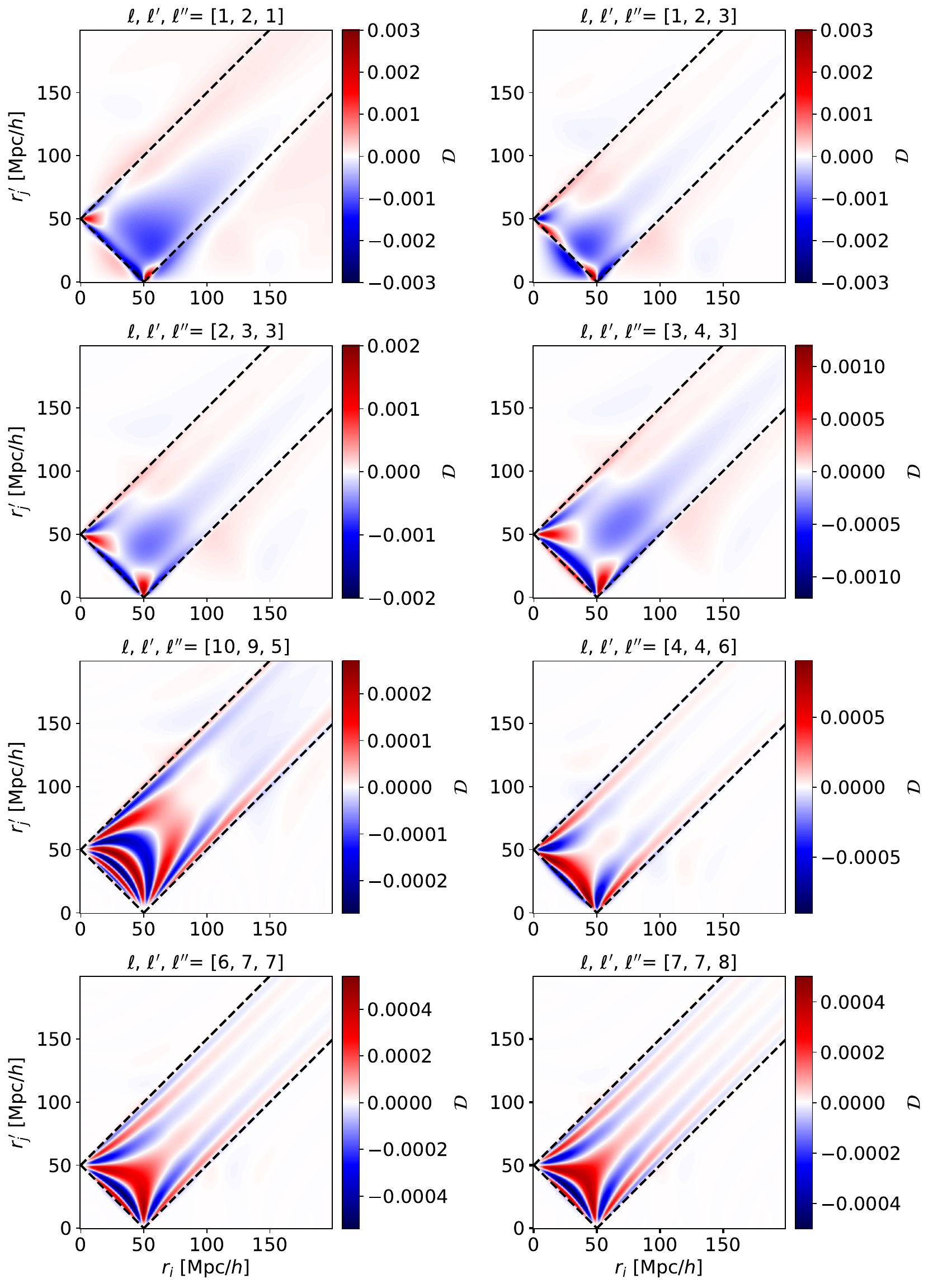}
    \caption{Here, we display the differences ($\mathcal{D}$, \pcref{eqn:D}) between the unweighted model and true $f$-integrals for the $\{\ell,\ell^{\prime},\ell^{\prime\prime}\}$ combinations displayed in \cref{fig:f_integral_opp_parity,fig:f_integral_high_ell}, at the fixed separation $s = \SI{50}{\per\hHubble\Mpc}$. The red regions show where the model overpredicts the true $f$-integrals; the blue regions show where the model underpredicts the true $f$-integrals. Similar to \cref{fig:Difference_f_integral}, the model $f$-integrals are accurate to the true ones, with the greatest deviations between them at small values of $r_i$ and $r^{\prime}_j$. Additionally, the final four subplots (those with higher sBF orders, \cref{fig:f_integral_high_ell}) have oscillatory structure that is not entirely captured by our model; however, the deviations here are of sub-percent order.}
    \label{fig:Difference2_f_integral}
\end{figure}

\begin{figure}
    \centering  
    \includegraphics[width=\textwidth]{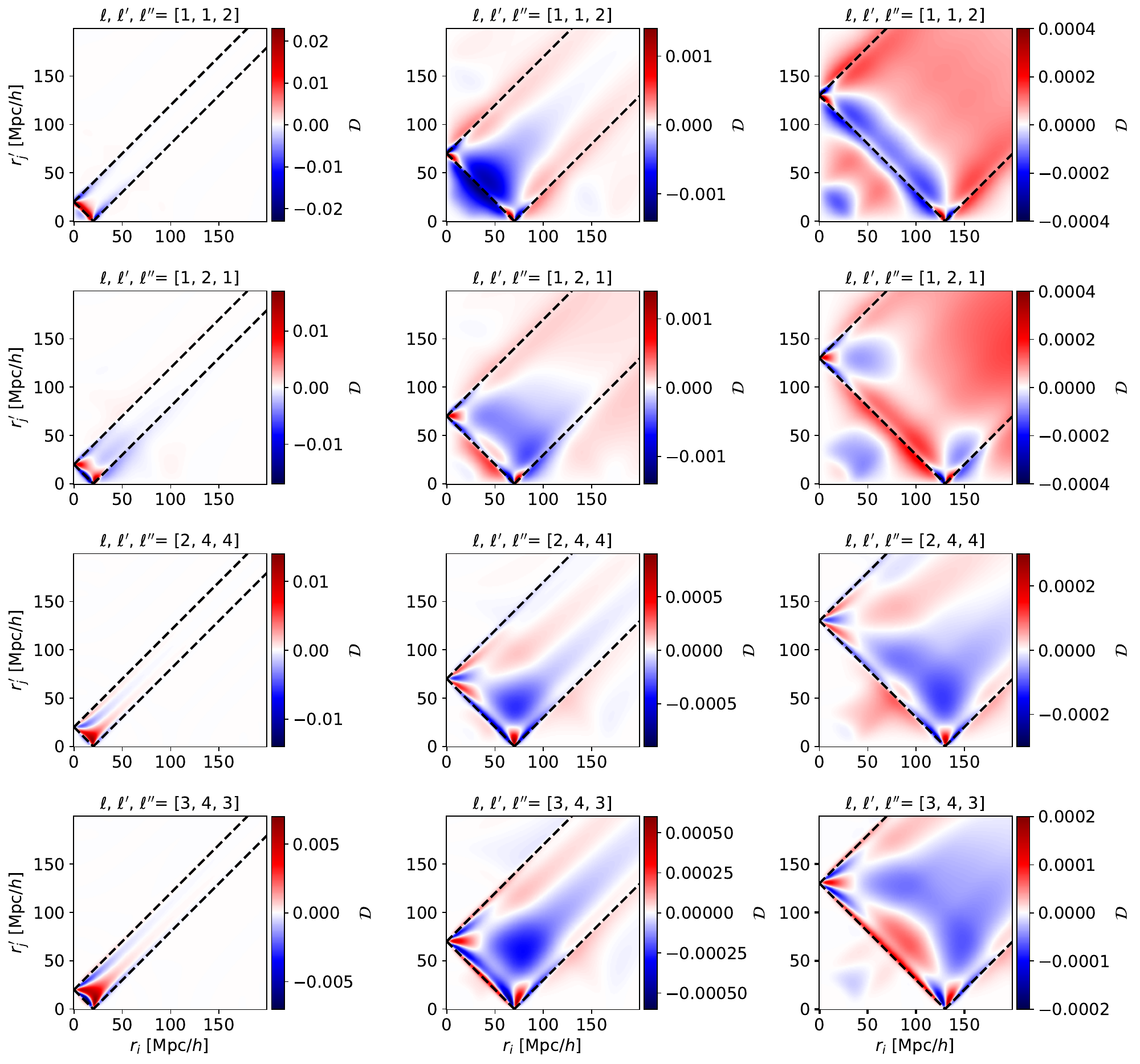}
    \caption{Here, we display the differences ($\mathcal{D}$, \pcref{eqn:D}) between the unweighted model and true $f$-integrals for the $\{\ell,\ell^{\prime},\ell^{\prime\prime}\}$ combinations and separations displayed in \crefrange{fig:f_integral_s=20}{fig:f_integral_s=130}. Each row shows the same combination of $\{\ell,\ell^{\prime},\ell^{\prime\prime}\}$. The \textit{left-hand column} displays subplots at $s = \SI{20}{\per\hHubble\Mpc}$, the \textit{middle column} at $s = \SI{70}{\per\hHubble\Mpc}$, and the \textit{right-hand column} at $s = \SI{130}{\per\hHubble\Mpc}$. The red regions show where the model overpredicts the true $f$-integrals; the blue regions show where the model underpredicts the true $f$-integrals. As the separation increases, the deviations between model and true $f$-integrals decrease; our model is more accurate at large separations. For $s = \SI{20}{\per\hHubble\Mpc}$, the greatest deviations are present at small values of $r_i$ and $r^{\prime}_j$, similar to \cref{fig:Difference_f_integral,fig:Difference2_f_integral}. For $s = \SI{70}{\per\hHubble\Mpc}$, the greatest deviations are present at mid-range values of $r_i$ and $r^{\prime}_j$. Finally, at $s = \SI{130}{\per\hHubble\Mpc}$, the deviations are not centralized to a certain region of $r_i$ and $r^{\prime}_j$.}
    \label{fig:Difference_f_integral_Diff_s}
\end{figure}

The differences $\mathcal{D}$ between the unweighted $f$-integrals at a fixed separation $s = \SI{50}{\per\hHubble\Mpc}$ (\crefrange{fig:f_integral_order_zero}{fig:f_integral_high_ell}) are displayed in \cref{fig:Difference_f_integral,fig:Difference2_f_integral}. \Cref{Tab:difference_f_int} provides a summary of these differences. For the cases where we allowed the separation to vary ($s = $ \SIlist{20; 70; 130}{\per\hHubble\Mpc}, \crefrange{fig:f_integral_s=20}{fig:f_integral_s=130}), we show the differences between the unweighted model and true $f$-integrals in \cref{fig:Difference_f_integral_Diff_s}, with a summary in \cref{Tab:difference_f_int_diff_s}. 

In each of \crefrange{fig:Difference_f_integral}{fig:Difference_f_integral_Diff_s}, the black dashed lines show the Heaviside functions from \cref{eqn:f_FINAL}. Since we are subtracting the true $f$-integrals from the model $f$-integrals, the red regions indicate where the model overpredicts the true $f$-integral while the blue regions show where the model underpredicts the true $f$-integral.

In \cref{fig:Difference_f_integral,fig:Difference2_f_integral}, we see that our model $f$-integrals accurately capture the true $f$-integrals at fixed separation $s = \SI{50}{\per\hHubble\Mpc}$. The greatest deviations between the two are present for low values of $r_i$ and $r^{\prime}_j$. In \cref{fig:Difference_f_integral_Diff_s}, for which we examine various separations, the model and true $f$-integrals once again agree. As the separation $s$ increases, the deviations between the model and true $f$-integrals decrease. Our model is more accurate at larger separations. Additionally, for $s = \SI{20}{\per\hHubble\Mpc}$, the largest deviations are present at small pair separations $r_i$ and $r^{\prime}_j$. For $s = \SI{70}{\per\hHubble\Mpc}$, the greatest deviations are at values in the mid-range of pair separations, and at $s = \SI{130}{\per\hHubble\Mpc}$, there is no notable pattern to which pair separations show the largest deviations.

\Cref{Tab:difference_f_int,Tab:difference_f_int_diff_s} display the mean, standard deviation, and maximum of the absolute value of all elements in each difference plot. For each case, we achieve single-digit percent level accuracy of the $f$-integrals, confirming that our power spectrum model provides a suitable alternative to the true power spectrum. Across all cases in \cref{Tab:difference_f_int,Tab:difference_f_int_diff_s}, the largest element found in the difference matrices is \SI{2.22}{\percent} away from zero.

\begin{table}
    \centering
    \begin{tabular}{@{\hspace{\tabcolsep}\phantom{1}} c S[table-format=+1.2e+1] *{2}{S[table-format=1.2e+1]}}
        \toprule
        {\boldmath$\{\ell,\ell^{\prime},\ell^{\prime\prime}\}$} & {\textbf{Mean}} & {\textbf{SD}} & {\textbf{Max(AbsVal)}}
        \\
        \midrule
        \{0,0,0\}  & 2.61e-3 & 8.14e-4 & 1.19e-2 \\
        \{1,1,0\}  & 2.53e-4 & 3.36e-4 & 2.76e-3 \\
        \{1,0,1\}  & 1.08e-5 & 6.79e-4 & 7.01e-3 \\
        \{2,0,2\}  & -8.79e-5 & 5.10e-4 & 7.03e-3 \\
        \{1,1,2\}  & -6.62e-5 & 3.73e-4 & 2.69e-3 \\
        \{1,3,4\}  & -1.43e-5 & 1.53e-4 & 2.97e-3 \\
        \{2,2,2\}  & -4.63e-5 & 2.07e-4 & 1.72e-3 \\
        \{2,4,4\}  & -9.95e-6 & 1.13e-4 & 1.77e-3 \\
        \{1,2,1\}  & -4.69e-5 & 2.71e-4 & 2.94e-3 \\
        \{1,2,3\}  & -3.58e-5 & 2.23e-4 & 2.83e-3 \\
        \{2,3,3\}  & -2.44e-5 & 1.47e-4 & 1.75e-3 \\
        \{3,4,3\}  & -1.84e-5 & 1.11e-4 & 1.11e-3 \\
        \phantom{\{}\llap{\{1}0,9,5\}  & -8.56e-7 & 4.21e-5 & 2.67e-4 \\
        \{4,4,6\}  & 1.48e-6 & 9.48e-5 & 9.01e-4 \\
        \{6,7,7\}  & 2.21e-6 & 6.59e-5 & 5.30e-4 \\
        \{7,7,8\}  & 2.71e-6 & 6.41e-5 & 4.98e-4 \\
        \bottomrule
    \end{tabular}

    \caption{We display the mean, standard deviation, and maximum of the absolute value of the elements in each difference plot between the model and true $f$-integrals. Here, we show various combinations of $\{\ell,\ell^{\prime},\ell^{\prime\prime}\}$ at the same separation, $s = \SI{50}{\per\hHubble\Mpc}$, corresponding to \cref{fig:Difference_f_integral,fig:Difference2_f_integral}. The rows are listed in the order that the figures appear in the paper. The mean values listed here are typically one order of magnitude smaller than the values of the unweighted $f$-integrals; the standard deviation values are typically of the same order of magnitude as the unweighted $f$-integrals. The mean and standard deviation values show that we are achieving sub-percent level accuracy between the model and true $f$-integrals. The maximum of the absolute value of the elements gives the largest value present in each difference plot; we see that across all such plots, the largest difference is \SI{1.19}{\percent} away from a perfect match between model and true $f$-integrals.}
    \label{Tab:difference_f_int}
\end{table}

\begin{table}
    \centering
    \begin{tabular}{c S[table-format=3.0] S[table-format=+1.2e+1] *{2}{S[table-format=1.2e+1]}}
        \toprule
        {\boldmath$\{\ell,\ell^{\prime},\ell^{\prime\prime}\}$} & {\textbf{\boldmath$s$ / \si[detect-all]{\per\hHubble\Mpc}}} & \textbf{Mean} & \textbf{SD} & \textbf{Max(AbsVal)}
        \\
        \midrule
        \{1,1,2\}  &  20 & 1.08e-5 & 1.08e-3 & 2.22e-2 \\
        \{1,2,1\}  &  20 & -6.64e-5 & 8.25e-4 & 1.67e-2 \\
        \{2,4,4\}  &  20 & 2.64e-5 & 6.08e-4 & 1.31e-2 \\
        \{3,4,3\}  &  20 & 1.90e-5 & 4.72e-4 & 6.95e-3 \\
        \{1,1,2\}  &  70 & -3.86e-5 & 2.24e-4 & 1.39e-3 \\
        \{1,2,1\}  &  70 & 4.10e-7 & 1.45e-4 & 1.42e-3 \\
        \{2,4,4\}  &  70 & -1.68e-5 & 8.58e-5 & 8.85e-4 \\
        \{3,4,3\}  &  70 & -2.06e-5 & 7.95e-5 & 5.55e-4 \\
        \{1,1,2\}  & 130 & 3.74e-5 & 7.04e-5 & 4.02e-4 \\
        \{1,2,1\}  & 130 & 4.44e-5 & 6.03e-5 & 3.96e-4 \\
        \{2,4,4\}  & 130 & -8.17e-6 & 3.58e-5 & 2.56e-4 \\
        \{3,4,3\}  & 130 & -7.60e-6 & 2.83e-5 & 1.56e-4 \\
        \bottomrule
    \end{tabular}

    \caption{We display the mean, standard deviation, and maximum of the absolute value of the elements in each difference plot between the model and true $f$-integrals. Here, we show various combinations of $\{\ell,\ell^{\prime},\ell^{\prime\prime}\}$ at the separations $s = \SIlist{20; 70; 130}{\per\hHubble\Mpc}$, which corresponds to \cref{fig:Difference_f_integral_Diff_s}. The rows are listed in the order that the figures appear in the paper. The mean values listed here are typically one order of magnitude less than the values of the unweighted $f$-integrals. The standard deviation values are typically of the same order of magnitude or one order smaller than the unweighted $f$-integral values. The mean and standard deviation values show that we are achieving sub-percent level accuracy between the model and true $f$-integrals. The maximum of the absolute value of the elements gives the largest value present in each difference plot; we see that across all such plots, the largest difference is \SI{2.22}{\percent} away from a perfect match between model and true $f$-integrals.} 
    \label{Tab:difference_f_int_diff_s}
\end{table}

\subsubsection{Implications for the Covariance}
\label{subsubsec:extrapolation_cov}

\begin{figure}
    \centering  
    \includegraphics[width=9.5cm]{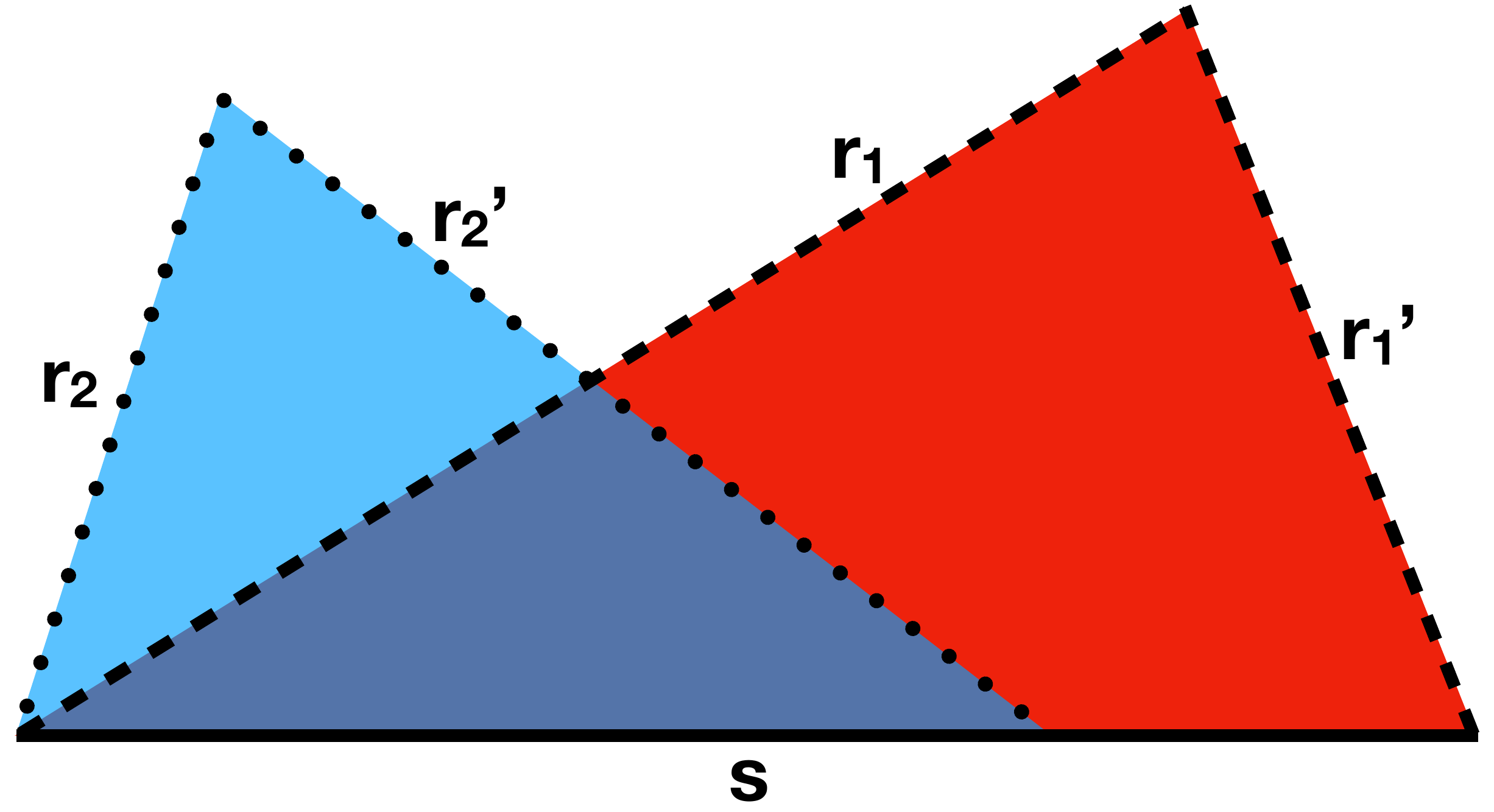}

    \caption{In \crefrange{fig:f_integral_order_zero}{fig:f_integral_s=130}, we showed that the $f$-integrals are near zero outside of the rectangular regions where $r_i$, $r^{\prime}_j$, and $s$ are able to form a closed triangle ($H(1-R_{+,ij})$ and $H(R_{-,ij}-1)$ terms of \pcref{eqn:f_FINAL}), and have the greatest contributions when $r_i$, $r^{\prime}_j$, and $s$ may form a closed triangle ($H(R_{+,ij}-1) H(1-R_{-,ij})$ term of \pcref{eqn:f_FINAL}). The covariance matrices of the 3PCF and 4PCF (\pcref{eqn:3pcf_cov_orig,eqn:c1_final,eqn:c2_orig}) are composed of products of $f$-integrals. This leads to sparse covariance matrices, with the greatest contributions from the intersection of the $f$-integrals' triangular regions. Here, we display this for the product $f_{\ell,
    \ell^{\prime},\ell^{\prime\prime}}(r_1,r^{\prime}_1,s)f_{\ell,
    \ell^{\prime},\ell^{\prime\prime}}(r_2,r^{\prime}_2,s)$. The red triangle (outlined by dashed and solid lines) represents the triangular region of $f_{\ell,
    \ell^{\prime},\ell^{\prime\prime}}(r_1,r^{\prime}_1,s)$. The blue triangle (outlined by dotted and solid lines) shows the triangular region of $f_{\ell,
    \ell^{\prime},\ell^{\prime\prime}}(r_2,r^{\prime}_2,s)$. Each of these regions will have nonzero contributions to the respective $f$-integrals. However, when they are multiplied together as needed for the covariance, the only significant contributions will arise from the overlapping triangular region. This overlapping region is shown in purple and bordered by a dashed line, dotted line, and solid line.}
    \label{fig:triagram}
\end{figure}

We recall that the covariance matrices of the 3- and 4PCF (\pcref{eqn:3pcf_cov_orig,eqn:c1_final,eqn:c2_orig}) contain integrals over $s$ of products of $f$-integrals. Thus, the trends present in the $f$-integrals shown in \crefrange{fig:f_integral_order_zero}{fig:f_integral_s=130} can be used to understand the covariance.

The regions bounded by the black dashed lines in \crefrange{fig:f_integral_order_zero}{fig:f_integral_s=130}, which correspond to the Heaviside functions in \cref{eqn:f_FINAL}, give insight into the configurations of $r_i$, $r^{\prime}_j$, and $s$ that control the covariance. The regions bounded on three sides by these Heaviside functions (\textit{i.\,e.}, the regions with the greatest contributions to the $f$-integrals) are where $|r_i-r_j^{\prime}| \leq s \leq r_i+r_j^{\prime}$ (dashed purple line in \cref{fig:heavisides}). Within these regions, the side lengths $r_i$ and $r^{\prime}_j$ are able to form a closed triangle with the separation $s$. We expect the greatest contributions to the covariance to stem from these regions of the $f$-integrals.

The upper left and lower right triangles in each subplot of \crefrange{fig:f_integral_order_zero}{fig:f_integral_s=130} represent the regions where $s<|r_i-r_j^{\prime}|$ (solid green line in \cref{fig:heavisides}). The lower left triangles in the subplots are given by $r_i+r_j^{\prime}<s$ (dotted orange line in \cref{fig:heavisides}). In both of these regions, a closed triangle may not be formed between $r_i$, $r^{\prime}_j$, and $s$. As one moves from the triangular region bounded by the Heaviside functions into these non-triangular regions, the $f$-integrals quickly become near-zero. The behavior of the $f$-integrals in triangular and non-triangular regions will be further explored in Moore and Slepian, in prep.

\cite{padmanabhan_sparse_precision} showed that precision matrices (inverse covariance matrices) are sparse; our analytic work here may help to understand that in the future. In \crefrange{fig:f_integral_order_zero}{fig:f_integral_s=130}, the $f$-integrals were shown to be near zero outside of the region where closed triangles between $r_i$, $r^{\prime}_j$, and $s$ can be formed, but have greater contributions where closed triangles are possible. The covariance matrices of the 3PCF and 4PCF (\pcref{eqn:3pcf_cov_orig,eqn:c1_final,eqn:c2_orig}) depend on products of $f$-integrals, each with different separations $r_i$ and $r^{\prime}_j$. This means that for each product, the $f$-integrals have different triangular regions (with greater contributions) and different non-triangular regions (with near-zero contributions). 

When the product of multiple $f$-integrals with different triangular regions is taken, the result will often be near zero. This is because although each $f$-integral will have significant contributions within its respective triangular region, these regions may not fully overlap; the product of a nonzero value (from one triangular region) with a near-zero value (from another non-triangular region) is near-zero. To achieve a significant nonzero value in the covariance, the $f$-integrals being multiplied need to have partially overlapping triangular regions. In these overlapping regions, the product of two nonzero values will provide a significant contribution to the covariance. As the non-overlapping triangular region is often larger than the overlapping triangular region for a product of $f$-integrals, the covariance matrix (and thus also the precision matrix) is sparse. This is depicted in \cref{fig:triagram}.

\subsubsection{Comparison of Model and True Covariance Matrices}
\label{subsubsec:comparison_of_cov}

\begin{figure}
    \centering
    \includegraphics[width=\textwidth]{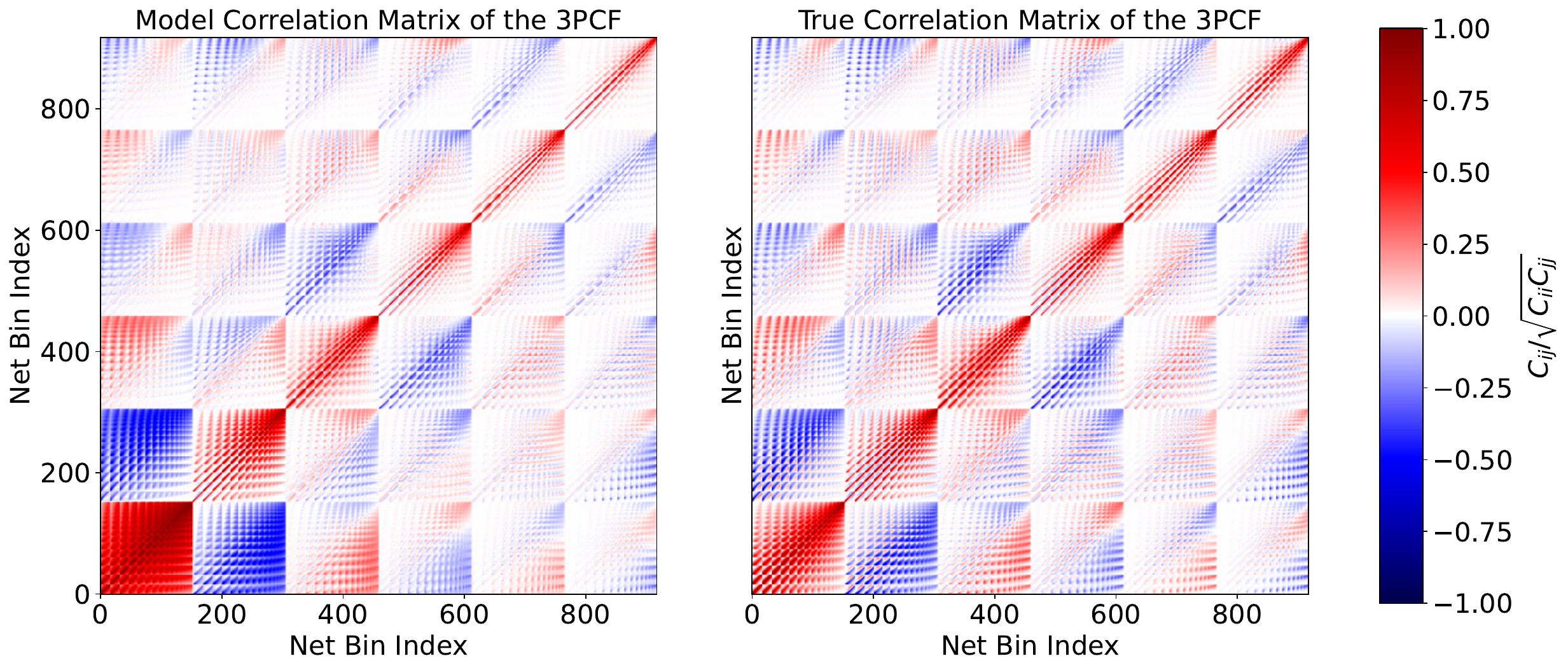}

    \caption{The \textit{left-hand panel} shows the correlation (normalized covariance) matrix of the 3PCF computed from our model power spectrum; the \textit{right-hand panel} is the correlation matrix from the \textsc{camb} power spectrum. Each index represents a unique combination of $\{r_1,r_2,\ell\}$, and similar for $\{r^{\prime}_1,r^{\prime}_2,\ell^{\prime}\}$. Within each sub-block, which corresponds to a unique $\{\ell,\ell^{\prime}\}$ channel, all possible combinations of both the unprimed and primed side lengths are displayed. The sub-blocks show how for each fixed channel, the covariance is affected by varying the side lengths. Although our model power spectrum leads to more nonzero elements in the covariance than expected, especially for lower values of $\{\ell,\ell^{\prime}\}$, it reproduces the expected structure.}
    \label{fig:3pcf_cov_fig}
\end{figure}

\begin{figure}
    \centering
    \includegraphics[width=10cm]{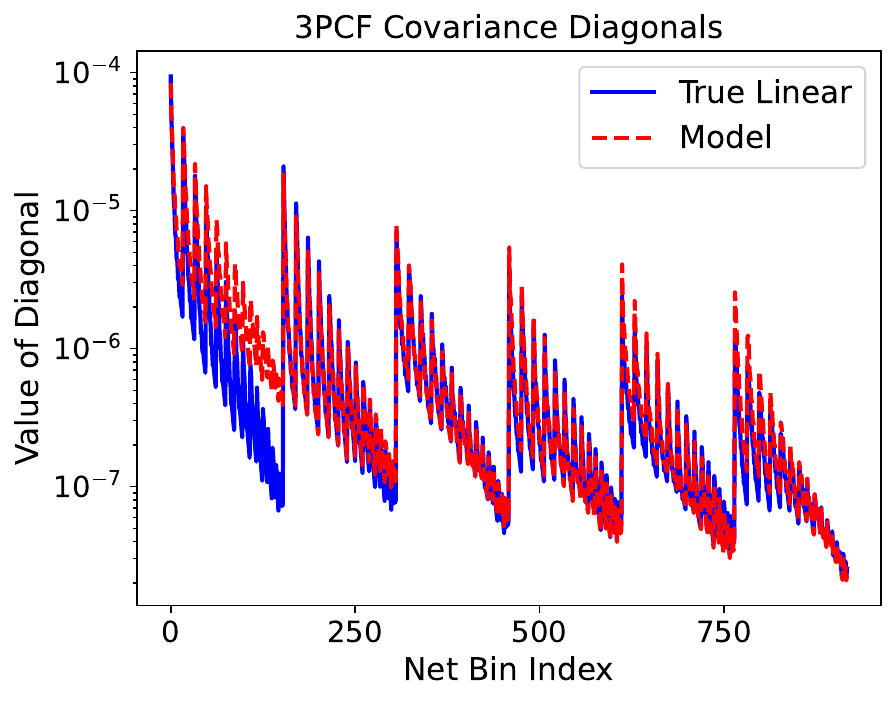}

    \caption{The diagonals of the model and true covariance matrices of the 3PCF (\cref{fig:3pcf_cov_fig}) are shown. The ratio of the diagonals, model/true, has a mean value of 1.43, median value of 1.06, and standard deviation of 0.981. We emphasize that the median shows good agreement between the diagonals. The mean and standard deviation are skewed by the first 153 indices, which correspond to $\{\ell,\ell^{\prime}\} = \{0,0\}$; over this region, the diagonals of the model covariance are greater than those of the true covariance.}
    \label{fig:3pcf_cov_diagonals}
\end{figure}

\begin{figure}
    \centering  
    \includegraphics[width=12cm]{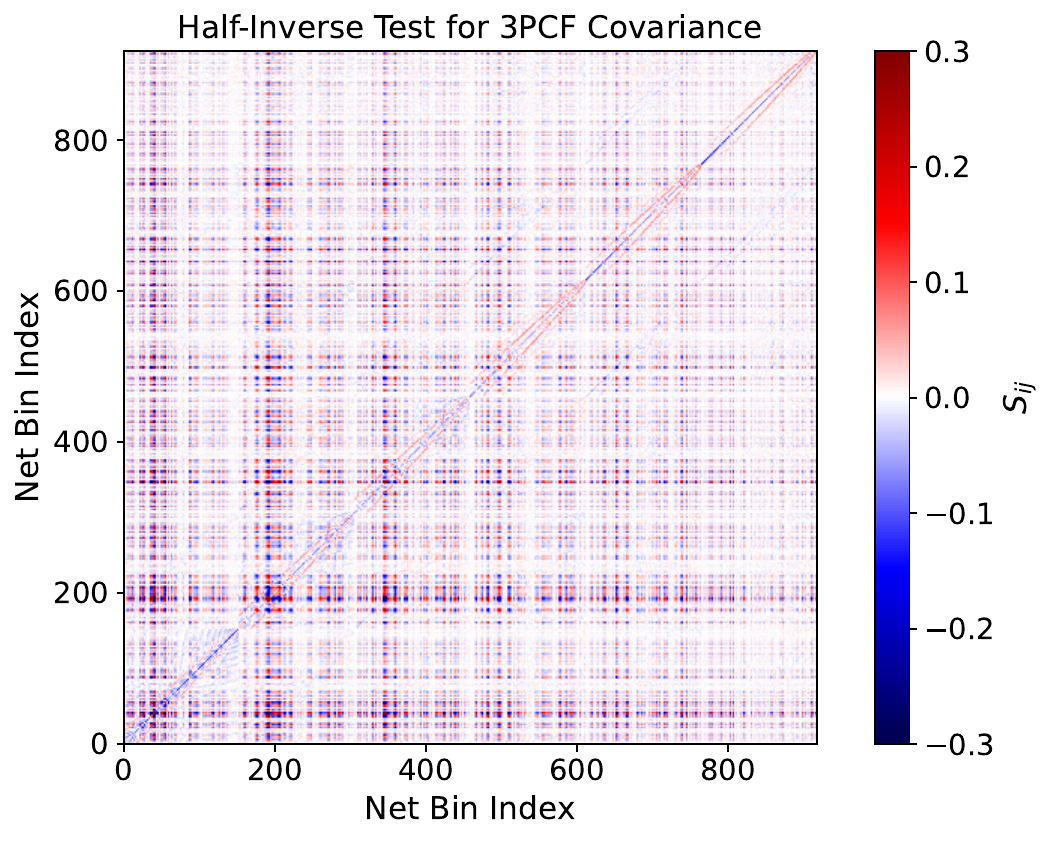}

    \caption{We display the half-inverse test (\pcref{eqn:half_inv_dup}) for the model and true covariance matrices of the 3PCF in \cref{fig:3pcf_cov_fig}. If the model covariance matrix exactly matched the true covariance matrix, the result of the half-inverse test would be the null matrix. The mean, median, and standard deviation of the half-inverse test are \numlist{-8.38e-6; 3.99e-5; 5.66e-2}.}
    \label{fig:3pcf_cov_halfinv_fig}
\end{figure}

\begin{table}
    \centering
    \begin{tabular}{c *{3}{S[table-parse-only]}}
        \toprule
        \textbf{Quantity} & \textbf{Mean} & \textbf{Median} & \textbf{SD}
        \\
        \midrule
        Model Covariance   & 4.07e-8 & 6.12e-12 & 5.37e-7 \\
        True Covariance    & 1.20e-8 & 2.80e-11 & 3.90e-7 \\
        Ratio of Diagonals & 1.43 & 1.06 & 0.981 \\
        Half-Inverse Test  & -8.38e-6 & 3.99e-5 & 5.66e-2 \\
        \bottomrule
    \end{tabular}

    \caption{We display the mean, median, and standard deviation values corresponding to the covariance matrices of the 3PCF. The model and true correlation (normalized covariance) matrices are shown in \cref{fig:3pcf_cov_fig}, the diagonals of each covariance matrix in \cref{fig:3pcf_cov_diagonals}, and the half-inverse test of the covariance in \cref{fig:3pcf_cov_halfinv_fig}. The model and true covariance matrices of the 3PCF are in good agreement.} 
    \label{Tab:3pcf_cov_table}
\end{table}

We now compare the full covariance of the 3PCF and 4PCF to those from our model power spectrum. We include linear bias with $b_1=2$, shot noise with $\bar{n} = \SI{3e-4}{\hHubble\cubed\per\Mpc\cubed}$, and set the survey volume to $\SI{2.5}{\per\hHubble\cubed\Gpc\cubed}$; these choices reflect a realistic survey such as BOSS. We compute the covariance matrices in 18 radial bins, ranging from \SIrange{20}{160}{\per\hHubble\Mpc}. The radial binning is weighted by $r^2$ in each $r$-bin, similar to equation A10 in \cite{connected}.

As shown in \cref{eqn:3pcf_cov_orig} and \cref{fig:3pcf_diagram}, the covariance of the 3PCF depends on $\{r_1,r_2,\ell\}$ and $\{r_1^{\prime},r_2^{\prime},\ell^{\prime}\}$. We allow $\{\ell,\ell^{\prime}\}$ to vary from 0 to 5. There are 918 unique combinations of $\{r_1,r_2,\ell\}$, each of which we map to a single unique index. We do the same for each unique combination of $\{r_1^{\prime},r_2^{\prime},\ell^{\prime}\}$.

\Cref{fig:3pcf_cov_fig} displays the correlation matrix (normalized covariance matrix) of the 3PCF. The left-hand panel is that computed from our model power spectrum while the right-hand panel uses the \textsc{camb} power spectrum. Each sub-block represents a unique $\{\ell,\ell^{\prime}\}$ channel. Within each sub-block, all possible binned combinations of $\{r_1,r_2,r^{\prime}_1,r^{\prime}_2\}$ are accounted for. The sub-blocks thus represent how the varying side lengths affect the covariance at a fixed channel. 

Our model power spectrum reproduces the true, sparse structure of the covariance of the 3PCF, which stems from the product of $f$-integrals within the covariance (as discussed in \cref{subsubsec:extrapolation_cov}). The covariance from our model power spectrum predicts more nonzero elements, particularly for low values of $\ell$ and $\ell^{\prime}$. This is consistent with the $f$-integrals displayed in \cref{subsec:f_plots_fixed_s,subsec:f_plots_diff_s}, particularly those in \cref{fig:f_integral_order_zero} (corresponding to the lower left sub-block of \cref{fig:3pcf_cov_fig}). In \cref{fig:3pcf_cov_diagonals}, we display the diagonals of the covariance matrices of the 3PCF. The ratio of the model diagonals to the true diagonals has a median value of 1.06.

To quantify the similarity between the model and true covariance matrices, we use a half-inverse test (\pcref{eqn:half_inv}, duplicated below):
\begin{align}
\label{eqn:half_inv_dup}
    \mathbf{S} &\equiv \mathbf{C}_{\mathrm{model}}^{-1/2}\mathbf{C}_{\mathrm{true}}\mathbf{C}_{\mathrm{model}}^{-1/2} - \mathbf{\mathds{1}}.
\end{align}
Here, $\mathbf{C}_{\mathrm{model}}$ is the covariance of the 3PCF from our model power spectrum, $\mathbf{C}_{\mathrm{true}}$ is the covariance of the 3PCF from the \textsc{camb} power spectrum, and $\mathbf{\mathds{1}}$ is the identity matrix. If the model and true covariance matrices match exactly, the result will be the null matrix.

We display the half-inverse test for the covariance matrices of the 3PCF in \cref{fig:3pcf_cov_halfinv_fig}. The mean of the half-inverse test, \num{-8.38e-6}, is two orders of magnitude greater than the mean values of the model and true covariance matrices; this is driven by the discrepancy between the model and true covariance matrices at low $\{\ell,\ell^{\prime}\}$, which is shown in \cref{fig:3pcf_cov_epsilon} and will be discussed in \cref{sec:inverse_cov}. To quantify the disparity between the null matrix and the half-inverse test, we measure the standard deviation, which is \num{5.66e-2}. This describes how the half-inverse test deviates from the null matrix, on average. We thus see that our model power spectrum allows us to achieve percent-level accuracy to the true covariance matrix of the 3PCF.

In \cref{Tab:3pcf_cov_table}, we provide a summary of the covariance of the 3PCF. This includes the model and true covariance matrices, diagonals of the covariance, and half-inverse test.

\begin{figure}
    \centering
    \includegraphics[width=\textwidth]{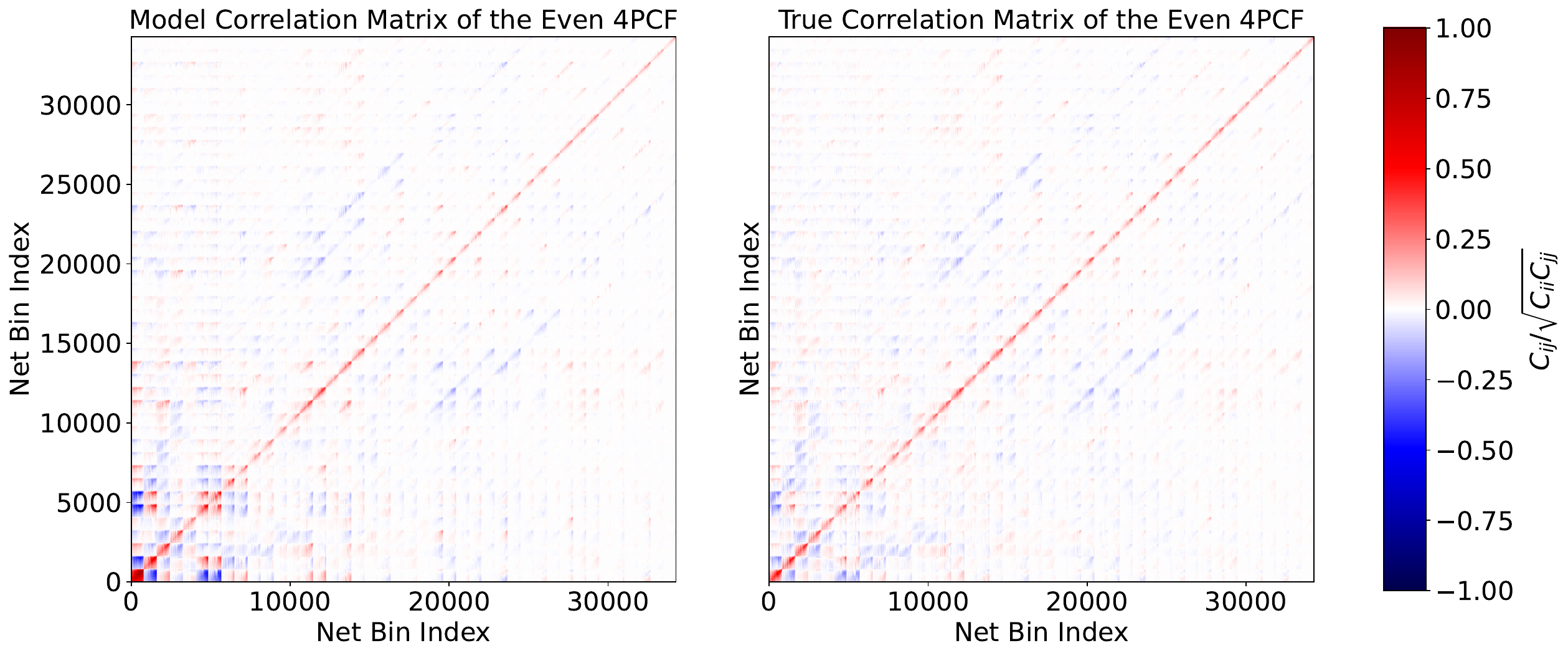}

    \caption{We display the correlation matrix of the even 4PCF. The \textit{left-hand panel} shows that computed from our model power spectrum, while the \textit{right-hand panel} uses the \textsc{camb} power spectrum. Each index denotes a unique combination of $\{r_1,r_2,r_3,\ell_1,\ell_2,\ell_3\}$, and similar for $\{r^{\prime}_1,r^{\prime}_2,r^{\prime}_3,\ell^{\prime}_1,\ell^{\prime}_2,\ell^{\prime}_3\}$. Here, we restrict the sums $\ell_1+\ell_2+\ell_3$ and $\ell_1^{\prime}+\ell_2^{\prime}+\ell_3^{\prime}$ to be even. The sub-blocks correspond to $\{\ell_i,\ell^{\prime}_j\}$ channels, and show how the covariance is affected as the side lengths vary. Similar to the correlation matrices of the 3PCF in \cref{fig:3pcf_cov_fig}, our model power spectrum overpredicts the covariance at low values of $\{\ell_i,\ell^{\prime}_j\}$. However, the structure of the covariance is accurately captured by our model, showing that the covariance of the 4PCF is sparse.}
    \label{fig:4pcf_corr_matrices_even}
%\end{figure}

%\begin{figure}
    \bigskip

    \centering  
    \includegraphics[width=\textwidth]{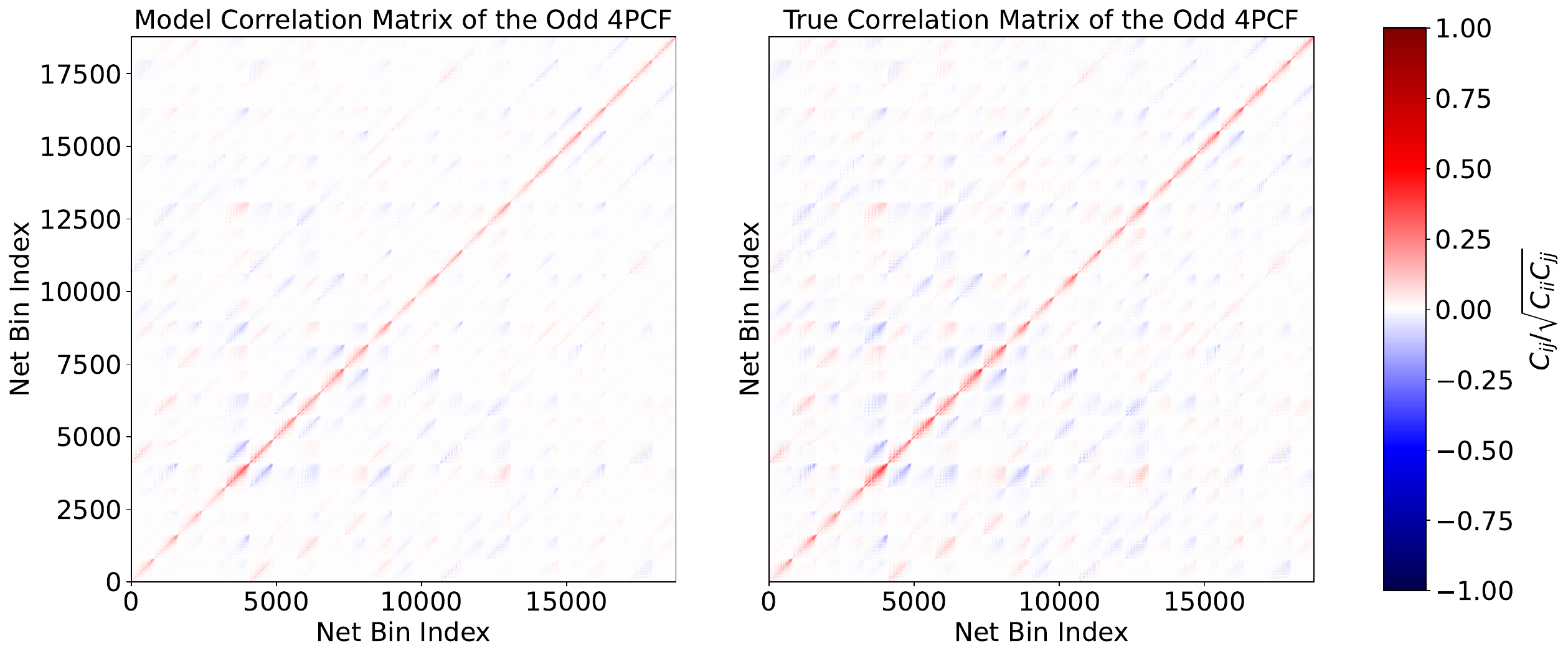}

    \caption{The odd modes of the correlation matrix of the 4PCF are shown. The \textit{left-hand panel} is computed from our model power spectrum and the \textit{right-hand panel} is from the \textsc{camb} power spectrum. Each index denotes a unique combination of $\{r_1,r_2,r_3,\ell_1,\ell_2,\ell_3\}$, and similar for $\{r^{\prime}_1,r^{\prime}_2,r^{\prime}_3,\ell^{\prime}_1,\ell^{\prime}_2,\ell^{\prime}_3\}$. Here, the sums $\ell_1+\ell_2+\ell_3$ and $\ell_1^{\prime}+\ell_2^{\prime}+\ell_3^{\prime}$ must be odd. Each sub-block represents a unique $\{\ell_i,\ell^{\prime}_j\}$ channel, showing how varying the side lengths affects the covariance. Our model power spectrum accurately reproduces the sparse structure of the covariance of the 4PCF.}
    \label{fig:4pcf_corr_matrices_odd}
\end{figure}

\Cref{eqn:c1_final,eqn:c2_orig,fig:4pcf_diagram} show that the covariance of the 4PCF depends on $\{r_1,r_2,r_3,\ell_1,\ell_2,\ell_3\}$ and $\{r^{\prime}_1,r^{\prime}_2,r^{\prime}_3,\ell^{\prime}_1,\ell^{\prime}_2,\ell^{\prime}_3\}$; we map each unique combination to an index. We bin the side lengths in 18 radial bins from \SIrange{20}{160}{\per\hHubble\Mpc}, and allow the $\{\ell_i,\ell^{\prime}_j\}$ to vary from 0 to 4. We split the 4PCF into even and odd modes. When the sum $\ell_1+\ell_2+\ell_3$ ($\ell_1^{\prime}+\ell_2^{\prime}+\ell_3^{\prime}$) is even, the 4PCF has even parity. When the sum is odd, the 4PCF is parity-odd. There are 34,272 net bin indices for the covariance of the even 4PCF, and 18,768 net bin indices for the covariance of the odd 4PCF. The 4PCF was computed using the public \textsc{Analytic4PC} code introduced in \cite{kendrick_no_parity}.

In \namecrefs{fig:4pcf_corr_matrices_even} \labelcref{fig:4pcf_corr_matrices_even} (even parity) and \labelcref{fig:4pcf_corr_matrices_odd} (odd parity), we display the correlation matrices of the 4PCF using the model power spectrum (left-hand panels) and the \textsc{camb} power spectrum (right-hand panels). Each sub-block is a unique $\{\ell_i,\ell^{\prime}_j\}$ channel, where all possible binned combinations of the side lengths are accounted for within each sub-block. This shows how the covariance depends on the side lengths at each fixed channel.

For both the even and odd modes, our model power spectrum accurately recovers the structure of the covariance of the 4PCF, showing that it is sparse. This sparsity stems from the product of $f$-integrals within the covariance, as was discussed in \cref{subsubsec:extrapolation_cov}. Similar to the correlation matrices of the 3PCF in \cref{fig:3pcf_cov_fig}, our model power spectrum overpredicts the correlation matrix of the even 4PCF at low values of $\{\ell_i,\ell^{\prime}_j\}$. This trend is not seen in the correlation matrix of the odd 4PCF.

\begin{figure}
    \centering
    \begin{subfigure}[b]{0.5\textwidth}
        \centering
        \includegraphics[width=\textwidth]{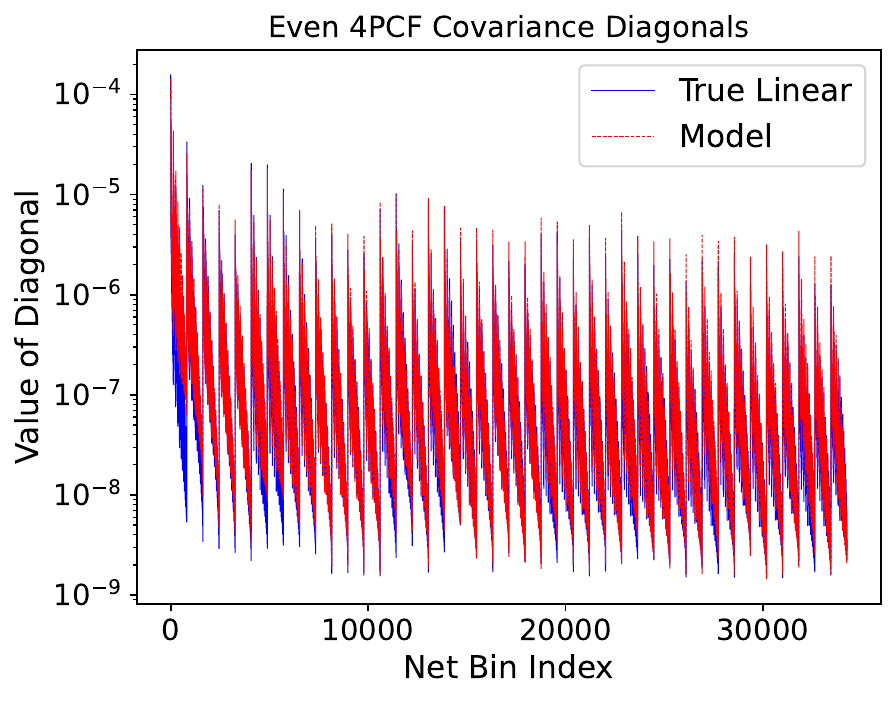}
    \end{subfigure}
    \hspace{-0.25cm}
    \begin{subfigure}[b]{0.5\textwidth}
        \centering
        \includegraphics[width=\textwidth]{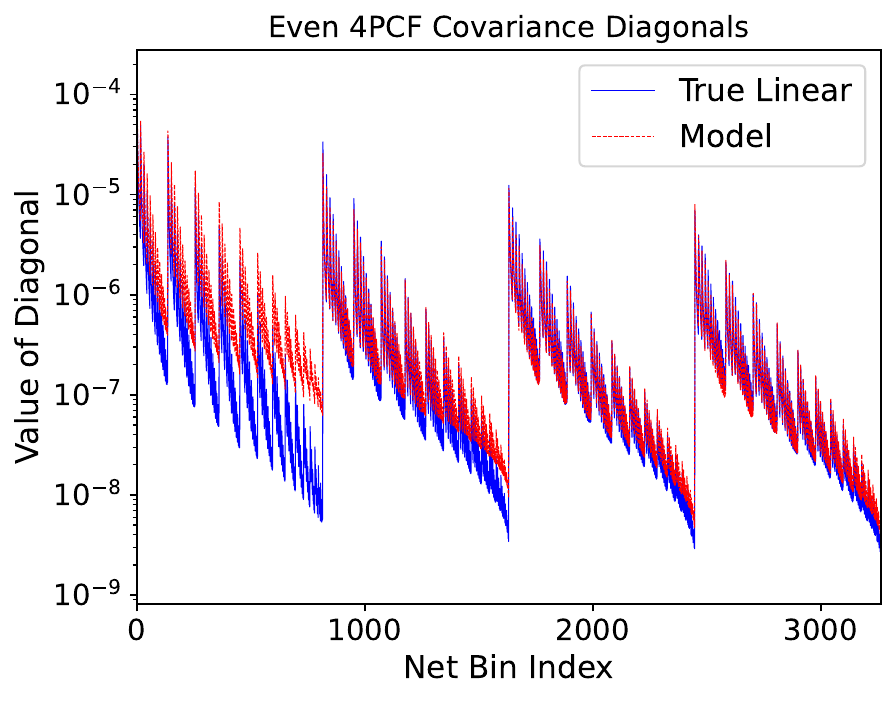}
    \end{subfigure}

    \caption{The diagonals of the model and true covariance matrices of the even 4PCF (\cref{fig:4pcf_corr_matrices_even}) are shown. The \textit{left-hand panel} displays all 34,272 net bin indices; the \textit{right-hand panel} is zoomed in to the first 3,264 indices. The ratio of the diagonals, model/true, has a mean value of 1.29. The median value of this ratio is 1.13 and the standard deviation is 0.787. We emphasize that the median shows good agreement between the model and true covariance diagonals. The standard deviation is skewed by the model overestimating the true diagonals at low net bin indices.}
    \label{fig:4pcf_cov_diagonals_even}
%\end{figure}

%\begin{figure}
    \bigskip

    \centering
    \begin{subfigure}[b]{0.5\textwidth}
        \centering
        \includegraphics[width=\textwidth]{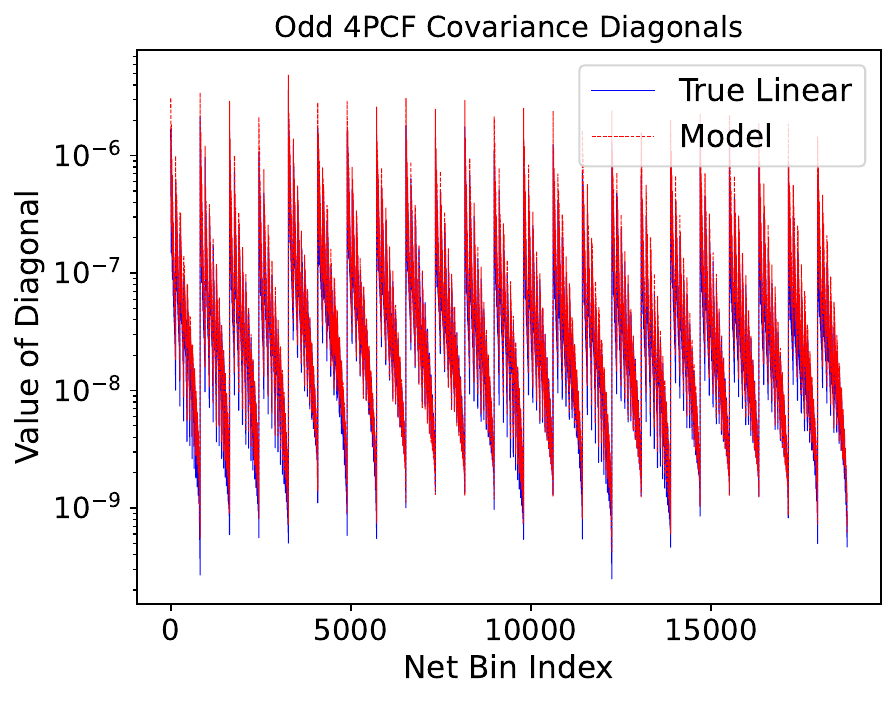}
    \end{subfigure}
    \hspace{-0.25cm}
    \begin{subfigure}[b]{0.5\textwidth}
        \centering
        \includegraphics[width=\textwidth]{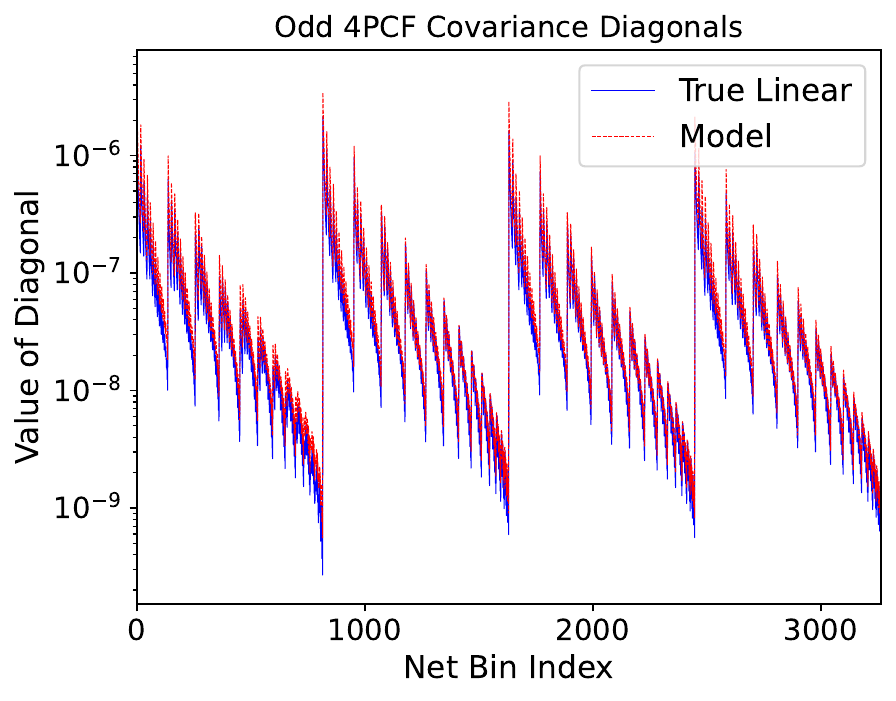}
    \end{subfigure}

    \caption{We display the diagonals of the model and true covariance matrices for the odd 4PCF (\cref{fig:4pcf_corr_matrices_odd}). All 18,768 net bin indices are shown in the \textit{left-hand panel}. In the \textit{right-hand panel}, only the first 3,264 indices are displayed. The ratio of the diagonals, model/true, has a mean value of 1.22, median value of 1.17, and standard deviation 0.222.}
    \label{fig:4pcf_cov_diagonals_odd}
\end{figure}

The diagonals of the covariance matrices of the 4PCF are shown in \namecrefs{fig:4pcf_cov_diagonals_even} \labelcref{fig:4pcf_cov_diagonals_even} (even parity) and \labelcref{fig:4pcf_cov_diagonals_odd} (odd parity). The left-hand panel of each of these figures shows the diagonals at all net bin indices; for clarity, the right-hand panels display the diagonals at the first 3,264 indices. On average, the model covariance has diagonals that are $1.29\times$ greater than the diagonals of the true covariance for even parity, and $1.22\times$ greater for odd parity. Both of these cases show better agreement between the model and true diagonals than was seen for the diagonals of the covariance of the 3PCF (\cref{fig:3pcf_cov_diagonals}).

\begin{figure}
    \centering  
    \includegraphics[width=11cm]{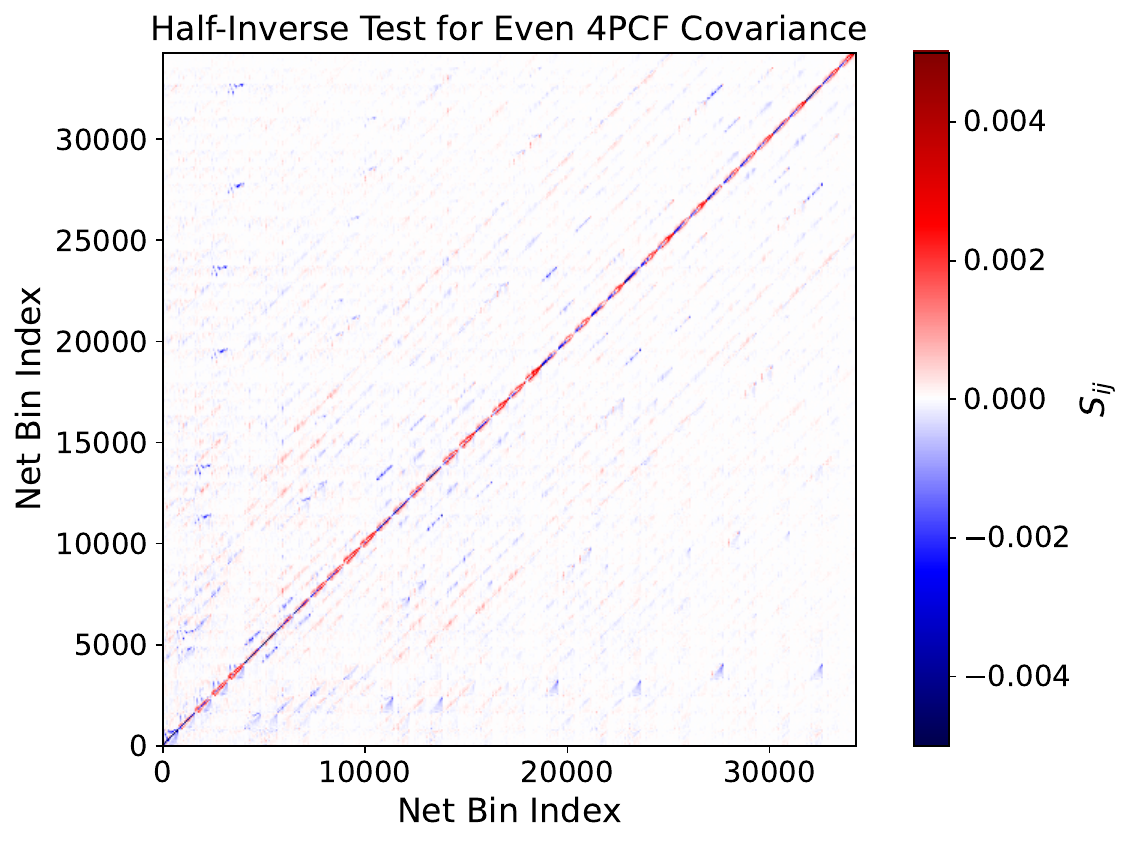}

    \caption{Here, we show the half-inverse test (\pcref{eqn:half_inv_dup}) for the model and true covariance matrices of the even 4PCF. In the case where the model covariance exactly matches the true covariance, the result of the half-inverse test would be the null matrix. Here, the half-inverse test has a mean of \num{-9.42e-7}, median of \num{-6.22e-8}, and standard deviation of \num{2.75e-3}.}
    \label{fig:4pcf_halfinv_even}
%\end{figure}

%\begin{figure}
    \bigskip

    \centering  
    \includegraphics[width=11cm]{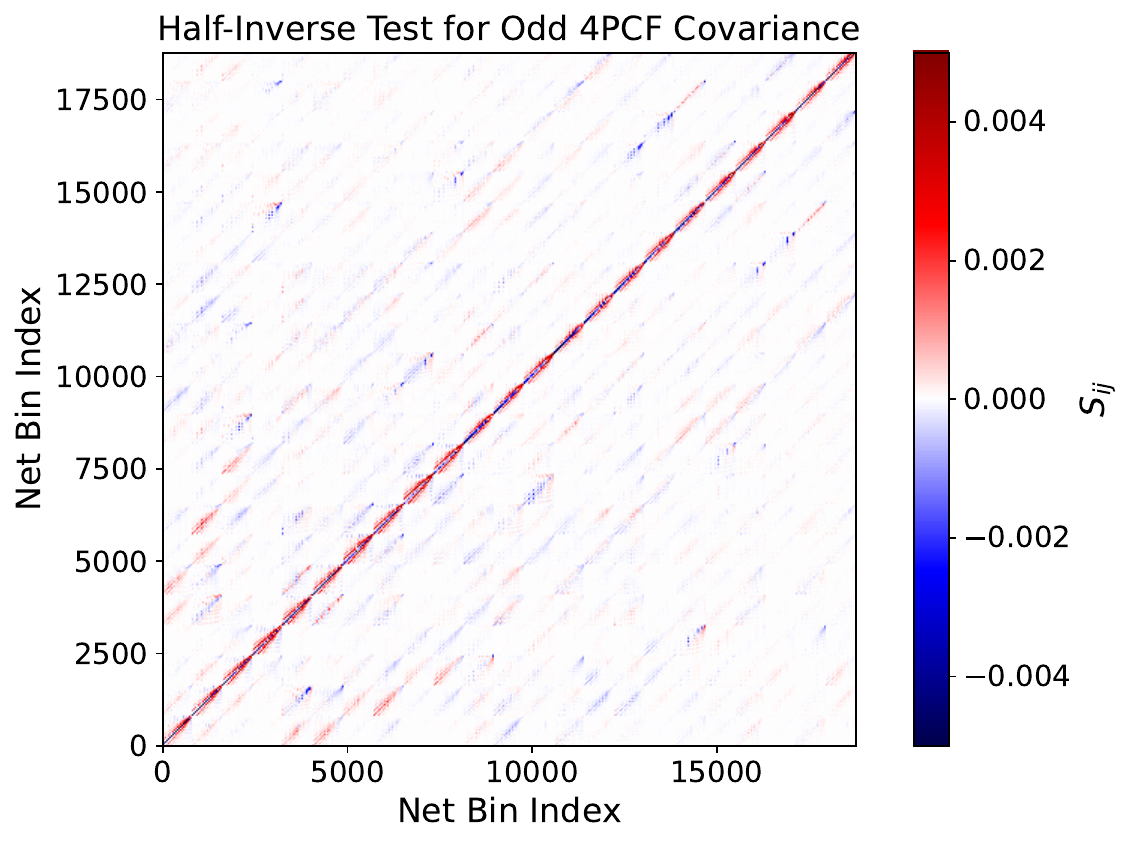}

    \caption{We display the half-inverse test (\pcref{eqn:half_inv_dup}) for the model and true covariance matrices of the odd 4PCF. The result of the half-inverse would be the null matrix if the model covariance exactly matched the true covariance. The mean, median, and standard deviation of the half-inverse test are \numlist{6.81e-6; 1.18e-7; 3.78e-3}.}
    \label{fig:4pcf_halfinv_odd}
\end{figure}

\begin{table}
    \centering
    \begin{tabular}{c *{3}{S[table-parse-only]}}
        \toprule
        \textbf{Quantity} & \textbf{Mean} & \textbf{Median} & \textbf{SD}
        \\
        \midrule
        Model Covariance   & 2.94e-10 & 2.21e-12 & 3.56e-8 \\
        True Covariance    & 1.23e-10 & 1.77e-12 & 2.34e-8 \\
        Ratio of Diagonals & 1.29 & 1.13 & 0.787 \\
        Half-Inverse Test  & -9.42e-7 & -6.22e-8 & 2.75e-3 \\
        \bottomrule
    \end{tabular}
    \caption{The covariance of the even 4PCF is summarized here. The correlation matrices, diagonals of the covariance, and half-inverse test of the covariance are shown in \cref{fig:4pcf_corr_matrices_even,fig:4pcf_cov_diagonals_even,fig:4pcf_halfinv_even}, respectively. There is better agreement between the model and true covariance matrices for the even 4PCF than for the 3PCF (summarized in \cref{Tab:3pcf_cov_table}).}
    \label{Tab:4pcf_even_cov_table}
\end{table}

\begin{table}
    \centering
    \begin{tabular}{c *{3}{S[table-parse-only]}}
        \toprule
        \textbf{Quantity} & \textbf{Mean} & \textbf{Median} & \textbf{SD}
        \\
        \midrule
        Model Covariance  & 4.19e-11 & -7.79e-15 & 2.89e-9 \\
        True Covariance   & 4.46e-11 & -1.57e-14 & 2.79e-9 \\
        Ratio of Diagonals & 1.22 & 1.17 & 0.222 \\
        Half-Inverse Test & 6.81e-6 & 1.18e-7 & 3.78e-3 \\
        \bottomrule
    \end{tabular}
    \caption{We summarize the covariance of the odd 4PCF here. \Cref{fig:4pcf_corr_matrices_odd,fig:4pcf_cov_diagonals_odd,fig:4pcf_halfinv_odd} display the correlation matrices of the odd 4PCF, diagonals of the covariance, and half-inverse test of the covariance, respectively. Overall, there is better agreement between the model and true covariance for the odd 4PCF than for the even 4PCF (\cref{Tab:4pcf_even_cov_table}) or 3PCF (\cref{Tab:3pcf_cov_table}).}
    \label{Tab:4pcf_odd_cov_table}
\end{table}

We display the half-inverse test (\pcref{eqn:half_inv_dup}) for the covariance matrix of the even 4PCF in \cref{fig:4pcf_halfinv_even}. The mean of the half-inverse test for even parity is \num{-9.42e-7}, which is three orders of magnitude greater than the mean values of the model and true covariance matrices. Similarly to the half-inverse test of the 3PCF, this is influenced by the mismatch between the model and true covariance matrices at low $\{\ell_i,\ell^{\prime}_j\}$, which is shown in \cref{fig:4pcf_cov_epsilon_even} and will be discussed in \cref{sec:inverse_cov}. The standard deviation of the half-inverse test, which measures the deviation from the null matrix, on average, is \num{2.75e-3}. This confirms that our model power spectrum accurately reproduces the covariance of the even 4PCF.

The half-inverse test (\pcref{eqn:half_inv_dup}) for the covariance matrix of the odd 4PCF is shown in \cref{fig:4pcf_halfinv_odd}. The mean value, \num{6.81e-6}, is five orders of magnitude greater than the mean values of the model and true covariance matrices. As will be shown in \cref{fig:4pcf_cov_epsilon_odd} and discussed in \cref{sec:inverse_cov}, this is driven by the discrepancies between the model and true covariance matrices. These discrepancies are not localized to low values of $\{\ell_i,\ell^{\prime}_j\}$, as they were for the covariance of the 3PCF and even 4PCF, but are distributed throughout the covariance, and are prominent on the main diagonal. The standard deviation of the half-inverse test is \num{3.78e-3}, which is of the same order as the standard deviation of the half-inverse test for the even 4PCF. We thus conclude that the model power spectrum used accurately reproduces the structure present within the covariance of the odd 4PCF.

We summarize the model and true covariance matrices of the even 4PCF, the ratio of their diagonals, and the half-inverse test in \cref{Tab:4pcf_even_cov_table}. The same is displayed for the covariance of the odd 4PCF in \cref{Tab:4pcf_odd_cov_table}.

\subsection{Inverse Covariance}
\label{sec:inverse_cov}
For cosmological analyses with many degrees of freedom, the covariance matrix is notably difficult to invert. This is because it is typically large and, if using mocks to compute it, there are often not enough mocks available.

The analytic formulae for covariance matrices derived throughout this work are, however, invertible. We suggest an efficient hybrid scheme where we combine these formulae with a correction matrix $\boldsymbol{\epsilon}$ that accounts for the differences between the analytic covariance and the true covariance, allowing accurate determination of the inverse covariance.

We decompose the true covariance, $\textbf{C}$, into the covariance from the (easily invertible) formulae given in this work, denoted as $\textbf{C}_{\mathrm{model}}$, and a correction matrix $\boldsymbol{\epsilon}$:
\begin{align}
\label{eqn:C_decomp}
    \textbf{C} &= \textbf{C}_{\mathrm{model}} + \boldsymbol{\epsilon}.
\end{align}

The inverse covariance is then given by the following matrix inversion lemma, as long as $\textbf{C}_{\mathrm{model}}$ and $\boldsymbol{\epsilon}$ are both nonsingular matrices with the same dimensions and $\boldsymbol{\epsilon}$ has rank one \cite{inverse_matrix}:
\begin{align}
\label{eqn:matrix_inverse_lemma}
    \textbf{C}^{-1} &= (\textbf{C}_{\mathrm{model}} + \boldsymbol{\epsilon})^{-1} \nonumber \\
    &= \textbf{C}_{\mathrm{model}}^{-1} - 1/\left[1+\mathrm{tr}(\boldsymbol{\epsilon}\textbf{C}_{\mathrm{model}}^{-1})\right]\textbf{C}_{\mathrm{model}}^{-1}\boldsymbol{\epsilon}\textbf{C}_{\mathrm{model}}^{-1}.
\end{align}
Here, ``$\mathrm{tr}$'' indicates the trace. Since $\textbf{C}_{\mathrm{model}} + \boldsymbol{\epsilon}$ is nonsingular, $\mathrm{tr}(\boldsymbol{\epsilon}\textbf{C}_{\mathrm{model}}^{-1}) \neq -1$ \cite{inverse_matrix}.

In the case where $\boldsymbol{\epsilon}$ does not have rank one, $\boldsymbol{\epsilon}$ may first be decomposed into a sum of matrices with rank one before the covariance is inverted. This is discussed in \cite{inverse_matrix}.

\begin{figure}
    \centering
    \begin{subfigure}[b]{0.51\textwidth}
        \centering
        \includegraphics[width=\textwidth]{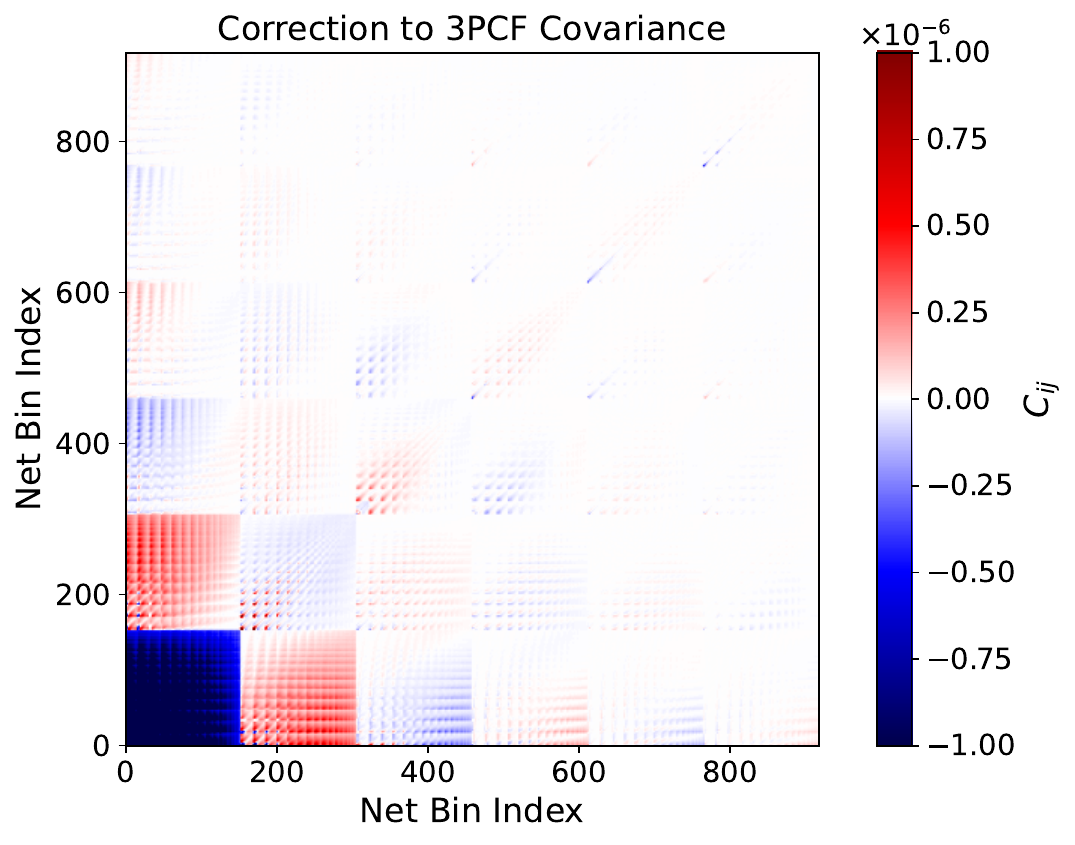}
    \end{subfigure}
    \hspace{-0.25cm}
    \begin{subfigure}[b]{0.49\textwidth}
        \centering
        \includegraphics[width=\textwidth]{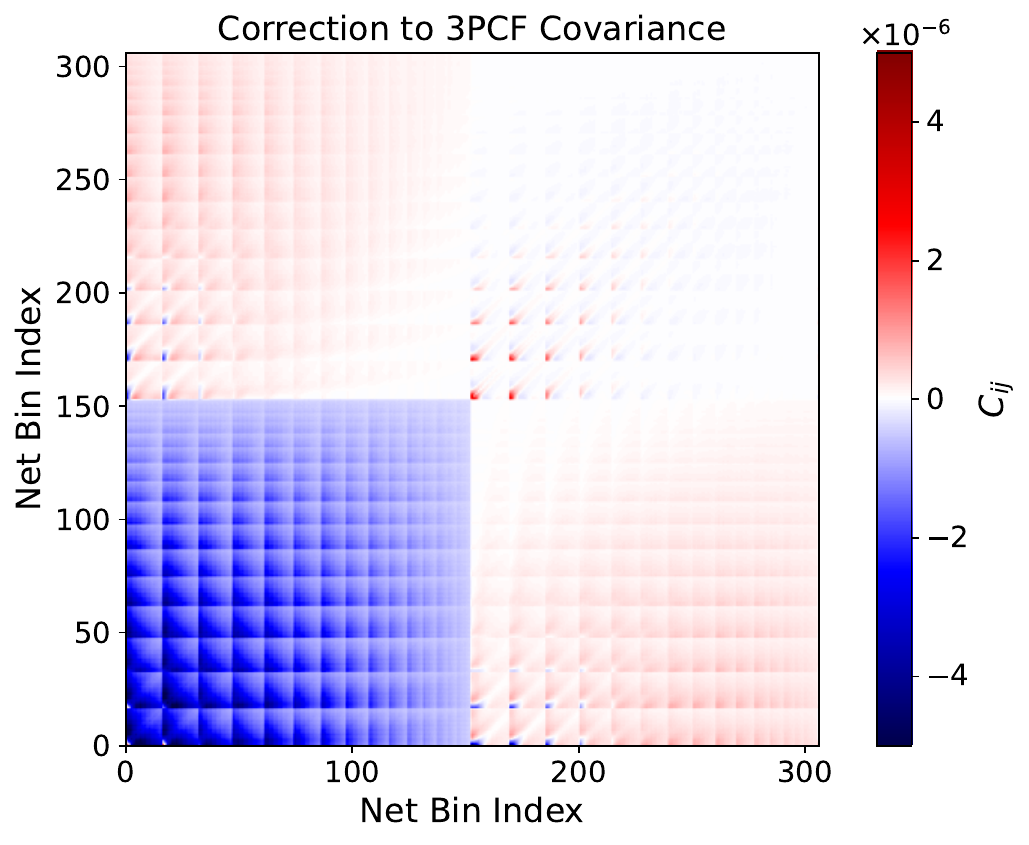}
    \end{subfigure}

    \caption{We display the correction matrix (\pcref{eqn:correction_matrix_for_plots}) for the covariance of the 3PCF (\cref{fig:3pcf_cov_fig}). The \textit{left-hand panel} shows the full correction matrix; the \textit{right-hand panel} displays the first 306 net bin indices, which provide the greatest contributions to the correction matrix. The red regions indicate where the model underpredicts the true covariance, while the blue regions show where the model overpredicts the covariance. The greatest discrepancy between the model and true covariance matrices occurs for low values of $\{\ell,\ell^{\prime}\}$ (blue region in the lower left corner of each panel). The mean of the full correction matrix (all 918 indices) is \num{-2.87e-8}, the median is \num{1.05e-11} and the standard deviation is \num{2.62e-7}.}
    \label{fig:3pcf_cov_epsilon}
\end{figure}

\begin{figure}
    \centering
    \begin{subfigure}[b]{0.505\textwidth}
        \centering
        \includegraphics[width=\textwidth]{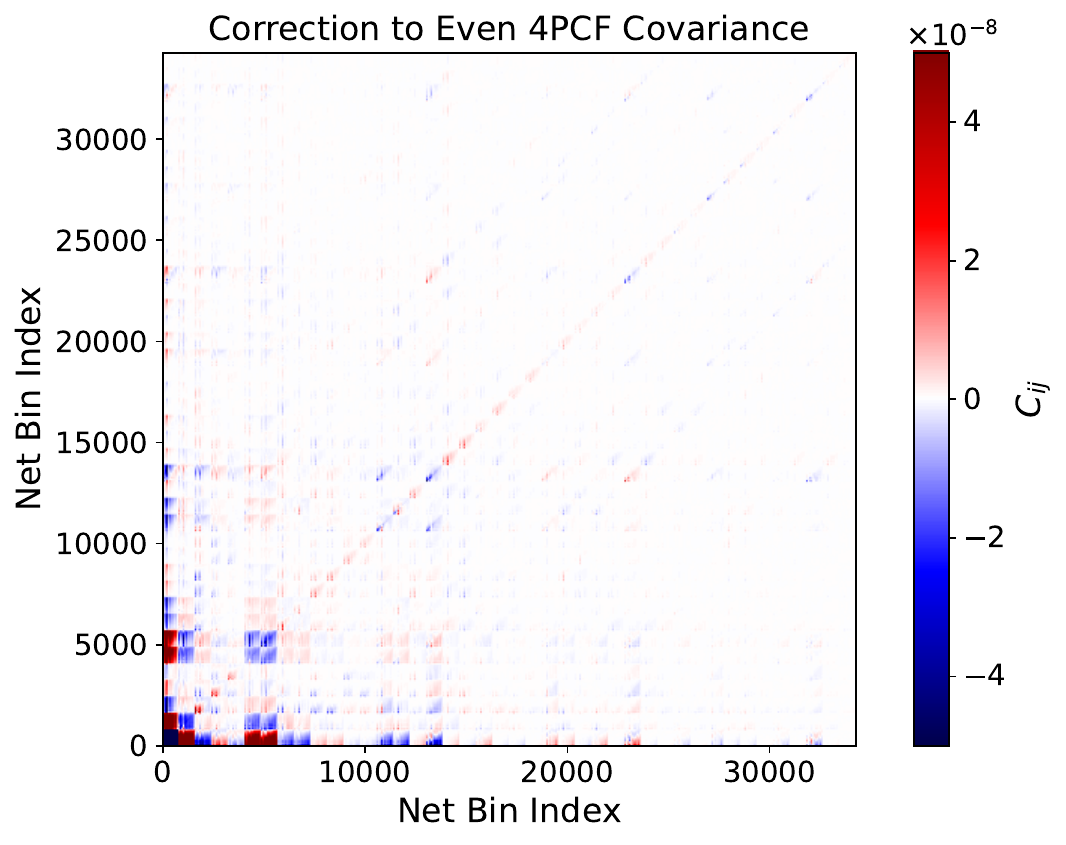}
    \end{subfigure}
    \hspace{-0.25cm}
    \begin{subfigure}[b]{0.495\textwidth}
        \centering
        \includegraphics[width=\textwidth]{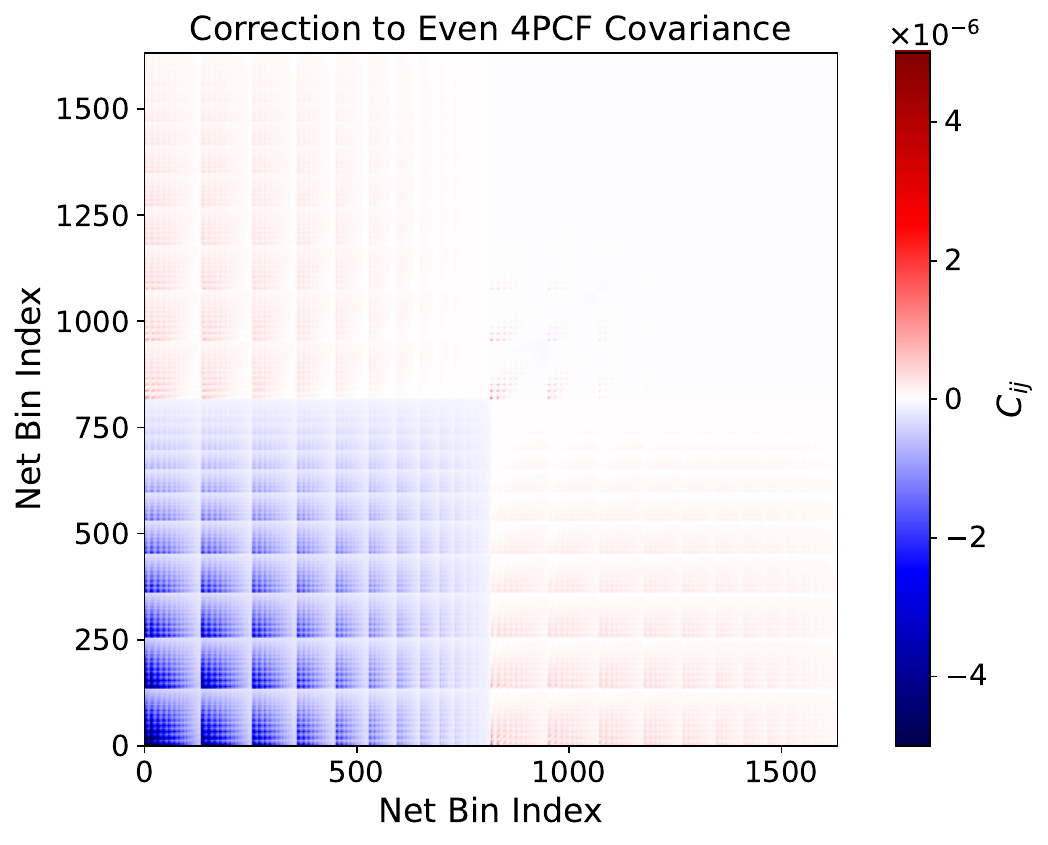}
    \end{subfigure}

    \caption{We display the correction matrix (\pcref{eqn:correction_matrix_for_plots}) to the covariance of the even 4PCF (\cref{fig:4pcf_corr_matrices_even}). The \textit{left-hand panel} shows the full correction matrix; the \textit{right-hand panel} zooms in on the first 1,632 net bin indices. These first 1,632 indices, which show low values of $\{\ell_i,\ell^{\prime}_j\}$, provide the greatest contributions to the correction matrix. In each panel, the blue region shows where the model overpredicts the covariance; the red regions are where the covariance is underpredicted by the model. The mean of the full correction matrix for the 4PCF with even parity (including all 34,272 even indices) is \num{-1.72e-10}, the median is \num{-2.21e-13}, and the standard deviation is \num{1.95e-8}.}
    \label{fig:4pcf_cov_epsilon_even}
\end{figure}

\begin{figure}
    \centering  
    \includegraphics[width=10cm]{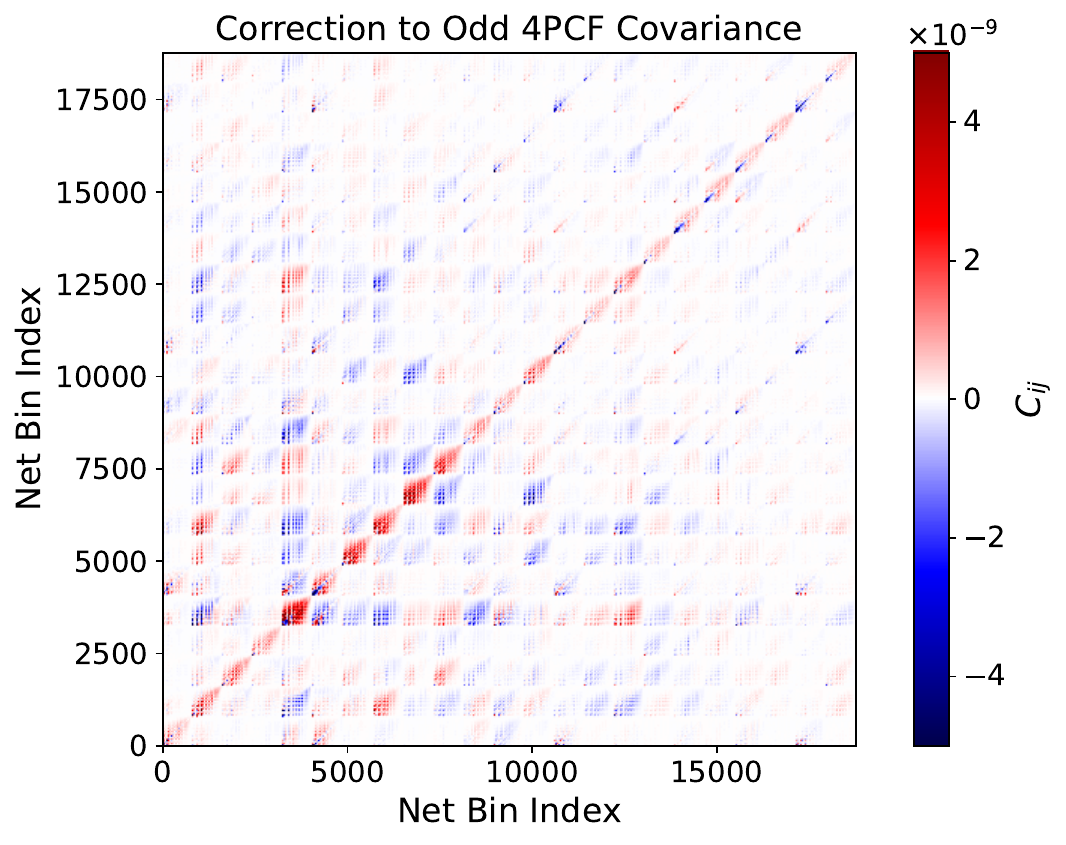}

    \caption{Since the correction matrix (\pcref{eqn:correction_matrix_for_plots}) to the covariance of the odd 4PCF (\cref{fig:4pcf_corr_matrices_odd}) is not dominated by contributions at low values of $\{\ell_i,\ell^{\prime}_j\}$, we display all 18,768 odd net bin indices. Areas where the model overpredicts the covariance are shown in blue. Red regions show where the model underpredicts the covariance. The mean of the odd correction matrix is \num{2.76e-12}, the median is \num{-5.56e-15}, and the standard deviation is \num{8.55e-10}.}
    \label{fig:4pcf_cov_epsilon_odd}
\end{figure}

We now discuss the correction matrix $\boldsymbol{\epsilon}$. To determine $\boldsymbol{\epsilon}$, we may compare our analytic formulae to the true covariance, which, for example, may be determined from mock catalogs. Here, we find $\boldsymbol{\epsilon}$ by taking the difference between the covariance matrices from the model and \textsc{camb} power spectra:
\begin{align}
    \boldsymbol{\epsilon} = \textbf{C}_{\textsc{camb}} - \textbf{C}_{\mathrm{model}}.
\label{eqn:correction_matrix_for_plots}
\end{align}
The correction matrices of the 3PCF, even 4PCF, and odd 4PCF are shown in \cref{fig:3pcf_cov_epsilon,fig:4pcf_cov_epsilon_even,fig:4pcf_cov_epsilon_odd}, respectively. These correction matrices are summarized in \cref{Tab:correction_matrices}.

\begin{table}
    \centering
    \begin{tabular}{c *{3}{S[table-format=+1.2e+2]}}
        \toprule
        \textbf{Correction matrix} & \textbf{Mean} & \textbf{Median} & \textbf{SD}
        \\
        \midrule
        3PCF Cov. Correction        & -2.87e-8 & 1.05e-11 & 2.62e-7 \\
        4PCF Cov. Correction (Even) & -1.72e-10 & -2.21e-13 & 1.95e-8 \\
        4PCF Cov. Correction (Odd)  & 2.76e-12 & -5.56e-15 & 8.55e-10 \\
        \bottomrule
    \end{tabular}
    \caption{We display the mean, median, and standard deviation of the correction matrices for the covariance of the 3PCF (\cref{fig:3pcf_cov_epsilon}), even 4PCF (\cref{fig:4pcf_cov_epsilon_even}), and odd 4PCF (\cref{fig:4pcf_cov_epsilon_odd}). Our model covariance matrices accurately reproduce all three of the true covariance matrices, leading to small correction matrices. Our correction matrices for the covariance of the 4PCF are smaller than that for the 3PCF. Additionally, the correction matrix for the covariance of the odd 4PCF is smaller than that of the even 4PCF.} 
    \label{Tab:correction_matrices}
\end{table}

\Cref{fig:3pcf_cov_epsilon} displays the correction matrix for the covariance of the 3PCF. The red regions indicate where the model underpredicts the covariance, while the blue regions show where the model overpredicts the covariance. The left-hand panel shows all 918 net bin indices. In the right-hand panel, we display the first 306 indices as the rest of the correction matrix is near zero, and the first 306 indices provide the greatest contributions to the correction matrix. The mean value of the full correction matrix (including all 918 indices) is \num{-2.87e-8}; this is of the same order as the mean values of the model and true covariance matrices.

In \cref{fig:4pcf_cov_epsilon_even,fig:4pcf_cov_epsilon_odd}, we show the correction matrix for the covariance of the even and odd modes of the 4PCF, respectively. Again, the red regions show where the model underpredicts the covariance, and the blue areas indicate where the model overpredicts the covariance. 

The covariance of the even 4PCF has 34,272 net bin indices, as shown in \cref{fig:4pcf_corr_matrices_even}. However, many of the elements in the correction matrix for even parity are near zero. We thus display the full correction matrix in the left-hand panel of \cref{fig:4pcf_cov_epsilon_even}, and the first 1,632 net bin indices in the right-hand panel. These 1,632 indices provide the greatest contributions to the correction matrix. The mean of the full even correction matrix (including all 34,272 indices) is \num{-1.72e-10}, which is of the same order as the mean values of the model and true covariance matrices for the even 4PCF.

For the correction matrix to the covariance of the odd 4PCF (\cref{fig:4pcf_cov_epsilon_odd}), we include all 18,768 net bin indices. The odd correction matrix is not dominated by contributions at low values of $\{\ell_i,\ell^{\prime}_j\}$, as the correction matrices to the 3PCF or even 4PCF are. The mean of the odd correction matrix is \num{2.76e-12}, which is an order of magnitude below the mean values of the model and true covariance matrices for the odd 4PCF.

The correction matrices for the covariance of both the even and odd 4PCF are smaller than that for the covariance of the 3PCF. This may be due to the covariance of the 4PCF (\pcref{eqn:c1_final,eqn:c2_orig}) having more products of $f$-integrals than the covariance of the 3PCF (\pcref{eqn:3pcf_cov_orig}). As seen in \crefrange{fig:f_integral_order_zero}{fig:f_integral_s=130}, the $f$-integrals composing the covariance are near-zero outside of the regions where the side lengths may form closed triangles. Thus, when more $f$-integrals are multiplied together, the overlapping triangular regions shrink (\cref{fig:triagram}). This leads to more near-zero elements in the covariance, which are accurately predicted by our model. 

Additionally, the correction matrix for the covariance of the odd 4PCF is smaller than that for the even 4PCF. From \cref{Tab:4pcf_even_cov_table,Tab:4pcf_odd_cov_table}, we see that the model and true covariance matrices of the odd 4PCF have mean values an order of magnitude smaller than those of the even 4PCF, and median values two to three orders of magnitude smaller than those of the even 4PCF. This may be indicative that the covariance of the odd 4PCF has more near-zero values than the covariance of the even 4PCF. In this case, since our model accurately predicts the near-zero values outside of the triangular regions (as seen in \crefrange{fig:f_integral_order_zero}{fig:f_integral_s=130}), our model would more accurately capture the covariance of the odd 4PCF than the even 4PCF. We caution that future work is needed to further investigate the root of this phenomenon.

\subsection{Computational Complexity}
\label{sec:computational_complexity}

The integrals required to compute covariance matrices consist of products of sBFs, which are highly oscillatory and computationally expensive. By choosing a power spectrum model that allowed analytic evaluation of such sBF integrals, we have bypassed the need to directly compute them. Although the primary intent of this work was to obtain a tractable covariance model, not to enable faster computation, it is worth exploring the computational complexity of our analytic covariance formulae.

We have found that the covariance of the 2PCF (\pcref{eqn:2PCF_FINAL}) depends on factors of the product or quotient of the pair separations ($g = \sqrt{rr^{\prime}}$ and $\chi=r/r^{\prime}$, respectively), the inverse hyperbolic tangent, and the Dirac $\delta$. The argument of the inverse hyperbolic tangent is $\chi$, while the Dirac $\delta$ argument is the difference between pair separations. As the Dirac $\delta$ is simply evaluated at a single point, it will not increase the computational complexity.

A direct integration of the covariance of the 2PCF (\pcref{eqn:2pcf_cov_orig}) would have computational complexity $N_kN_R$, assuming $N_k$ points in $k$ and $N_R$ combinations of $\{r,r^{\prime}\}$, where $r^{\prime} \geq r$. To use our analytic result (\pcref{eqn:2PCF_FINAL}), we may set up a vector consisting of all possible combinations of $\{r,r^{\prime}\}$ (with $r^{\prime} \geq r$), which scales as $N_R$. However, there are different combinations of $r$ and $r^{\prime}$ that will map to the same value of $\chi$, and a similar argument holds for $g$. By mapping the $\{r,r^{\prime}\}$ combinations to $\chi$ and $g$, then evaluating \cref{eqn:2PCF_FINAL} at the unique values of them, our analytic covariance formula scales as $N_R$, at most.

The covariance matrices of the 3- and 4PCF required evaluation of two sBF integrals: $f_{\ell,0,\ell}(r_i,0,s)$ (\pcref{eqn:f_int_zero_eval}) and $f_{\ell,\ell^{\prime},\ell^{\prime\prime}}(r_i,r^{\prime}_j,s)$ (\pcref{eqn:f_FINAL}). Both $f_{\ell,0,\ell}(r_i,0,s)$ and $f_{\ell,\ell^{\prime},\ell^{\prime\prime}}(r_i,r^{\prime}_j,s)$ depend on factors of the side lengths and separation between primary vertices, $\Gamma$-functions, and either Gauss $\twoFone$ or $\threeFtwo$ hypergeometric functions. $f_{\ell,0,\ell}(r_i,0,s)$ also relies on a Dirac $\delta$; additionally, $f_{\ell,\ell^{\prime},\ell^{\prime\prime}}(r_i,r^{\prime}_j,s)$ requires computation of Wigner 3-$j$ and 6-$j$ symbols, binomial coefficients, and Meijer $G$-functions. The Wigner symbols, binomial coefficients, and $\Gamma$-functions are determined by the orders of the sBFs. The arguments of all other functions listed above are combinations of the side lengths and/or separation between primary vertices in the shapes of which we are seeking the covariance.

The functions dependent on the orders of the sBFs ($\Gamma$-functions, Wigner 3-$j$ and 6-$j$ symbols, and binomial coefficients) are straightforward to compute. Below, we describe the complexity of computing functions that depend on the side lengths and distances between galaxies. All functions may be evaluated with \textsc{Python} libraries such as \textsc{SciPy} \cite{scipy}, \textsc{SymPy} \cite{sympy}, or \textsc{mpmath} \cite{mpmath}. 

The direct integration for $f_{\ell,0,\ell}(r_i,0,s)$ (\pcref{eqn:f_int_zero_unequal_orders}) scales as $N_kN_{r_i}N_s$, assuming $N_k$ points in $k$, $N_{r_i}$ points in $r_i$, and $N_s$ points in $s$. In practice, the separations $r_i$ are binned, and the direct integration scales as $N_kN_{\mathrm{bins}}N_s$, where $k$ and $s$ are sampled on finer grids than the radial bins. The analytic evaluation of this integral (\pcref{eqn:f_int_zero_eval}) depends on Gauss hypergeometric functions with arguments $(r_i/s)^2$ or $(s/r_i)^2$, factors of the product or quotient of $r_i$ and $s$, and a Dirac $\delta$ with argument $r_i-s$. Since the Dirac $\delta$ is evaluated at a single point, it adds no computational complexity. The analytic result for $f_{\ell,0,\ell}(r_i,0,s)$ will be dominated by the Gauss hypergeometric functions.

We may map each combination of $r_i$ and $s$ to $\alpha \equiv r_i/s$, where $\alpha^2$ or its inverse are the arguments of the hypergeometric functions. Thus, by creating a one-dimensional grid of $\alpha$ values, we may precompute the hypergeometric functions for each unique $\alpha$, which is more efficient than a direct integration. A naive interpretation of $\alpha$ would scale at most as $N_{r_i}N_s$; however, since various $r_i$ and $s$ combinations map to the same $\alpha$, the scaling will actually be less than this. When binning the separations $r_i$, the computational complexity becomes $N_{\mathrm{bins}}N_s$. Since $s$ is sampled on a finer grid than the $r_i$ bins, the complexity of $f_{\ell,0,\ell}(r_i,0,s)$ is dominated by $N_s$.

Finally, for $f_{\ell,\ell^{\prime},\ell^{\prime\prime}}(r_i,r^{\prime}_j,s)$ (\pcref{eqn:f_int}), a direct integration has computational complexity $N_kN_{r_i}N_{r^{\prime}_j}N_s$. $\{N_k,N_{r_i},N_{r^{\prime}_j},N_s\}$ are, respectively, the number of points being sampled for $\{k,r_i,r^{\prime}_j,s\}$. However, $r_i$ and $r^{\prime}_j$ will be evaluated over the same set of radial bins, so the scaling becomes $N_kN_{\mathrm{bins}}^2N_s$. This is dominated by $N_k$ and $N_s$, since $k$ and $s$ are sampled on finer grids than the radial bins.

The analytic result for $f_{\ell,\ell^{\prime},\ell^{\prime\prime}}(r_i,r^{\prime}_j,s)$ (\pcref{eqn:f_FINAL}) depends on various factors of $r_i$, $r^{\prime}_j$, and $s$, but the computational complexity will be controlled by $\threeFtwo$ hypergeometric functions and Meijer $G$-functions. These special functions both have arguments $\pm R_{\pm,ij}^2 = \pm \big(|r_i \pm r^{\prime}_j|/s\big)^2$. To evaluate them, we may define a three-dimensional space by $r_i$, $r^{\prime}_j$, and $s$. Each constant value of $|r_i \pm r^{\prime}_j|/s$ then defines a manifold in this space, which maps to $\pm R_{\pm,ij}^2$. The special functions can be evaluated for each unique $\pm R_{\pm,ij}^2$, which is more efficient than a direct integration. Since different combinations of $r_i$, $r^{\prime}_j$, and $s$ will map to the same $\pm R_{\pm,ij}^2$, and we sample the $r_i$ and $r^{\prime}_j$ over the same set of radial bins, our analytic result will scale as $N_{\mathrm{bins}}N_s$, at most. This is controlled by $N_s$, as $s$ is sampled on a finer grid than the radial bins.

\section{Conclusion}
\label{sec:conclusion}

As more spectroscopic surveys come online, it is imperative to have a clearer idea of how to handle the inversion of covariance matrices for NPCF analyses with many degrees of freedom. In this work, we used a power law model of the linear matter power spectrum to obtain a closed-form result for the covariance of the 2PCF in terms of physical quantities. We also used this power spectrum model to evaluate the fundamental building blocks ($f$-integrals) of the covariance of the 3- and 4PCF in the GRF approximation, which captures the leading-order covariance. These $f$-integrals may be split into triangular and non-triangular configurations, leading to sparse covariance matrices. The results obtained here present a clearer picture of the structure of the covariance, which is useful in guiding future NPCF analyses with data from spectroscopic surveys, especially in cases where there are too many degrees of freedom and too few mock catalogs to smoothly invert the covariance. 

In \cref{sec:pk}, we showed that our power law model for the power spectrum is a good representation of the true power spectrum, and can be used to accurately determine the covariance of the 2PCF. We then used this power law model to find a closed-form solution for the covariance of the 2PCF in terms of physically relevant quantities such as the geometric mean ($g$) and ratio ($\chi$) of pair separations, number of galaxies within a sphere with radius set by $g$ ($N_{\mathrm{g}}$), total number of galaxies in the sample ($N$), and the scale of cosmic variance-shot noise equality ($\eta$). This was displayed in \cref{sec:2pcf_structure}.

The covariance of higher-order NPCFs depends on integrals of the power spectrum with three sBFs, which we define as $f$-integrals (\pcref{eqn:f_int}).  We used our model power spectrum to obtain closed-form results for these $f$-integrals throughout \cref{sec:f_int}, with the final results given in \cref{eqn:f_int_zero_eval,eqn:f_FINAL}. These closed-form solutions are easily separated into different regions (\cref{fig:heavisides}), depending on whether the side lengths of the shapes ($r_i$ and $r^{\prime}_j$) within the covariance are able to form a closed triangle with the separation $s$ between the shapes.

In \cref{sec:sparse_matrices}, we compared the $f$-integrals obtained with our model power spectrum to those obtained from the \textsc{camb} power spectrum (\crefrange{fig:f_integral_order_zero}{fig:f_integral_s=130}) and found that our model accurately captures the structure of the $f$-integrals (\crefrange{fig:Difference_f_integral}{fig:Difference_f_integral_Diff_s}). For each $f$-integral, the region in which $r_i$, $r^{\prime}_j$, and $s$ may form a closed triangle has greater contributions than the non-triangular regions, where the contributions drop close to zero. Since the covariance matrices of the 3PCF and 4PCF depend on products of $f$-integrals, each with different triangular and non-triangular regions, the covariance will be sparse. This was depicted in \cref{fig:triagram}, where we showed that the most significant nonzero contributions to the covariance arise when the triangular regions of multiple $f$-integrals overlap. As the covariance of the 4PCF depends on more products of $f$-integrals than the covariance of the 3PCF does, the covariance of the 4PCF will be more sparse. 

In \cref{subsubsec:comparison_of_cov}, we compared the covariance matrices of the 3PCF, even 4PCF, and odd 4PCF from our model power spectrum to those computed with the \textsc{camb} power spectrum, showing that our model accurately captures the sparse structure of the true covariance. In all cases, we showed good agreement between the diagonals of the covariance matrices. For the half-inverse tests of the even and odd 4PCF covariance, the greatest deviations from the null matrix were on the main diagonals.

The closed-form results for the covariance of the 2PCF and the $f$-integrals are invertible. As such, we may decompose the true covariance as the sum of our analytic expressions and a correction matrix. In doing so, we may easily evaluate the inverse covariance, avoiding the problem of having too few mock catalogs for smooth inversion. In \cref{sec:inverse_cov}, we outlined this procedure and determined the correction matrices. We showed that with our model, the correction terms needed for the covariance of the 3PCF and 4PCF have many near-zero elements.

Finally, although not the primary objective of this work, our closed-form results allow for more efficient computation of the covariance (\cref{sec:computational_complexity}). We have bypassed the need to compute highly oscillatory sBF integrals by choosing a power spectrum model that allows for analytic evaluation. The closed-form expressions found may be radially binned, as is typically done in practice for the covariance matrix. This efficient computation, combined with the decomposition of the covariance discussed in \cref{sec:inverse_cov}, allows for easier inversion of the covariance.

\appendix

\section{Truncated Power Spectrum}
\label{app_trunc_pk}
\begin{figure}
    \centering
    \vspace{0cm}
    \begin{subfigure}[b]{0.5\textwidth}
        \centering
        \includegraphics[width=\textwidth]{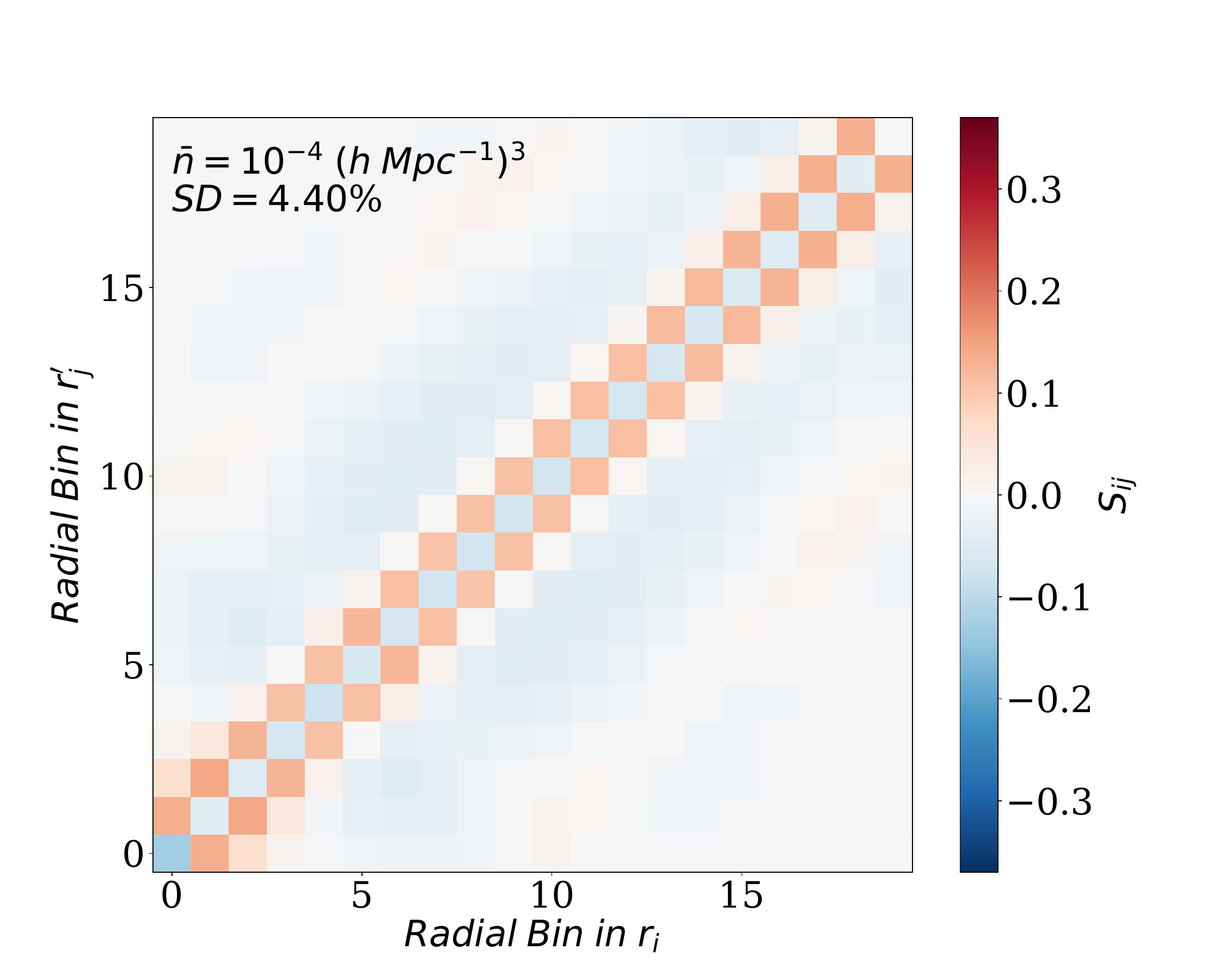}
    \end{subfigure}
    \hspace{-0.25cm}
    \begin{subfigure}[b]{0.5\textwidth}
        \centering
        \includegraphics[width=\textwidth]{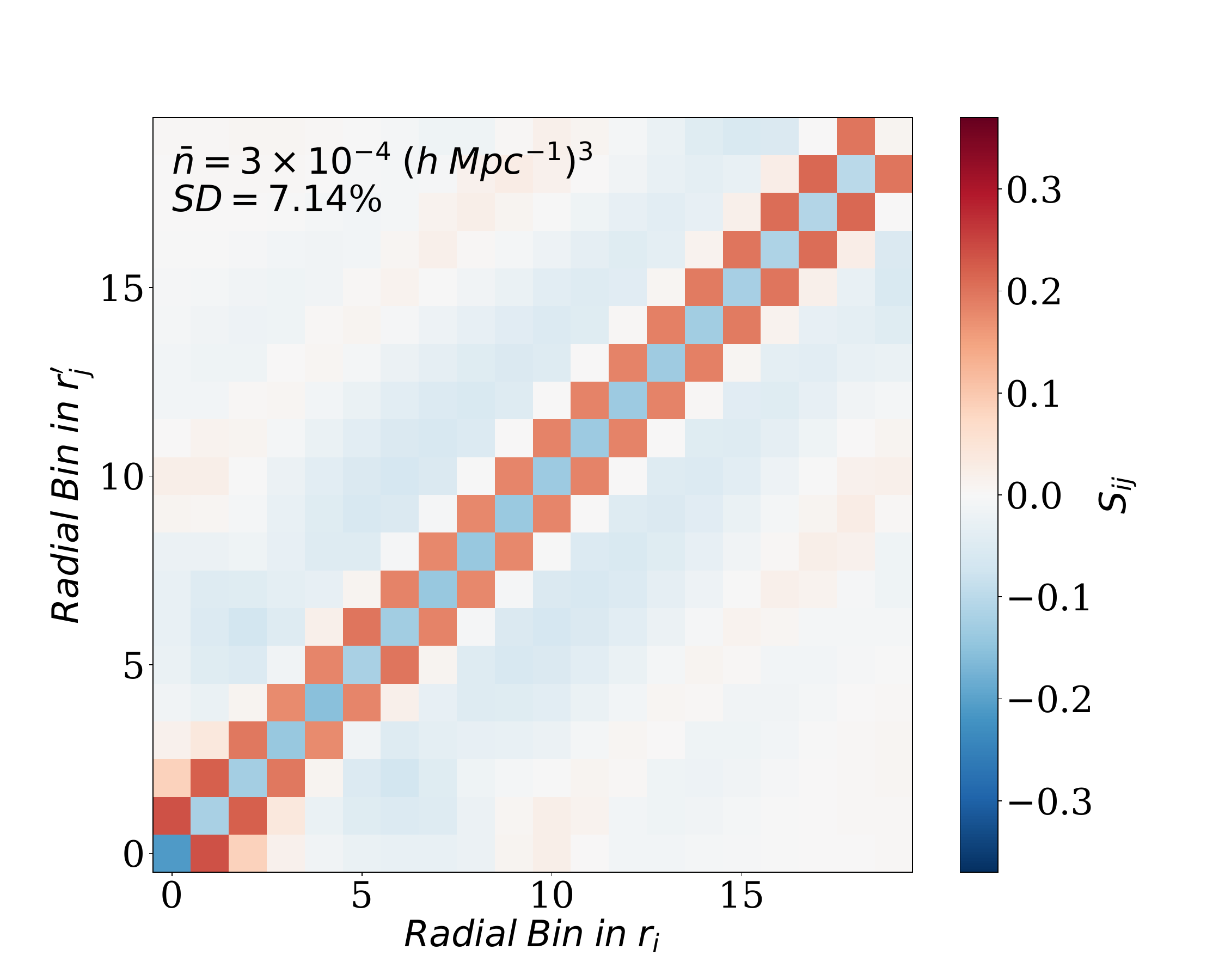}
    \end{subfigure}

    \vspace*{-1.2ex}

    \begin{subfigure}[b]{0.5\textwidth}
        \centering
        \includegraphics[width=\textwidth]{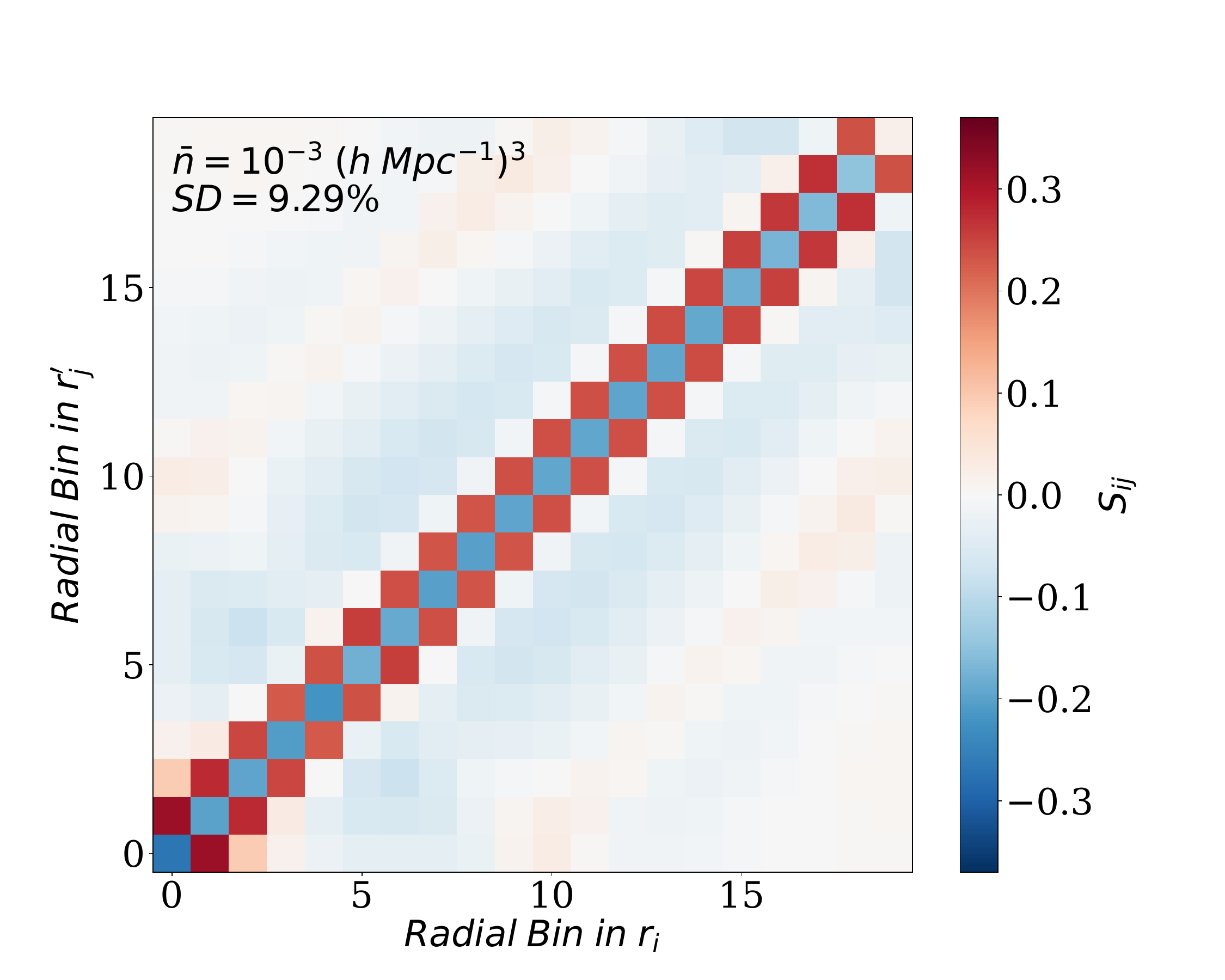}
    \end{subfigure}
    \hspace{-0.25cm}
    \begin{subfigure}[b]{0.5\textwidth}
        \centering
        \includegraphics[width=\textwidth]{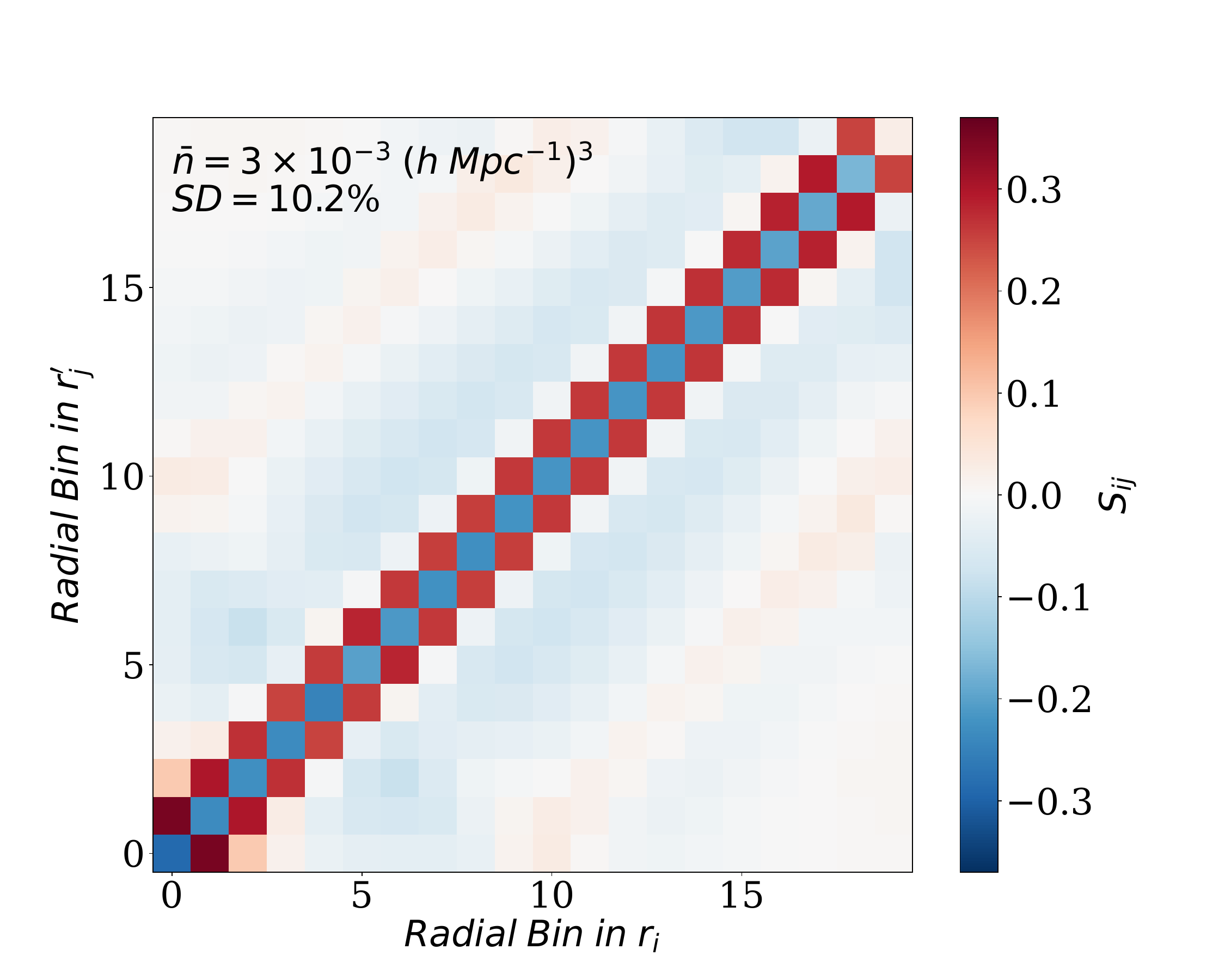}
    \end{subfigure}

    \vspace*{-1.2ex}

    \begin{subfigure}[b]{0.5\textwidth}
        \centering
        \includegraphics[width=\textwidth]{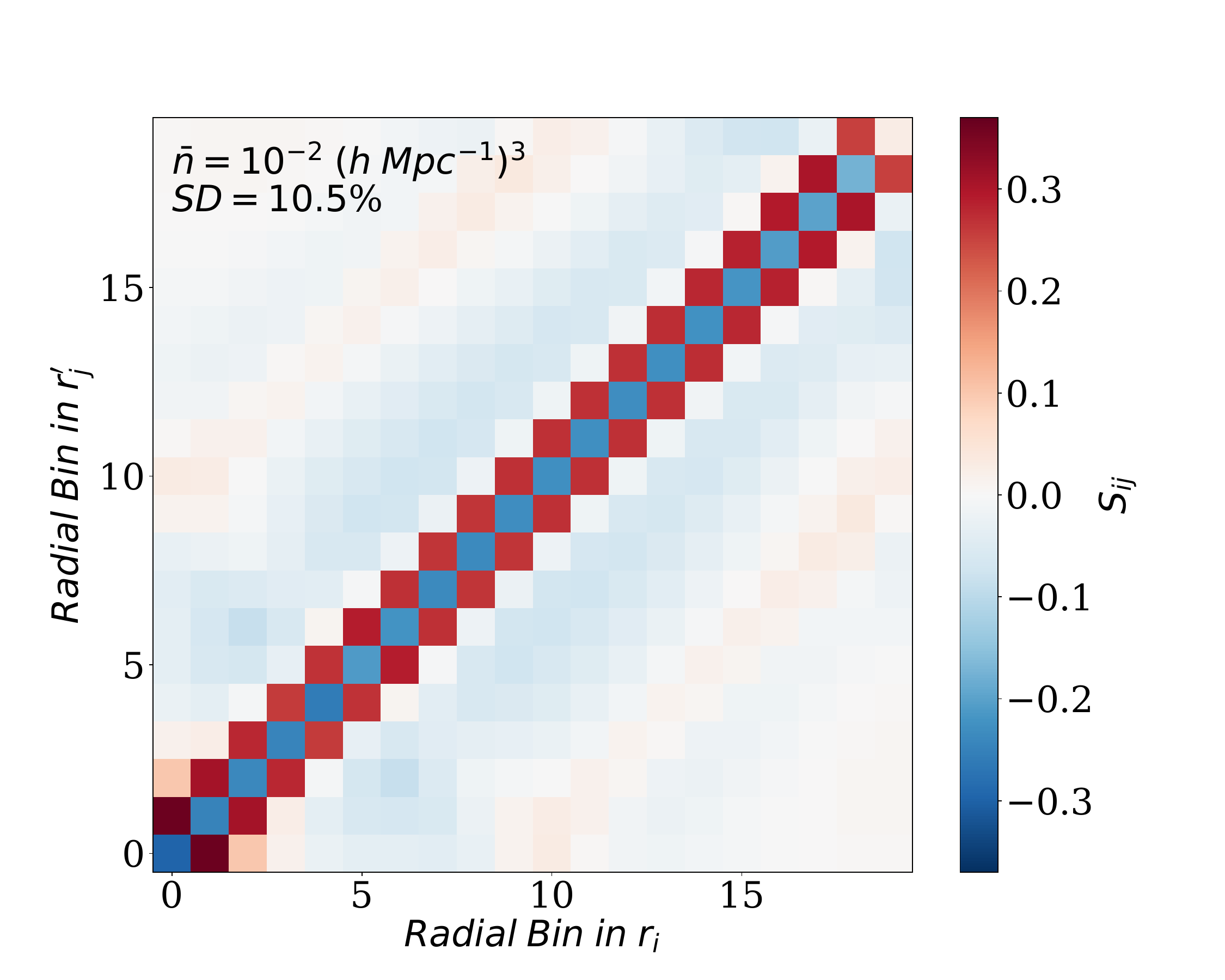}
    \end{subfigure}
    \hspace{-0.25cm}
    \begin{subfigure}[b]{0.5\textwidth}
        \centering
        \includegraphics[width=\textwidth]{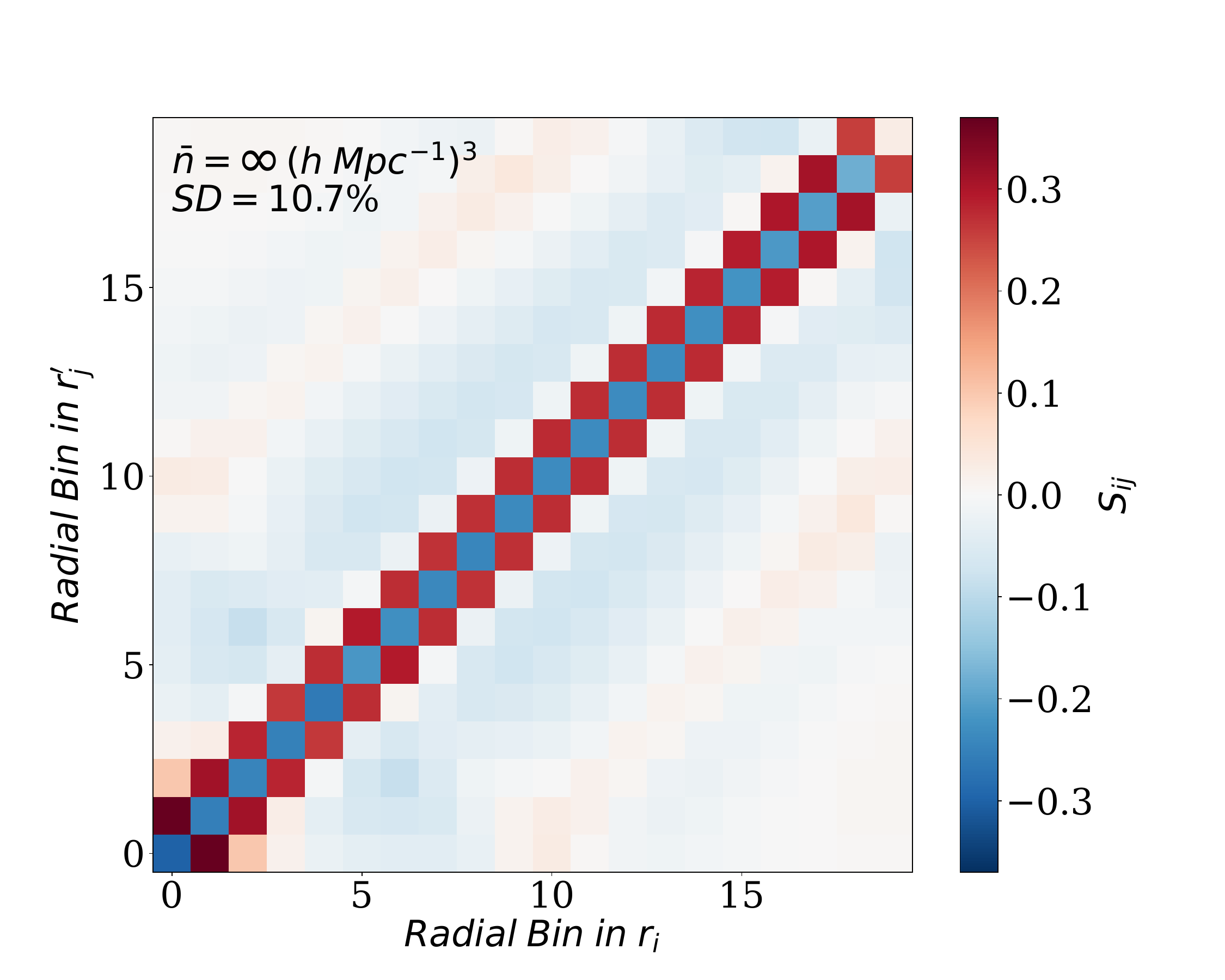}
    \end{subfigure}

    \caption{Similarly to \cref{fig:half_inv}, we display the half-inverse test ($\mathbf{S} \equiv \mathbf{C}_{\mathrm{model}}^{-1/2}\mathbf{C}_{\mathrm{true}}\mathbf{C}_{\mathrm{model}}^{-1/2} - \mathbf{\mathds{1}}$) of the covariance of the 2PCF for various number densities. $\mathbf{C}_{\mathrm{model}}$ is computed from the model power spectrum while $\mathbf{C}_{\mathrm{true}}$ is from the \textsc{camb} power spectrum. Here, we limit both power spectra to $k \geq \SI{0.2}{\hHubble\per\Mpc}$, which is the region over which they agree. If the two covariance matrices matched exactly, the half-inverse test would result in the null matrix. The SD displayed in each subplot quantifies the average deviation of the half-inverse test from the null matrix. As in \cref{fig:half_inv}, the model and true covariance matrices show better agreement for lower number densities.} 
    \label{fig:half_inv_trunc}
\end{figure}

In \cref{sec:pk}, we used the half-inverse test to quantify the similarity between the covariance of the 2PCF from the model power spectrum and from the \textsc{camb} power spectrum (\cref{fig:half_inv}). This was done over the range $k = \SIrange{e-3}{10}{\hHubble\per\Mpc}$; however, in \cref{fig:pk_vs_k}, the two power spectra are shown to deviate at low $k$. To examine the effect of this deviation on the inverse covariance, we computed the half-inverse tests using the region over which the two power spectra agree: $k \geq \SI{0.2}{\hHubble\per\Mpc}$. These truncated half-inverse tests are shown in \cref{fig:half_inv_trunc}. Their corresponding SD are given in \cref{Tab:rms}.

\begin{figure}
    \centering
    \vspace{0cm}
    \begin{subfigure}[b]{0.5\textwidth}
        \centering
        \includegraphics[width=\textwidth]{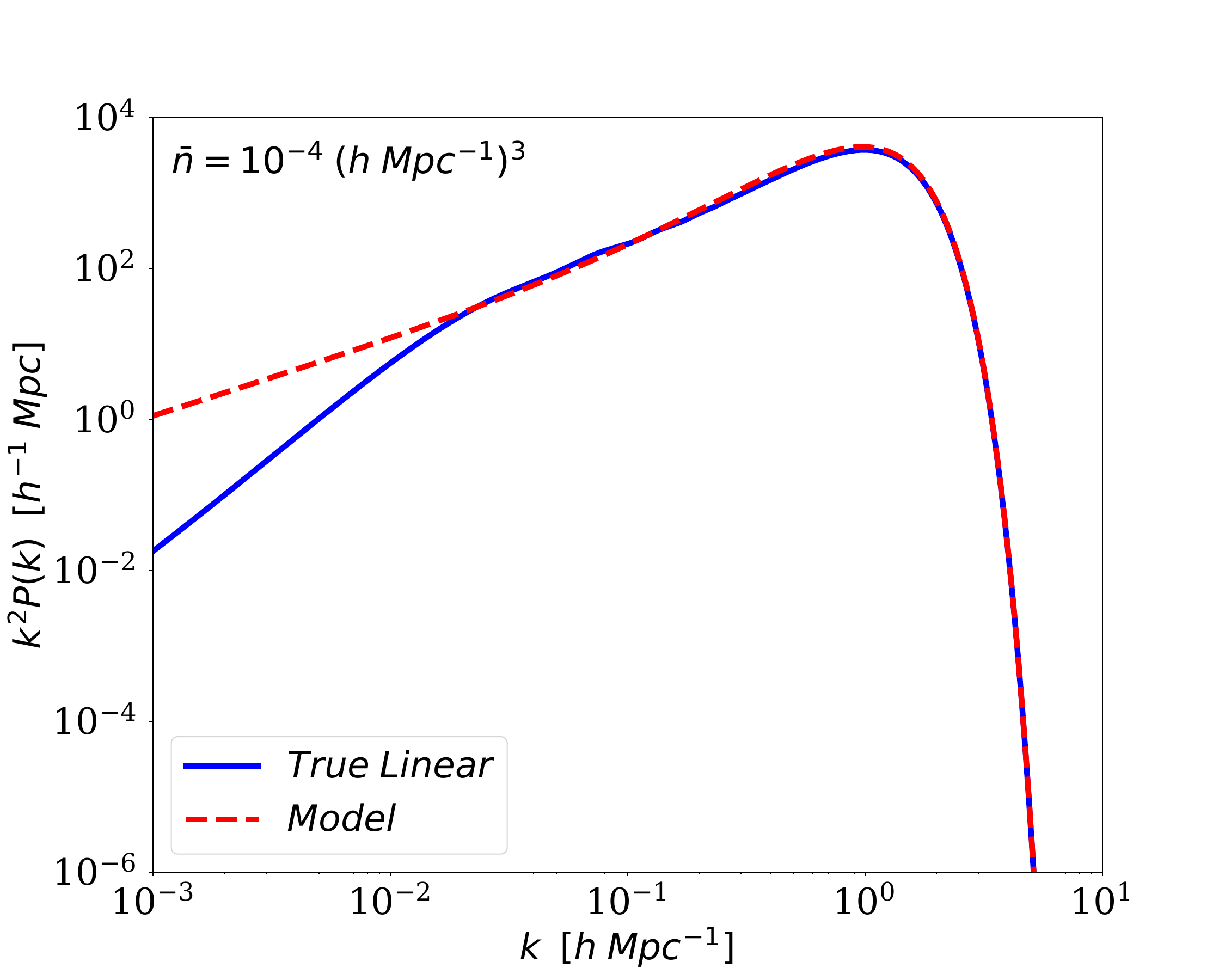}
    \end{subfigure}
    \hspace{-0.25cm}
    \begin{subfigure}[b]{0.5\textwidth}
        \centering
        \includegraphics[width=\textwidth]{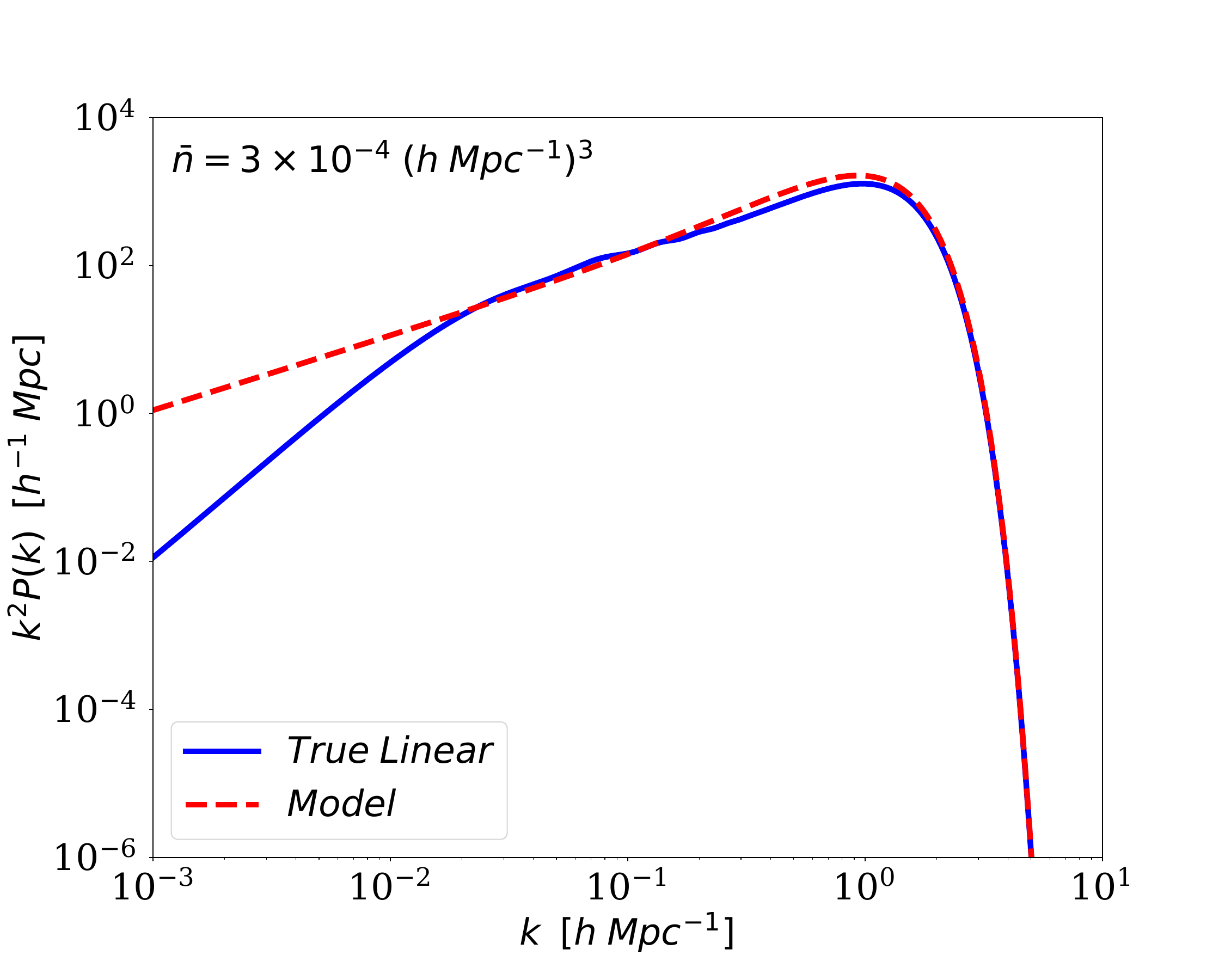}
    \end{subfigure}

    \begin{subfigure}[b]{0.5\textwidth}
        \centering
        \includegraphics[width=\textwidth]{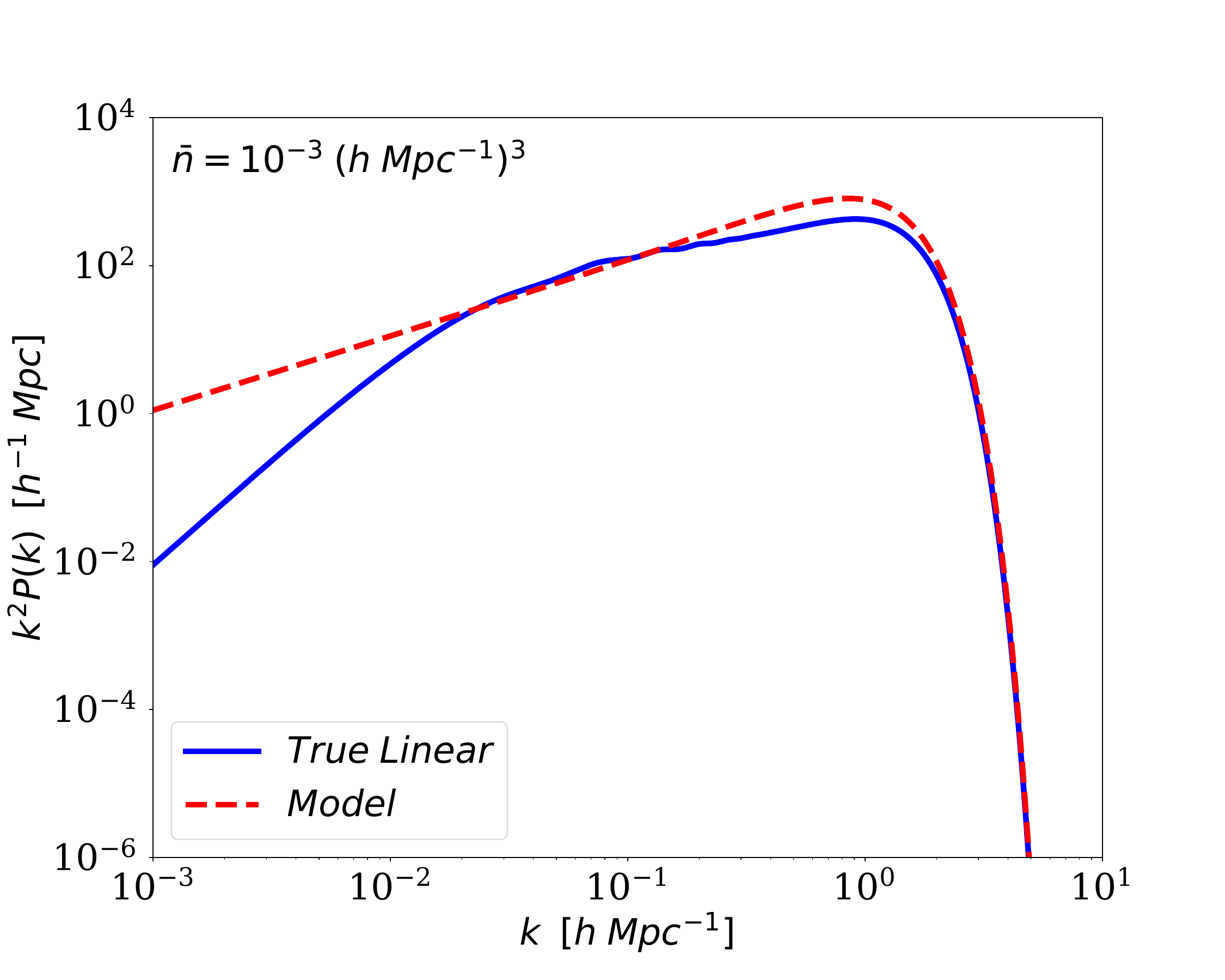}
    \end{subfigure}
    \hspace{-0.25cm}
    \begin{subfigure}[b]{0.5\textwidth}
        \centering
        \includegraphics[width=\textwidth]{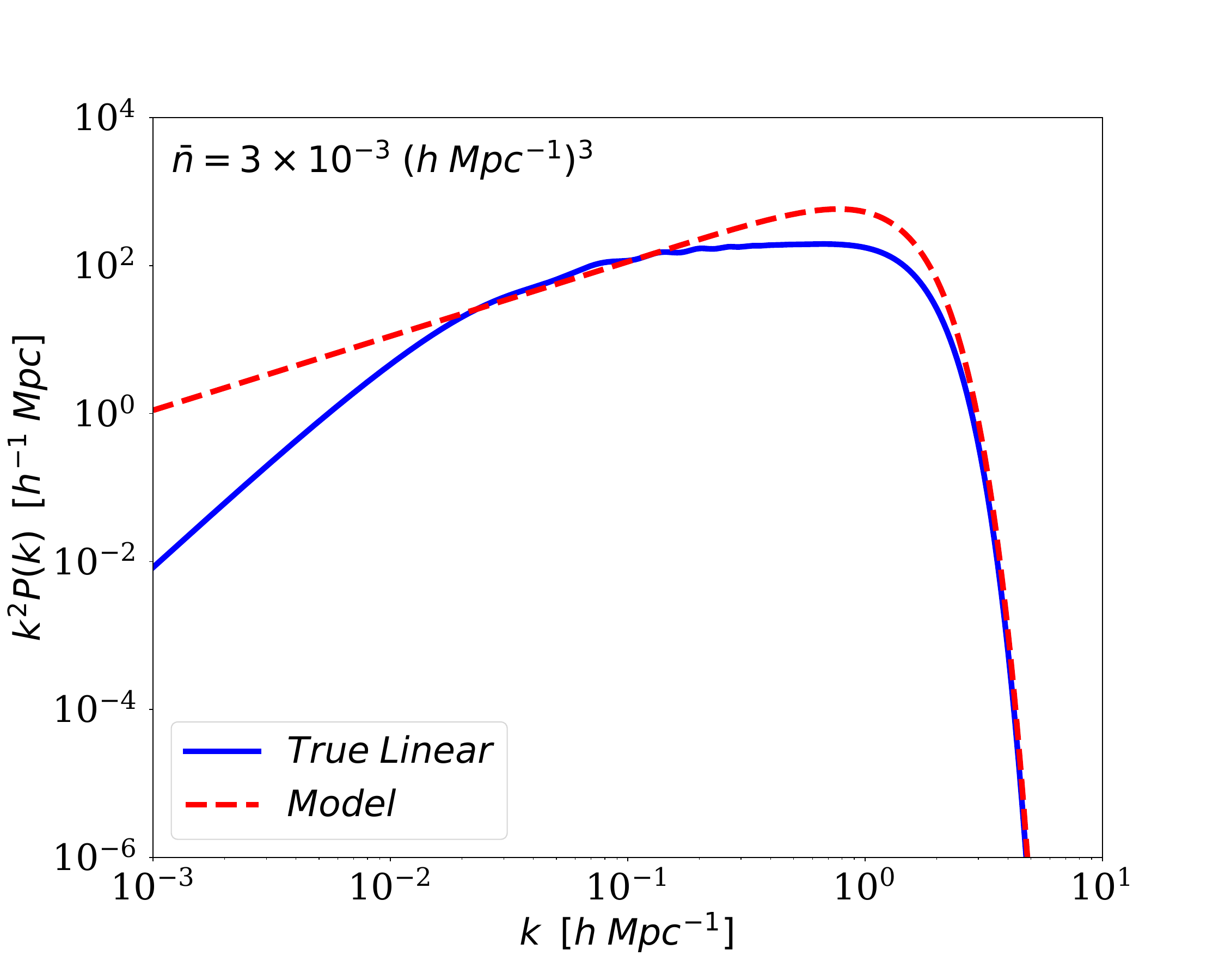}
    \end{subfigure}

    \begin{subfigure}[b]{0.5\textwidth}
        \centering
        \includegraphics[width=\textwidth]{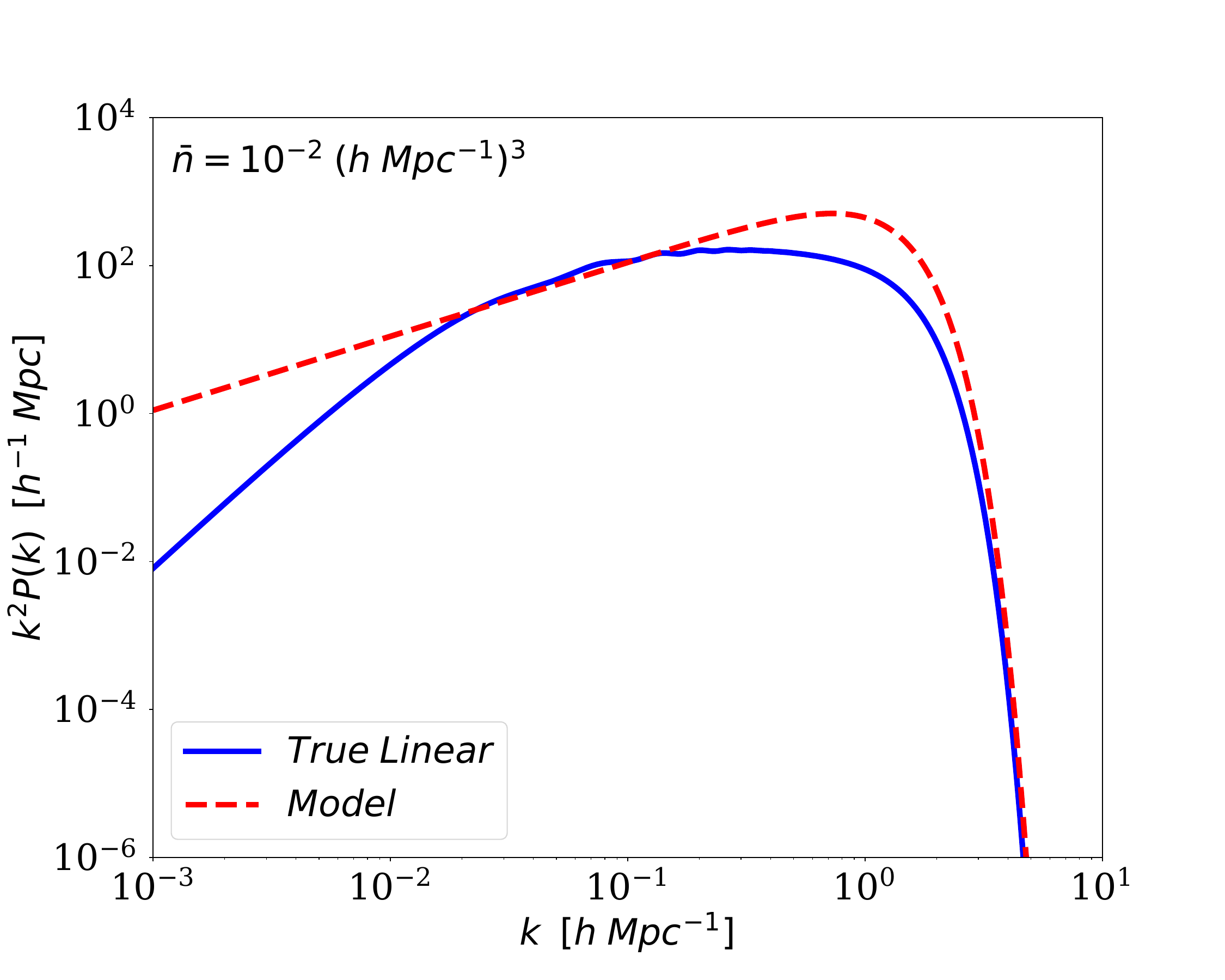}
    \end{subfigure}
    \hspace{-0.25cm}
    \begin{subfigure}[b]{0.5\textwidth}
        \centering
        \includegraphics[width=\textwidth]{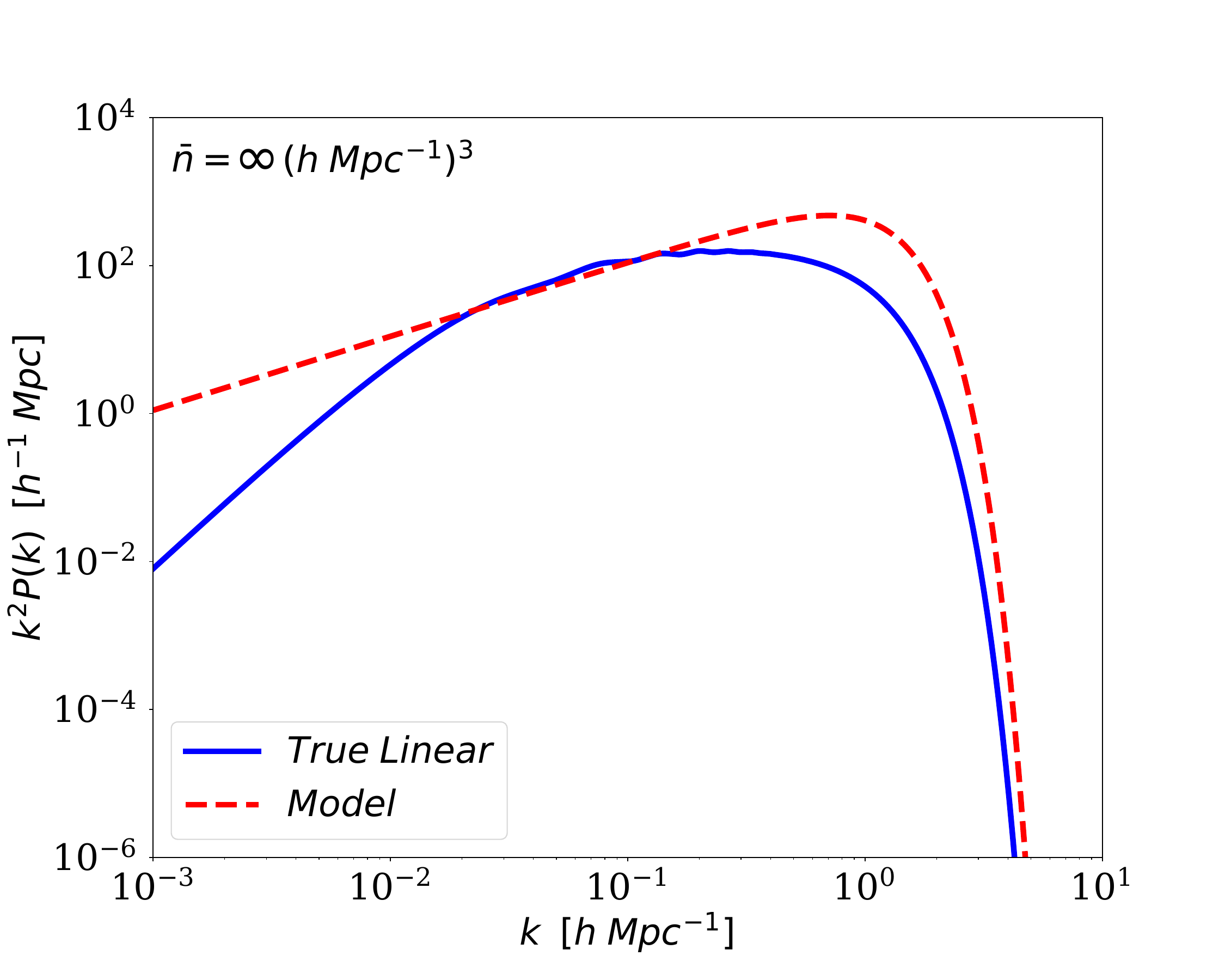}
    \end{subfigure}

    \caption{Here, we display the Jacobian-weighted power spectrum for various number densities, similarly to the \textit{lower left panel} of \cref{fig:pk_vs_k}. As number density increases, the deviation between the peaks of the model and true power spectra increases. This deviation is the leading contribution to the nonzero SD for the half-inverse test. When the power spectra are truncated to include only $k \geq \SI{0.2}{\hHubble\per\Mpc}$, the deviation in the peaks of the power spectra remains. Thus, truncating the power spectra to exclude low $k$ does not significantly improve the SD of the half-inverse test.} 
    \label{fig:square_weight_trunc}
\end{figure}

\Cref{Tab:rms} shows that limiting the power spectra to $k \geq \SI{0.2}{\hHubble\per\Mpc}$ does not significantly improve the inverse covariance. As number density increases, the improvement to the half-inverse SD gained from truncating the power spectrum decreases. This is because at greater number density, the physical part of the power spectrum is more dominant over the shot noise piece, which is exact. Therefore, as number density increases, we see a greater deviation between the model and true power spectra. This deviation is peaked near $k = \SI{1}{\hHubble\per\Mpc}$, as shown in \cref{fig:square_weight_trunc}, and contributes more to the half-inverse SD than the deviation at low $k$ that is removed with truncation.

We note that in each subplot shown, the amplitude of the model power spectrum is set to $A = \SI{277}{\per\hHubble\squared\Mpc\squared}$. This amplitude was chosen to minimize the deviation between the model and true power spectra for $\bar{n} = \SI{3e-4}{\hHubble\cubed\per\Mpc\cubed}$. In practice, the amplitude should be fine-tuned for each number density, rather than using the same amplitude for all.

\section{Integral of Gauss Hypergeometric Function Against Power of the Argument}
\label{integral_hyp_to_meijer}
In \cref{eqn:I_3lin_change_var}, the final two integrals were of a Gauss hypergeometric function multiplied by that function's argument raised to an integer power. Here we describe how such integrals may be evaluated.

We begin with the integral
\begin{align}
\label{eqn:starting_J}
    \mathcal{J}(a,b,c;p) &\equiv \int_{z_{\mathrm{min}}}^{z_{\mathrm{max}}} \dl z \;z^p\twoFone\mleft(a,b;c;z\mright).
\end{align}
For positive or zero $p$, this integral has a known result in terms of the hypergeometric function $\threeFtwo$ (as we show in \pcref{eqn:int_2f1_x}); however, here we will allow $p$ to also explore negative values. For the integral to converge, it is only required that none of the $\{a,b,c\}$ are equal to zero or a negative integer, and that $\{1-a,1-b,2-a+p,2-b+p\}$ are not positive integers. These conditions were satisfied in \cref{eqn:int_2f1_y}, so that we were justified in using the integral result we now derive in our covariance calculation.

The hypergeometric function of \cref{eqn:starting_J} can be rewritten as a Meijer $G$-function (\cite{GR} equation 9.34.8):
\begin{align}
\label{eqn:J_as_G2122}
    \mathcal{J}(a,b,c;p) &= \frac{\Gamma(c)}{\Gamma(a)\Gamma(b)}\int_{z_{\mathrm{min}}}^{z_{\mathrm{max}}}\dl z \;z^p\;\MeijerGTwoOneTwoTwo{-\frac{1}{z}}{1}{c}{a}{b}.
\end{align}
If any of $\{a,b,c\}$ are equal to zero or a negative integer, the corresponding $\Gamma$-function in \cref{eqn:J_as_G2122} will not be analytic. For $1 \leq k \leq n$ and $1 \leq j \leq m$, the Meijer $G$-function $G^{m,n}_{p,q}$ is defined when none of the possible values of $a_k - b_j$ are equal to a positive integer, using the common definition where the $a_k$ constitute the top row of parameters and the bottom row of parameters are given by $b_j$ (\cite{NIST} equation 16.17.1). The Meijer $G$-function in \cref{eqn:J_as_G2122} thus requires that $\{1-a,1-b\}$ are not positive integers.

We now use the change of variable $x = -1/z$ to rewrite \cref{eqn:J_as_G2122} so that the integration is over the variable constituting the argument of the Meijer $G$-function:
\begin{align}
\label{eqn:J_power_and_G}
    \mathcal{J}(a,b,c;p) &= (-1)^p\frac{\Gamma(c)}{\Gamma(a)\Gamma(b)}\int_{-1/{z_{\mathrm{min}}}}^{-1/z_{\mathrm{max}}}\dl x\;x^{-(p+2)}\;\MeijerGTwoOneTwoTwo{x}{1}{c}{a}{b}.
\end{align}

A Meijer $G$-function multiplied by a power of its argument may be rewritten as a Meijer $G$-function with parameters related to the power (\cite{GR} equation 9.31.5). We use this to rewrite \cref{eqn:J_power_and_G} as an integral over a Meijer $G$-function with no power law in front of it:
\begin{align}
\label{eqn:J_G_int}
    \mathcal{J}(a,b,c;p) &= (-1)^p\frac{\Gamma(c)}{\Gamma(a)\Gamma(b)}\int_{-1/{z_{\mathrm{min}}}}^{-1/z_{\mathrm{max}}}\dl x\;\MeijerGTwoOneTwoTwo{x}{-(p+1)}{c-(p+2)}{a-(p+2)}{b-(p+2)}.
\end{align}
As in \cref{eqn:J_as_G2122}, the Meijer $G$-function in \cref{eqn:J_G_int} requires that $\{1-a,1-b\}$ are not positive integers.

We may now perform the Meijer $G$-integral:\footnote{\url{https://functions.wolfram.com/HypergeometricFunctions/MeijerG/21/01/01/}}
\begin{align}
\label{eqn:J_eval}
    \mathcal{J}(a,b,c;p) &= (-1)^p\frac{\Gamma(c)}{\Gamma(a)\Gamma(b)}\;\MeijerGTwoTwoThreeThree{x}{1}{-p}{c-(p+1)}{a-(p+1)}{b-(p+1)}{0}\Bigg|_{-1/z_{\mathrm{min}}}^{-1/z_{\mathrm{max}}}.
\end{align}
The Meijer $G$-function in \cref{eqn:J_eval} imposes two new constraints in addition to those previously stated: $\{2-a+p,2-b+p\}$ may not be positive integers. 

In summary, the constraints that must be satisfied to use \cref{eqn:J_eval} as the solution to the integral of \cref{eqn:starting_J} are as follows: $\{a,b,c\}$ may not be equal to zero or a negative integer, and $\{1-a,1-b,2-a+p,2-b+p\}$ may not be a positive integer. \Cref{eqn:J_eval} was used to evaluate the final two integrals of \cref{eqn:I_3lin_change_var}, where all of the stated conditions are met.

\acknowledgments

We thank all members of the Slepian research group for many useful discussions. We especially thank Robert Cahn and Jiamin Hou for additional helpful conversations regarding this work.

\bibliographystyle{JHEP}
\bibliography{main_jcap}

\end{document}